%%%%%%%%%%%%%%%%%       FORMATO

\magnification=\magstep1\hoffset=0.cm
\voffset=-1.5truecm\hsize=16.5truecm    \vsize=24.truecm
\baselineskip=14pt plus0.1pt minus0.1pt \parindent=12pt
\lineskip=4pt\lineskiplimit=0.1pt      \parskip=0.1pt plus1pt

\def\st{\scriptstyle}

%%%%%%%%%%%%%%%% GRECO

\let\a=\alpha \let\b=\beta  \let\d=\delta \let\e=\varepsilon
 \let\g=\gamma     \let\k=\kappa \let\l=\lambda
\let\m=\mu       
\let\r=\rho \let\s=\sigma  
  \let\z=\zeta
\let\D=\Delta  \let\G=\Gamma \let\L=\Lambda 
    \let\X=\Xi

%%%%%%%%%%%%%%%% EQUAZIONI CON NOMI SIMBOLICI
%%% per assegnare un nome simbolico ad una equazione basta
%%% scrivere \Eq(...) o, in \eqalignno, \eq(...) o,
%%% nelle appendici, \Eqa(...) o \eqa(...):
%%% dentro le parentesi e al posto dei ...
%%% si puo' scrivere qualsiasi commento; per avere i nomi
%%% simbolici segnati a sinistra delle formule si deve
%%% dichiarare il documento come bozza, iniziando il testo con
%%% \BOZZA. Sinonimi \Eq,\EQ.
%%% All' inizio di ogni paragrafo si devono definire il
%%% numero del paragrafo e della prima formula dichiarando
%%% \numsec=... \numfor=... (brevetto Eckmannn).

\global\newcount\numsec\global\newcount\numfor
\gdef\profonditastruttura{\dp\strutbox}
\def\senondefinito#1{\expandafter\ifx\csname#1\endcsname\relax}
\def\SIA #1,#2,#3 {\senondefinito{#1#2}
\expandafter\xdef\csname #1#2\endcsname{#3} \else
\write16{???? il simbolo #2 e' gia' stato definito !!!!} \fi}
\def\etichetta(#1){(\veroparagrafo.\veraformula)
\SIA e,#1,(\veroparagrafo.\veraformula)
 \global\advance\numfor by 1
% \write15{@def@equ(#1){\equ(#1)} \%:: ha simbolo= #1 }
 \write16{ EQ \equ(#1) ha simbolo #1 }}
\def\etichettaa(#1){(A\veroparagrafo.\veraformula)
 \SIA e,#1,(A\veroparagrafo.\veraformula)
 \global\advance\numfor by 1\write16{ EQ \equ(#1) ha simbolo #1 }}
\def\BOZZA{\def\alato(##1){
 {\vtop to \profonditastruttura{\baselineskip
 \profonditastruttura\vss
 \rlap{\kern-\hsize\kern-1.2truecm{$\scriptstyle##1$}}}}}}
\def\alato(#1){}
\def\veroparagrafo{\number\numsec}\def\veraformula{\number\numfor}
\def\Eq(#1){\eqno{\etichetta(#1)\alato(#1)}}
\def\eq(#1){\etichetta(#1)\alato(#1)}
\def\Eqa(#1){\eqno{\etichettaa(#1)\alato(#1)}}
\def\eqa(#1){\etichettaa(#1)\alato(#1)}
\def\equ(#1){\senondefinito{e#1}$\clubsuit$#1\else\csname e#1\endcsname\fi}

%%%%%%%%%%%%%%%% GRAFICA

\def\bb{\hbox{\vrule height0.4pt width0.4pt depth0.pt}}\newdimen\u
\def\pp #1 #2 {\rlap{\kern#1\u\raise#2\u\bb}}

\def\ins #1 #2 #3 {\rlap{\kern#1\u\raise#2\u\hbox{$#3$}}}

\def\pallina{{\kern-0.4mm\raise-0.02cm\hbox{$\scriptscriptstyle\bullet$}}}
\def\palla{{\kern-0.6mm\raise-0.04cm\hbox{$\scriptstyle\bullet$}}}
\def\pallona{{\kern-0.7mm\raise-0.06cm\hbox{$\displaystyle\bullet$}}}

%%%%%%%%%%%%%%%% PIE PAGINA

\def\data{\number\day/\ifcase\month\or gennaio \or febbraio \or marzo \or
aprile \or maggio \or giugno \or luglio \or agosto \or settembre
\or ottobre \or novembre \or dicembre \fi/\number\year}

\setbox200\hbox{$\scriptscriptstyle \data $}

\newcount\pgn \pgn=1
\def\foglio{\number\numsec:\number\pgn
\global\advance\pgn by 1}
\def\foglioa{a\number\numsec:\number\pgn
\global\advance\pgn by 1}

\footline={\rlap{\hbox{\copy200}\ $\st[\number\pageno]$}\hss\tenrm \foglio\hss}

%%%%%%%%%%%%%%% DEFINIZIONI LOCALI

\def\sqr#1#2{{\vcenter{\vbox{\hrule height.#2pt
\hbox{\vrule width.#2pt height#1pt \kern#1pt
\vrule width.#2pt}\hrule height.#2pt}}}}
\def\square{\mathchoice\sqr34\sqr34\sqr{2.1}3\sqr{1.5}3}

\let\ciao=\bye 
\def\pagina{{\vfill\eject}} \def\\{\noindent}

  \let\io=\infty 

\let\dpr=\partial

\def\tende#1{\vtop{\ialign{##\crcr\rightarrowfill\crcr
              \noalign{\kern-1pt\nointerlineskip}
              \hskip3.pt${\scriptstyle #1}$\hskip3.pt\crcr}}}
\def\otto{{\kern-1.truept\leftarrow\kern-5.truept\to\kern-1.truept}}
\def\arm{{}}
\font\bigfnt=cmbx10 scaled\magstep1

\BOZZA
\vglue2.truecm
{\centerline{\bigfnt Renormalization group approach to 
interacting}}
{\centerline{\bigfnt polymerised 
manifolds}}
\vglue1.5truecm
{\centerline{ P. K. Mitter }}
{\centerline{Laboratoire de Physique math\'ematique\footnote{${}^*$}
{\arm Laboratoire Associ\'e au CNRS, UMR 5825}}}
{\centerline{Universit\'e Montpellier 2}}
{\centerline{Place E. Bataillon, Case 070}}
{\centerline{34095 Montpellier Cedex 05 France}}
\vglue1.5truecm
{\centerline{ B. Scoppola}}
{\centerline{ Dipartimento di Matematica\footnote{${}^{**}$}
{\arm Partially supported by CNR, G.N.F.M. and Short Term Mobility Program}}}
{\centerline{  Universit\'a ``La Sapienza'' di Roma}}
{\centerline{ Piazzale A. Moro 2}}
{\centerline{00185 Roma}}

\pagina
\vglue2.truecm

{\centerline{ ABSTRACT}}
\vglue1.truecm
\\We propose to study the infrared behaviour of
polymerised (or tethered) random manifolds of
dimension $D$ interacting via an exclusion condition
with a fixed impurity in $d$-dimensional Euclidean
space in which the manifold is embedded. In this paper
we take $D=1$, but modify the underlying free Gaussian
covariance (thereby changing the canonical
scaling dimension of the Gaussian random field) so as
to simulate a polymerised manifold with fractional dimension
$D:\ 1<D<2$. We prove rigorously, via methods of Wilson's
renormalization group, the convergence to a non Gaussian fixed point 
for $\e>0$, sufficiently small. Here, $\e=1-\b{d\over 2}$, where
$-\b/2$ is the canonical scaling dimension of the Gaussian embedding 
field. Although $\e$ is small, our analysis is non-perturbative
in $\e$. A similar model was studied earlier [CM] in the
hierarchical approximation.
\pagina

\vglue1.5truecm

\\{\centerline{\it TABLE OF CONTENTS}}
\vglue.5truecm

\\\S0: Introduction [4-7]
\vglue.3truecm

\\\S1: The model, polymer gas representation and RG tranformations. [8-18]
\vglue.3truecm

\\\S2: Polymer activity norms and basic Lemmas. [19-34]
\vglue.3truecm

\\\S3: Estimates for a generic RG step. Remainder estimates. [35-43]
\vglue.3truecm

\\\S4: RG step to second order. Relevant and irrelevant terms. Estimates.
[44-60]
\vglue.3truecm

\\\S5: RG action on the remainder. Relevant and irrelevant terms. Estimates.
[61-73]
\vglue.3truecm

\\\S6: Extraction estimates. [74-80]
\vglue.3truecm

\\\S7: Convergence to a non Gaussian fixed point. [81-87]
\vglue.5truecm
\\Appendix: Stability of the large field regulator. [88-92]
\pagina
\vglue1.5truecm\numsec=0\numfor=1\pgn=1
{\hfill ``God is in the details'': F. Dyson, Disturbing the Universe}
\vglue1.5truecm
\\\S0: {\it Introduction}
\vglue.5truecm
\\Consider the Euclidean action

$$S(\phi)={1\over 2}\int d^Dx(\phi(x),-\D\phi(x))+
g\int d^Dx\d^{(d)}(\phi(x))\Eq(0.1)$$

\\where $\phi$ has values in ${\bf R}^d$, $(\ ,\ )$ is scalar
product in ${\bf R}^d$, and the corresponding formal partition
function:

$$Z=\int d\m_C(\phi)e^{-g\int d^Dx\d^{(d)}(\phi(x))}\Eq(0.2)$$

\\$\m_C$ is a Gaussian measure with covariance $C=(-\D)^{-1}$.

\\$\phi$ can be considered as the embedding function of a $D$
dimensional tethered or polymerised manifold (for tethered
manifolds see [NPW]) in $d$-dimensional Euclidean space, and
for $g>0$, we have a repulsive interaction with a fixed
impurity in the embedding space. It is easy to see that
the coupling constant $g$ has canonical (engineering) dimension

$$[g]=\e=D-(2-D){d\over 2}\Eq(0.3)$$

\\$\phi$ has canonical scaling dimension: $[\phi]=-(2-D)/2$
and the upper critical dimension of the embedding space is

$$d_c={2D\over 2-D}\Eq(0.4)$$

\\Such a model was studied in [DDG1,2] (the model
was first considered in [D2]). It was shown in [DDG1,2]
for $1<D<2$ and $\e>0$ sufficiently small, that there exists
an $\e$-expansion in renormalised perturbation series,
and that the infrared behaviour is governed by a non-Gaussian
fixed point. The model with an ultraviolet cutoff was reconsidered in
[CM] in the hierarchical approximation to Wilson's Renormalization
group (henceforth called RG). It was shown, under the same conditions,
that the iteration of the hierarchical RG transformations converge
to a non-Gaussian fixed point independent of the $\e$-expansion.

\\In the present paper we consider a version of the above model
and study the iterations of the exact RG with 
an ultraviolet cutoff in a finite but
large volume eventually tending to infinity.
The precise definition of the model and the RG iterations
will be found in section 1.1. We shall simulate a fractional dimension $D$,
$1<D<2$, by choosing $D=1$ and modifying the covariance $C$ to be

$$C=(-\D)^{(\d-1)}F(-\D)\Eq(0.5)$$

\\with $0<\d<1/2$. Here $F$ is an ultraviolet cutoff function which is
positive, and in the momentum space $F(p^2)$ is of fast decrease.

\\The covariance $C$ is so chosen (see section 1.1) that $C(x-y)$
is smooth and of compact support. In finite volume, the zero 
momentum mode $p=0$ is automatically taken care of
(for details see later). The delta function interaction is replaced
(see [CM]) by its regularised version in finite volume:

$$V(\phi)=\int_{\L} dx\ v(\phi(x))\Eq(0.6)$$

\\with

$$v(\phi(x))=\left({\l\over 2\pi}\right)^{d\over 
2}e^{-{\l\over 2}\vert\phi (x)\vert^{2}}\Eq(0.7)$$

\\where $|\cdot |$ is the norm in ${\bf R}^d$.

\\It is easy to see that the canonical scaling dimension of the field
is

$$[\phi]=-\b/2\Eq(0.7.1)$$

\\where

$$\b=1-2\d>0\Eq(0.8)$$

\\since $0<\d<1/2$ by choice.

\\The coupling constant $g$ has dimension

$$[g]=\e=1-\b{d\over 2}\Eq(0.9)$$

\\The upper critical dimension is

$$d_c= {2\over\b}$$

\\Note that the canonical scaling of the field is such that it simulates
a Gaussian random field with covariance $(-\D)^{-1}$ in dimension 
$D=1+2\d$. We have $1<D<2$, and for $\d$ close to 1/2, $D$ is close to 
2. The idea of simulating fractional dimensions 
by changing the covariance is not new. Even in a rigorous
framework, it figures for
example in [GK] and in the recent paper of [BDH3].

\bigskip
\\We shall hold $\b> 0$ and very small and $d$ very large       
so that $\e>0$ is sufficiently small. Clearly in this configuration the
critical embedding dimension $d_c$ is very large and $d$ close to $d_c$
from below. Our main result
is that {\it the exact RG iteractions converge to a non-Gaussian
fixed point close to the unstable 
Gaussian fixed point}. The precise statement is to be found in Theorem 7.5
of section 7. We will sketch here the various steps of the proof.

\\We show  that the second order flow of the RG is under control, 
and it gives an approximate  non trivial fixed point. 
We then prove that the remainder is also under
control and that it gives a negligible correction to
the second order RG flow. Next we prove that there exists
an invariant small neighborhood of the approximate fixed point.
The renormalisation group transformations are contractive in
this domain and this permits us to prove that there exists a 
true attractive non-trivial fixed point of the exact RG.

\\Here negligible is something rigorous, i.e. we
bound at every scale the remainder to second order perturbation
theory, and
we show that a suitable function of the coupling
constant and the fields (i.e. the polymer activity, defined later)
evolves under the action of RG in an analytic way, and it gives
a relevant contribution to the flow of the coupling which is 
under control in the same sense. The partition function
density with respect to the gaussian measure is competely
parametrised by the couple (coupling constant, Polymer activity)
and this couple converges to a non-trivial fixed point.

\\Note that a direct control of the perturbative series
is difficult due to
the fact that some non trivial cancellations occur,
and expanding naively in series of $g$ such cancellations 
are difficult to exploit.

\\The formulation of RG iterations
in terms of a polymer gas representation, as well as the method of
analysis employed in this paper, have been much influenced by
the original paper of Brydges-Yau [BY], the Lausanne lectures
of Brydges [B, Laus.], together with developments due to Brydges,
Dimock and Hurd [BDH1,2,3]. 
This technique has the
advantage, with respect to the naive perturbative expansion,
that the polymer activities as functions
of the coupling are not expanded in series
when the expansion is unnecessary. 

\\In this paper, however, there are definite simplifications and differences
with respect to [BY], [BDH1,2,3]. The
simplifications stem from the use of ``compact covariances'', an idea suggested
to us by David Brydges. This enables us to dispense with cluster or
Mayer expansions. All polymer activities appearing in this paper are
based on connected polymers. We can dispense with analyticity
norms. As to the differences, they stem from the special form of 
the interaction. As a consequence we have that the growth of 
polymer activities is measured by a norm which employs a large
fields regulator quite specific to this problem. Relevant terms
are also extracted in a special way appropriate to this model.
These matters are explained in detail in the subsequent sections.

\\Some further remarks are in order. The reader may wonder why
we interpolated in the covariance starting with D=1 instead of 
D=2. The reason is technical and stems from the scaling properties
of the fields which for D=1 permits us to exploit with advantage
the simple large field regulator that we have devised for the
construction of norms in which convergence is proved. For the case
D=2 the large field behaviour is not yet under control. This problem 
deserves more attention. Needless to say, this problem is not seen 
in perturbation theory.

\\Finally, we note that much progress has been made in the study
of self-avoiding polymerised manifolds via
perturbative $\epsilon$ expansions, see [DDG3,4], [DW1,2,3,4] 
and for earlier work [NPW], [KN], [D1], [DHK], [H].  

\pagina
\vglue1.5truecm\numsec=1\numfor=1\pgn=1
\\\S1: {\it  The model, polymer gas representation and RG tranformations.}
\vglue.5truecm
\\{\it 1.1. The model}
\vglue.3truecm
\\Let $\L_N$ be the closed interval $[-{L^{N+1}\over 2},{L^{N+1}\over 2}]$
of length ${L^{N+1}}$.
\\The model is
described by
the partition function 

$$Z_0(\L_N)=\int d\m_{{\bf C}_N}z_0(\L_N,\phi)\Eq(11.1)$$

where

$$z_0(\L_N,\phi)=e^{-V_0(\L_N,\phi)}\Eq(11.1.1)$$

$$V_0(\L_N,\phi )=g_{0}V_*(\L_N,\phi )=g_{0}\int_{\L_N} dx
\ v_*(\phi (x))=
g_{0}\int_{\L_N} dx\left({\l_{*}\over 2\pi}\right)^{d\over 
2}e^{-{\l_{*}\over 2}\vert\phi (x)\vert^{2}}\Eq(11.3)$$

\\where each $\phi\in {\bf R}^d$ and 
$\l_{*}>0$ will be fixed later (see below).

\\$\m_{{\bf C}_N}$ is a gaussian measure with mean 0 and
covariance
${\bf C}_N$. The components $\phi_j,\ 1\le j\le d$, are independent
gaussian random variables, and each component has covariance
$C_N$. $d\m_{{\bf C}_N}(\phi)=\otimes_{j=1}^dd\m_{{ C}_N}(\phi_j)$.
\vglue.3truecm
\\We now describe the covariance $C_N$.

\\Let $g(x)$ be a $C^{\io}$ function of compact support:

$$g(x)=0\quad\forall\ |x|\ge{1\over 2}\Eq(11.8.11)$$

Choose for definiteness

$$g(x)=\cases{e^{-{1\over 1-4x^2}}&for $|x|\le 1/2$\cr
0&elsewhere\cr}\Eq(11.8.2)$$

\\Define

$$u(x)=(g*g)(x)\Eq(11.8.1)$$

\\Then $u(x)$ is $C^{\io}$ and of compact support:

$$u(x)=0\quad\forall\ |x|\ge{1}\Eq(11.8.00)$$

\\Define

$$\G_{L}(x)=\int_{1}^{L}{dl\over l} l^{\b}u\left({x\over
{l}}\right)\Eq(11.8)$$

\\Note that $\G_{L}(x)$  is $C^{\io}$ and of compact support:

$$\G_{L}(x)=0\quad\forall\ |x|\ge L\Eq(11.7.3)$$

\\Finally we define the covariance $C_N$ by a truncated
multiscale decomposition:

$$C_N(x)=\sum_{j=0}^N L^{j\b}\G_L(x/L^j)\Eq(11.7)$$

\\with $\b=1-2\d>0$ as in the introduction.

\\From\equ(11.7) $C_N$ is $C^{\io}$ and of compact support:

$$C_{N}(x)=0\quad\forall |x|\ge L^{N+1}\Eq(11.7.3.1)$$

\\From \equ(11.8) we have

$$\G_{L}(x)=\int{dp\over 2\pi}e^{ipx}
\int_{1}^{L}{dl}\ l^{\b}{\hat u}(lp)\Eq(11.8.4)$$

\\${\hat u}$ is of fast decrease, since $u$ is $C^\io$ with compact support.
Moreover, since $u=g*g, {\hat u}=|{\hat g}^2|$, 
so that ${\hat\G_{L}}(p)\ge 0$.
It follows that $\G_{L}(x)$ defines a positive definite function:

$$\sum_{i,j=1}^{n}\G_{L}(x_{i} -x_{j})a_{i}{\bar a}_{j}\ge 0\quad \forall 
n,\ x_{1},\ldots,x_{n}\in {\bf R},\ a_{1},\ldots,a_{n}\in {\bf C}
\Eq(11.7.2)$$

\\From the definition of $C_N$ in \equ(11.7), using \equ(11.7.2) we have

$$\sum_{i,j=1}^{n}C_N(x_{i} -x_{j})a_{i}{\bar a}_{j}\ge 0\quad
\forall  n,\ x_{1},\ldots,x_{n}\in {\bf R},\ a_{1},\ldots,a_{n}\in {\bf C}
\Eq(11.7.1)$$

\\Thus $C_N(x)$ also defines a positive definite function.
The positive definiteness of  $C_N(x)$ together with its smoothness
implies that there exists a gaussian measure of mean zero and covariance
$C_N$ which we call $\mu_{C_N}$, realized on a Sobolev space $H_s(\L_N)$,
with $s>1/2+\s$ for any positive integer $\s$. The Sobolev embedding
theorem implies that the sample fields $\phi$ are $\s$-times
differentiable. For our purposes it is enough to fix $\s=2$.
\vglue.5truecm
\\There exists another formula for $C_N$, derived from its definition,
which is useful because it makes contact with the cutoff function
$F_N$: from \equ(11.8) it follows that

$$L^{j\b}\G_{L}(x/L^j)=\int_{L^j}^{L^{j+1}}{dl\over l} l^{\b}u\left({x\over
{l}}\right)\Eq(11.8.22)$$

\\Introducing this in  \equ(11.7) we get

$$C_N(x)=\int_{1}^{L^{N+1}}{dl\over l} l^{\b}u\left({x\over
{l}}\right)\Eq(11.8.33)$$

\\which we can rewrite as

$$C_N(x)=\int{dp\over 2\pi}e^{ipx}{F_N(p^2)\over p^{2(1-\d)}}\Eq(11.8.44)$$

\\where the UV cutoff function 
in finite volume $F_N$ has now the following form
$$F_N(p^2)=\int_{|p|}^{|p|L^{N+1}} dl \ l^\b {\hat u}(l)=
p^{2(1-\d)}\int_1^{L^{N+1}} dl\ l^\b {\hat u}(l |p|)\Eq(11.8.3)$$

\\By our choice of $g$ , ${\hat g}$ 
is an even function, and hence so is  ${\hat u}= | {\hat g}|^2$ . This 
justifies the replacement of $p$ by $p^2$ in the above formula

\\$F_N(p^2)$ is of fast decrease because ${\hat u}$ is of fast decrease.
We also see that in our finite volume covariance the zero mode at $p=0$
is automatically regularized

\\To complete the definition of the model we specify the constant $\l_*$
as

$$\l_*={\b\over u(0)}\Eq(11.8.66)$$

\\Note that  

$$\g=\G_L(0)={u(0)\over\b}(L^\b-1)\Eq(11.8.67)$$ 

\\we have

$$\l_*={L^\b-1\over\g}\Eq(11.8.77)$$

\\{\it Remark}: if we had started with an arbitrary $\l$ in the
interaction \equ(11.3), and we had performed a Renormalization
Group transformation 
(defined below), then we would have obtained in the absence of 
quantum corrections:

$$\l'={L^\b\l\over 1+\g\l}$$
Since $\b>0$, $\l_*$ is an attractive fixed point. Choosing
$\l=\l_*$ from the beginning is an updating of the Renormalization
Group trajectory
which simplifies the subsequent analysis.
\vglue.3truecm
\\From now on, a bound of the form $O(1)$ will mean a bound independent 
of $L$.

\\We have the following bound on the derivatives of $\G_L(x)$:

\vglue.5truecm
\\LEMMA 1.1.1
\vglue.3truecm
\\For $0\le \b\le 1/4$ and all $k\ge 1$ we have

$$\sup_x|\dpr^k\G_L(x)|\le O(1)\Eq(11.A)$$

$$\int_{\bf R}dx\ |\dpr^k\G_L(x)|^2\le O(1)\Eq(11.B)$$

\vglue.3truecm
\\{\it Proof}

\\The proof of \equ(11.A) 
follows directly from \equ(11.8) and from the fact that
$$\sup_x|\dpr^ku(x)|\le O(1)$$
One has, taking $k$ 
derivatives
$$\sup_x|\dpr^k\G_L(x)|\le O(1){1-L^{\b-k}\over k-\b}\le O(1)$$

\\To prove \equ(11.B) we have

$$\int_{\bf R}dx\ |\dpr^k\G_L(x)|^2=2\int_1^L{dl_1\over l_1}l_1^{\b-k}
\int_1^{l_1}{dl_2\over l_2}l_2^{\b-k}
\int_{-l_2}^{l_2}dx\ (\dpr^ku)(x/l_1)(\dpr^ku)(x/l_2)$$

\\Using \equ(11.A) the last integral can be bounded by $O(1)l_2$.
Therefore we have

$$\int_{\bf R}dx\ |\dpr^k\G_L(x)|^2\le O(1)\int_1^L{dl_1\over l_1}l_1^{\b-k}
\int_1^{l_1}{dl_2\over l_2}l_2^{\b-k}l_2$$

\\and it is easy to see by direct computation that the integrals
are bounded again by $O(1)$.
$$\eqno Q.E.D.$$

\\From now on we drop for simplicity the suffix $L$ from $\G$.

\\It is also natural to define the rescaled propagator

$${\cal R}\G(y)={L^{-\b}}\G(yL)\Eq(11.8.88)$$

\\We will also use sometimes the following notation

$${\bar\G}(y)=\l_*{L^{-\b}}\G(yL)=\l_*{\cal R}\G(y)\Eq(11.8.89)$$

\\Note that

$${\bar\G}(0)=1-L^{-\b}\Eq(11.8.99)$$

\vglue.5truecm
\\{\it 1.2. Renormalization Group transformation}
\vglue.3truecm

\\It is easy to see from \equ(11.7) that

$$C_N(x)=\G(x)+L^{\b}C_{N-1}(x/L)\Eq(11.9)$$

\\Define the rescaled field ${\cal R}\phi(x)$

$${\cal R}\phi(x)=L^{\b/2}\phi(x/L)\Eq(11.9.1)$$

\\From \equ(11.9) and \equ(11.9.1) we see that we can write

$$\int d\m_{{\bf C}_N}(\phi)z_0(\L_N,\phi)=
\int d\m_{{\bf C}_{N-1}}(\phi)z_1(\L_{N-1},\phi)\Eq(11.1.01)$$

\\where

$$z_1(\L_{N-1},\phi)=
\int d\m_{\G}(\z)z_0(\L_N,\z+{\cal R}\phi)
\Eq(11.1.02)$$

\\This constitutes our Renormalization Group (RG) transformation,
which can be iterated.

\\After $n$ steps $(0\le n\le N)$ we get

$$\int d\m_{{\bf C}_N}(\phi)z_0(\L_N,\phi)=
\int d\m_{{\bf C}_{N-n}}(\phi)z_n(\L_{N-n},\phi)\Eq(11.1.03)$$

\\where

$$z_n(\L_{N-n},\phi)=
\int d\m_{\G_L}(\z)z_{n-1}(\L_{N-n+1},\z+{\cal R}\phi)
\Eq(11.1.04)$$

\\After $N$ steps we get $z_n(\L_{0},\phi)$ and measure
$\m_{{\bf C}_{0}}(\phi)$. 
Note that $\L_0$ is the closed interval $[-L/2,L/2]$ and $C_0=\G$.
Then we want to pass to the $N\rightarrow\io$
limit.
We would like to study the convergence of the iteration \equ(11.1.04).
This is awkward because the volume is changing with 
the iterations (see \equ(11.1.03)). We can take $N$, $n$, $N-n$ very large.
Then the RG iterations can be viewed as those of a fixed map. The precise
sense of the convergence of these iterations will be explained later.

\vglue.3truecm
\\{\it 1.3. The master formula for fluctuation integrals.}
\vglue.3truecm

\\By a simple gaussian integration it is easy to see that,
choosing suitably the constants $\b$ and $d$,
the direction defined by the initial interaction
\equ(11.3) is relevant under the action of RG: with our
definition of $\l_*$ we have that

$$\int d\m_\G(\z) v_*(\z(x)+{\cal R}\phi(x))=
L^{-\a}v_*(L^{-\b/2}{\cal R}\phi(x))$$ 

\\with $\a=\b d/2$ chosen in such a way that $\a<1$.
Changing variables $x'=x/L$ we get

$$L^{-\a}\int dx\ v_*(L^{-\b/2}{\cal R}\phi(x))=
L^\e\int dx\ v_*(\phi(x))$$

\\with $\e=1-\a>0$.

\\We will see later that this direction is actually the only
relevant one. In order to prove this it is often
useful to define a modified fluctuation integration such that the result 
of this integration is
already multiplied by a factor 
$L^{-\a}v_*(L^{-\b/2}{\cal R}\phi(x))$. This is
given by the following equality 
\vglue.5truecm
\\LEMMA 1.3.1 ({\it Master formula})
\vglue.3truecm

$$\int d\m_\G(\z) v_*(\z(x)+{\cal R}\phi(x))F(\z+{\cal R}\phi)=$$
$$=L^{-\a}v_*(L^{-\b/2}{\cal R}\phi(x))\int 
d\m_{\Sigma^x}(\z) F(\z+L^{-\b}T^x{\cal R}\phi)
\Eq(11.13)$$

\\where

$$\Sigma^x(y-z)=\G(y-z)-\l_*\G(x-z)\G(x-y)\Eq(11.14)$$

$$T^{{ x}}{\cal R}\phi(y)=L^{\b}{\cal R}\phi(y)-
\l_*\G(y-{x}){\cal R}\phi({x})$$ 

\\Note that
by trivial algebraic manipulation,
and by
$L^{\b}-1=\l_*\G(0)$ we obtain

$$L^{-\b}T^{{x}}{\cal R}\phi(y)= [{\cal R}\phi(y)-{\cal R}\phi({
x})]+L^{-\b}(1+\l_*(\G(0)-\G(x-y))) {\cal R}\phi({ x})\Eq(11.15)$$ 

$$L^{-\b}T^{{x}}{\cal R}\phi(x)= L^{-\b} {\cal R}\phi({x})\Eq(11.16)$$ 

\\The proof of Lemma 1.3 is obtained simply by gaussian integration.
\vglue.3truecm
\\{\it 1.4. Polymer expansion}, [BY].
\vglue.3truecm
\\The polymerised version of the partition function 
\equ(11.1) is obtained by expressing the
volume $\Lambda$ as a union of closed blocks $\D$ of size 1;
since $\L$ is one-dimensional, the blocks are actually
closed unit intervals.

\\We define first of all a {\it polymer} $X$ as a union of
blocks $\D$.
A {\it cell} may be the interior of a block, 
i.e. an open block, or a point of its boundary.

\\Then we define a commutative product, denoted $\circ$,
on functions of sets containing
polymers and cells, in the following way

$$(F_1\circ F_2)(X)=\sum_{Y,Z:Y{\circ\atop\cup}Z=X}F_1(Y)F_2(Z)
\Eq(1.16.1)$$

\\where $X=Y{\circ\atop\cup}Z$ iff $X=Y\cup Z$ and $Y\cap Z=\emptyset$.
The $\circ$ identity ${\cal I}$ is defined by

$${\cal I}=\cases{1&if $X=\emptyset$\cr 0&otherwise\cr}\Eq(1.16.2)$$

\\The ${\cal E}$xponential is defined by

$${\cal E}xp(K)={\cal I}+K+K\circ K/2!+...\Eq(1.16.3)$$

This is the usual series for an exponential except that the product
has been replaced by the $\circ$ product. The ${\cal E}$xponential
with the $\circ$ product satisfies the usual properties of an exponential. 

\\Moreover we define a {\it space filling} function $\square$
as

$${\square}=\cases{1&if $X$ is a cell\cr 0&otherwise\cr}
\Eq(1.16.4)$$

\\Finally we denote, for $X$ polymer or cell

$$V_0(X,\phi)=
g_0\int_X d^Dxv_*(\phi (x))\Eq(11.17)$$ 

\\With these notations it is clear that, since

$${\cal E}xp(\square)(X)=1\Eq(11.17.1)$$

one has

$$Z_0(\L_N)=\int d\m_{{\bf C}_N}e^{-V_0(\phi,\L_N)}
{\cal E}xp(\square )(\L_N)\Eq(11.18)$$ 

\\In what follows it will be understood that in an expression
of the form

$$ e^{-V(\L)}{\cal E}xp(\square+K)(\L)\Eq(11.19)$$
or
$${\cal E}xp(\square +{\hat K})(\L)\Eq(11.20)$$

\\the functions $K,{\hat K}$,  called {\it activities}, are
supported only on polymers and vanish on cells.
In particular an expression like \equ(11.20) is often called
{\it polymer gas}, because it can be written in the form

$${\cal E}xp(\square +{\hat K})(\L)=
\sum_{N=0}^\io (1/N!)\sum_{X_1,...,X_N}{\hat K}(X_1)...{\hat K}(X_N)
\Eq(11.20.1)$$

\\where $X_1,...,X_N$ are all disjoint polymers in $\Lambda$.
Since these are closed, they are separated by a distance of 
at least one.

\\We will consider often activities defined on connected 
polymers. For such activities the decomposition \equ(11.20.1)
is on connected subsets of $X$. Such activities will be called
{\it connected} activities.

\\The polymer expansion given above is borrowed from [BY].

\\Let us conclude this subsection
stating two useful lemmas about manipulations on ${\cal E}$xp.
The (easy) proofs can be found in [BY] and [B,Laus.].
\vglue.5truecm
\\LEMMA 1.4.1 
\vglue.3truecm
\\For any pair of polymer activities $A,B$

$${\cal E}xp(\square+A){\cal E}xp(\square+B)=
{\cal E}xp(\square+A+B+A\vee B)\Eq(11.20.2)$$

\\where the polymer activity $(A\vee B)$ is defined
by

$$(A\vee B)(X)=\sum_{\{X_i\},\{Y_j\}\rightarrow X}
\prod_iA(X_i)\prod_jB(Y_j)\Eq(11.20.3)$$

\\with $\{X_i\},\{Y_j\}\rightarrow X$ meaning that
the sum is over
the families of polymers $\{X_i\},\{Y_j\}$ such that
$(\cup_iX_i)\cup(\cup_jY_j)=X$, the $X$'s are disjoint,
the $Y$'s are disjoint, but the two families are overlap 
connected.

\vglue.5truecm
\\LEMMA 1.4.2 
\vglue.3truecm
\\Let us define, for a polymer activity A, the quantity

$$A^+(X)=\sum_{N\ge 1}{1\over N!}\sum_{X_1...X_N\rightarrow X}
\prod_{j=1}^NA(X_j)\Eq(11.20.5)$$

\\where ${X_1...X_N\rightarrow X}$ means that ${X_1...X_N}$
are distinct, overlap connected and such that $\cup_iX_i=X$.

\\Then we have

$$\prod_{X\subset Z}e^{A(X)}={\cal E}xp(\square+(e^A-1)^+)(Z)
\Eq(11.20.6)$$

\vglue.3truecm
\\{\it 1.5. RG strategy.}
\vglue.3truecm
\\A single RG step has for us four parts. We describe them
briefly here and we will fill in the details in later sections [3-5].
We begin with an expression of the form

$$ e^{-V(\phi,\L)}{\cal E}xp(\square+K)(\phi,\L)\Eq(11.21)$$

\\Here $V$ is of the form \equ(11.17), with $g_0$ replaced by $g$.
Althought $K$ is absent initially, it is necessarily generated
in RG operations.

\\The structure of the activity $K$, together with bounds,
will be exhibited in later sections. Suffice to say at this stage
that $K$ consists of an exact second order perturbation theory
contribution plus a remainder.

\\Before we proceed further, let us rewrite \equ(11.21) in the form

$$e^{-V(\phi,\L)}{\cal E}xp(\square +K)(\phi,\L)
={\cal E}xp(\square +{\hat K})(\phi,\L)\Eq(11.22)$$

\\${\hat K}$ is a functional of $K$ and $V$, given by
a standard formula, given later (see section 3).
\vglue.3truecm
\\{\it Step 1: Reblocking}
\vglue.3truecm
\\We reblock \equ(11.22) using the reblocking operator ${\cal B}$.
The reblocking operator was introduced in [BY]. So far $\L$ has been
paved with closed 1-blocks. Introduce a compatible paving of $\L$
on the next scale by closed $L$-blocks. Each closed $L$-block
is a union of closed 1-blocks. For any 1-polymer $X$, let 
${\bar X}$ be the smallest $L$-polymer containing $X$. If
$Z$ is a 1-polymer, $LZ$ denotes a $L$-polymer. Then

$$({\cal B}{\hat K})(LZ)=\sum_{N\ge 1}{1\over N!}\sum_
{X_1...X_N\ {\rm disjoint}\atop{\{{\bar X}_j\}\ {\rm overlap\ connected}
\atop\cup{\bar X}_j=LZ}}\prod_{j=1}^N{\hat K}(X_j)\Eq(11.22.1)$$

\\We then have

$${\cal E}xp(\square +{\hat K})(\L,\phi)=
{\cal E}xp_L(\square_L +{\cal B}{\hat K})(\L,\phi)\Eq(11.23)$$

\\where all the operations 
on the r. h. s. of \equ(11.23) are on scale $L$.

\\We shall now let the RG act on \equ(11.23). The RG action,
as outlined before (see \equ(11.1.02)), consists of a convolution
with respect to the measure $\mu_{\G_L}$ (called the fluctuation
integration) followed by rescaling.
\vglue.3truecm
\\{\it Step 2: Fluctuation integration}
\vglue.3truecm

\\This is

$$(\mu_\G*{\cal E}xp_L(\square_L +{\cal B}{\hat K}))(\L,{\cal R}\phi)$$

\\with ${\cal R}\phi$ defined by \equ(11.9.1)

\\The expansion of ${\cal E}xp_L(\square_L +{\cal B}{\hat K})$ 
gives a sum over products of $L$-polymer activities, where the
$L$-polymers are closed and disjoint. They are thus separated
from each other by a distance greater or equal to $L$.
The fluctuation covariance $\G$ is of compact support by construction:

$$\G(x-y)=0\quad\forall\ |x-y|\ge L$$

\\As a consequence

$$(\mu_\G*{\cal E}xp_L(\square_L +{\cal B}{\hat K}))(\L,{\cal R}\phi)=
{\cal E}xp_L(\square_L +\mu_\G*{\cal B}{\hat K})(\L,{\cal R}\phi)
\Eq(11.24)$$

\\This leads to a considerable simplification in the RG analysis.

\\Our next step is
\vglue.3truecm
\\{\it Step 3: Rescaling}
\vglue.3truecm

\\We recall here \equ(11.9.1) the definition of the rescaling
operator ${\cal R}$ acting on the field $\phi$:

$${\cal R}\phi(x)=L^{\b/2}\phi(x/L)$$

\\We have already defined also 
(see \equ(11.8.88)) the rescaled fluctuation covariance

$${\cal R}\G(y)={L^{-\b}}\G(yL)$$

\\For a polymer activity $K$ we define

$${\cal R}K(L^{-1}X,\phi)=K(X,{\cal R}\phi)\Eq(11.24.1)$$

\\Note that

$$(\mu_\G*K)(X,{\cal R}\phi)=(\mu_{{\cal R}\G}*{\cal R}K)
(L^{-1}X,\phi)\Eq(11.24.2)$$

\\as it is easy to see.

\\Define

$${\cal S}={\cal R}{\cal B}\Eq(11.24.3)$$

\\Then we have from \equ(11.24)

$${\cal E}xp_L(\square_L +\mu_\G*{\cal B}{\hat K})(\L,{\cal R}\phi)
={\cal E}xp(\square+({\cal S}{\hat K})^{\natural})(L^{-1}\L,\phi)
\Eq(11.25)$$

\\Here $\natural$ denotes the convolution operation with respect
to $\mu_{{\cal R}\G}$.

\\On the r.h.s. of \equ(11.25), $L^{-1}\L$ stands for $\L$
shrunk by $L^{-1}$ and paved by closed 1-blocks.

\\Our final step is the {\it extraction}:

\vglue.3truecm
\\{\it Step 4: Extraction}
\vglue.3truecm

\\This consists of picking up relevant parts $F$ from
$({\cal S}{\hat
K})^{\natural}$ and
exponentiating them, in such a way that

$${\cal E}xp(\square+({\cal S}{\hat K})^{\natural})(L^{-1}\L,\phi)=
e^{-V'(L^{-1}\L,\phi)}{\cal E}xp(\square +K')(L^{-1}\L,\phi)\Eq(11.26)$$

\\We will find that 

$$V'(F)(X)=g'\int_X dx\ v_*(\phi(x))\Eq(11.26.1)$$

\\with a new coupling constant $g'$.

\\$K'$ is a functional of ${\hat K}$ and of the relevant part $F$:

$$K'={\cal E}({\hat K},F)\Eq(11.26.2)$$

\\An explicit formula for the extraction operator ${\cal E}$
is given in [B,Laus.] and we will put it to good use.
The aim of the RG analysis is to control the discrete flows
obtained by a large number of iterations
$(V,K)\rightarrow(V',K')$.

\pagina

\vglue.5truecm\numsec=2\numfor=1\pgn=1
\\\S2. {\it Polymer activity norms and basic Lemmas.}
\vglue.5truecm
\\We want to define in this section the basic properties
that we need on the activities $K$, and the appropriate norms
to control them.
\vglue.3truecm
\\{\it 2.1. Decay in $X$: the large set regulator $\G$.}
\vglue.3truecm
\\Let $K(X)$ be a connected polymer activity (with possible $\phi$ 
dependence suppressed). The decay of $K$ in the ``size''
of $X$ is controlled by a norm of the following type:

$$\Vert K\Vert_{\G_n}=\sup_{\D}\sum_{X\supset\D
\atop X\ {\rm connected}}|K(X)|\G_n(X)\Eq(22.1)$$

\\Where the {\it large set regulators} are defined by

$$\G_n(X)=2^{n|X|}\G(X)\Eq(22.2)$$

$$\G(X)=L^{(D+2)|X|}\Eq(22.3)$$

\\and $|X|$ denotes the number of blocks in $X$.
Because our fluctuation covariance is compactly supported it is 
sufficient to define the norm of the $K$ only for connected $X$. 
This simplifies the definition of $\G$ with respect to [BY].

\\We define a {\it small set} as follows: a connected
polymer $X$ is a small set if $|X|\le 2^D$.

\\Recall that the {\it $L$-closure ${\bar X}$} of a polymer $X$ is defined
to be the smallest union of $L$-blocks containing $X$

\\The main result about  $\G$ that we need in the next
sections is the following statement

\vglue.5truecm
\\LEMMA 2.1.1 
\vglue.3truecm
\\For each $p=0,1,2,...$ there is an O(1) constant $c_p$ such that for
$L$ sufficiently large and for any polymer $X$

$$\G_p(L^{-1}{\bar X})\le c_p\G(X)\Eq(22.4)$$

\\For any large set $X$ a stronger bound is valid

$$\G_p(L^{-1}{\bar X})\le c_pL^{-D-1}\G(X)\Eq(22.5)$$

\vglue.3truecm
\\{\it Proof}

\\For $X$ small set one has (using $D=1$) 
$|X|\le 2,\ |L^{-1}{\bar X}|\le|X|$.
This proves the \equ(22.4) with $c_p=2^{2p}$

For large sets we note that, for $L\ge 3$,
$|X|\ge 3,\ |L^{-1}{\bar X}|\le{2\over 3}|X|$.
These relations imply:
$$\G_p(L^{-1}{\bar X})=
2^{p|L^{-1}{\bar X}|}L^{(D+2)|L^{-1}{\bar X}|}\le
2^{p{2\over 3}|X|}L^{{2\over 3}(D+2)|X|}\le
2^{p{2\over 3}|X|}L^{-{1\over 3}(D+2)|X|}\G(X)\le$$
$$\le
2^{p{2\over 3}|X|}L^{-{1\over 3}|X|}L^{-{1\over 3}(D+1)|X|}\G(X)\le
2^{p{2\over 3}|X|}L^{-{1\over 3}|X|}L^{-D-1}\G(X)\eqno Q.E.D.$$
\vglue.3truecm
\\{\it 2.2. Smoothness in the fields.}
\vglue.3truecm
\\Functionals of $\phi$ are defined on the Banach space
${\cal C}^r(\L)$ of $r$ times continuously differentiable
fields with the norm

$$\Vert f\Vert_{{\cal C}^r}=
\sum_{l=0}^r\sup_x |\dpr^lf(x)|\Eq(22.6)$$

\\A derivative of a functional with respect to $\phi$
in the direction $f$ is a linear functional
$f\rightarrow D^K(X,\phi;f)$ on this
Banach space defined by

$$\left.{\dpr\over\dpr s}
K(X,\phi+ sf)\right|_{s=0}
=DK(X,\phi;f)$$

\\The size of a functional derivative is naturally measured by the norm

$$\Vert DK(X,\phi)\Vert=
\sup[|DK(X,\phi;f)|:f\in{\cal C}^r(X),
\Vert f\Vert_{{\cal C}^r(X)}\le 1]$$

\\ and $\Vert K(X,\phi)\Vert=\vert K(X,\phi)\vert$.

\\In the proof of the main theorem we will need to introduce
the norm

$$\Vert K(X)\Vert_1=\Vert K(X,\phi)\Vert+
\Vert DK(X,\phi)\Vert\Eq(22.7)$$

\\We have the obvious 
property:

\vglue.5truecm
\\LEMMA 2.2.1 
\vglue.3truecm

\\For any 
polymers $X_1,X_2$ and for any activities $K_1,K_2$

$$\Vert K_1(X_1)K_2(X_2)\Vert_1\le \Vert K_1(X_1)\Vert_1
\Vert K_2(X_2)\Vert_1\Eq(22.8)$$

\vglue.3truecm
\\{\it 2.3. Growth in the fields: the large fields regulator $G$.}
\vglue.3truecm

\\The growth of $K(X,\phi)$ as a function of $\phi$ and 
derivatives of $\phi$ is controlled by a 
{\it large fields regulator} $G(X,\phi)$.

\\The natural norm defined by $G$ has the form

$$\Vert K(X)\Vert_G=\sup_{\phi\in{\cal C}^r}\Vert K(X,\phi)\Vert
G^{-1}(X,\phi)\Eq(22.9)$$

\\The functional $G(X,\phi)$ is chosen so as to
satisfy the following inequality

$$G(X\cup Y,\phi)\ge G(X,\phi)G(Y,\phi)\quad {\rm if}\ X\cap Y=\emptyset
\Eq(22.10)$$

\\The form of our interaction suggests the use of the following
regulator

$$G_{\r,k}(X,\phi)={1\over |X|}\int_Xdxe^{-(\l_*/2)(1-\r)|\phi(x)|^2}
e^{\k\Vert\phi\Vert^2_{X,1,\s}}\Eq(22.11)$$

\\with $0<\r <1$, $k >0$ and 

$$\Vert\phi\Vert^2_{X,1,\s}=\sum_{1\le \a\le\s}\Vert\dpr^\a\phi\Vert^2_X
\Eq(22.12)$$

\\where $\Vert\phi\Vert_X$ is the $L^2$ norm. We take $\s$ large enough
so that this norm can be used in Sobolev inequalities to control
$\dpr\phi$ pointwise.

\\Let us show that \equ(22.10) is true for this choice.
\vglue.5truecm
\\LEMMA 2.3.1
\vglue.3truecm
$G_{\r,k}$ satisfies \equ(22.10).
\vglue.3truecm
\\{\it Proof}

\\It is enough to show

$${1\over |X\cup Y|}\int_{X\cup Y}dxe^{-(\l_*/2)(1-\r)|\phi(x)|^2}\ge$$
$$\ge{1\over |X|}\int_Xdxe^{-(\l_*/2)(1-\r)|\phi(x)|^2}
{1\over |Y|}\int_Ydxe^{-(\l_*/2)(1-\r)|\phi(x)|^2}$$

\\Define:

$$a={1\over |X|}\int_Xdxe^{-(\l_*/2)(1-\r)|\phi(x)|^2}$$

$$b={1\over |Y|}\int_Ydye^{-(\l_*/2)(1-\r)|\phi(y)|^2}$$

$$p={|X\cup Y|\over |X|} \  , \  q={|X\cup Y|\over |Y|}$$

\\Note that 

$$0\le a,b \le 1$$ 

$${1\over p}+{1\over q}=1$$

\\We have :

$${1\over |X\cup Y|}\int_{X\cup Y}dxe^{-(\l_*/2)(1-\r)|\phi(x)|^2}=
{1\over p}a+{1\over q}b=$$
$$={1\over p}(a^{1\over p})^p + {1\over q}(b^{1\over q})^q$$
$$\ge a^{1\over p}  b^{1\over q} \ge ab$$

where to go to the last line we have used $0\le a,b \le 1$.

$$\eqno Q.E.D.$$

\\For the norm \equ(22.9) 
to be useful, we will need further properties for the regulator $G$.
In particular to control the fluctuation
step we will need that $G$ is {\it stable} in the sense of the 
following lemma:
\vglue.5truecm
\\LEMMA 2.3.2 ({\it Stability of the large field regulator})
\vglue.3truecm
\\Let $0<\r < 1/8 \quad \r=O(1)$ 
and $\k>0\quad \k=O(1)$ be both sufficiently small and independent
of $L$. Let $\k/\r<1$ and $L$ be sufficiently large.

\\Then

$$(\mu_\G*G_{\r,\k})(X,{\cal R}\phi)\le G^\sharp_{\r,\k}(X,{\cal R}\phi)
\Eq(22.12.2)$$

\\with

$$G^\sharp_{\r,\k}(X,{\cal R}\phi)=O(1)2^{|X|}{L^{-\a}\over |X|}
\int_Xdxe^{-(\l_*/2)(1-\r/L^{\b/2})|L^{-\b/2}{\cal R}\phi(x)|^2}
e^{4\k\Vert{\cal R}\phi\Vert^2_{X,1,\s}}\Eq(22.13)$$

\\The proof of this lemma, which is straightforward but rather
long, is presented in Appendix A.

\\It is useful to note that from the scaling property of the field
$\phi$ and the definition of $\b$ in (0.9) we have

$$\Vert{\cal R}\phi\Vert^2_{X,1,\s}\le L^{\b-1}
\Vert\phi\Vert^2_{L^{-1}X,1,\s}$$

\\and for $L$ sufficiently large 

$$L^{\b-1}4<1$$

\\since $\b>0$ but very small.

\\Note also that

$$L^{D-\a}=L^{\e}=O(1)$$

\\for $\e$ sufficiently small ( depending on $L$).
Using these two facts it is easy to see that

$$G^\sharp_{\r,\k}(X,{\cal R}\phi)\le O(1)2^{|X|}L^{-D}G(L^{-1}X,\phi)$$

\\which is the original form of the regulator up to the contractive factor
$L^{-D}$ and a vacuum energy contribution depending on the size of X.

\\Finally we remark that the stability of the large
fields regulator can be stated 
analogously using the master formula: actually in section 5
the following form of the stability of the regulator will be used
for $0<\r<1/32$ and $\k/\r <1$ and $L$ sufficiently large
$$\int d\mu_{\Sigma^{{\bar x}}}(\z) e^{(\l_*/2)4\r|\z({\bar x})
+L^{-\b}{\cal R}\phi({\bar x})|^2}
e^{4\k\Vert\z+L^{-\b}T^{{\bar x}}{\cal R}\phi\Vert^2_{X,1,\s}}
\le$$
$$\le e^{(\l_*/2)(4\r/L^{\b/2})L^{-\b}|{\cal R}\phi({\bar x})|^2}
e^{8\k\Vert{\cal R}\phi\Vert^2_{L^{-1}X,1,\s}}
\Eq(1)$$

\\This can be derived from 
 \equ(22.12.2), \equ(22.13) using the master formula in the following way
$$\int d\mu_{\Sigma^{{\bar x}}}(\z) e^{(\l_*/2)4\r|\z({\bar x})
+L^{-\b}{\cal R}\phi({\bar x})|^2}
e^{4\k\Vert\z+L^{-\b}T^{{\bar x}}{\cal R}\phi\Vert^2_{X,1,\s}}
=[L^{-\a}v_*(L^{-\b/2}{\cal R}\phi({\bar x}))]^{-1}.$$
$$.\int  d\mu_\G(\z)v_*(\z({\bar x})+{\cal R}\phi({\bar x}))
e^{(\l_*/2)4\r|\z({\bar x})
+{\cal R}\phi({\bar x})|^2}
e^{4\k\Vert\z+{\cal R}\phi\Vert^2_{X,1,\s}}=$$
$$=[L^{-\a}v_*(L^{-\b/2}{\cal R}\phi({\bar x}))]^{-1}v_*(0)
\int d\mu_\G(\z)e^{-(\l_*/2)(1-4\r)|\z({\bar x})
+{\cal R}\phi({\bar x})|^2}
e^{4\k\Vert\z+{\cal R}\phi\Vert^2_{X,1,\s}}\Eq(2)$$
The right hand side of \equ(2) is controlled using 
\equ(22.13):
$$[L^{-\a}v_*(L^{-\b/2}{\cal R}\phi({\bar x}))]^{-1}v_*(0)
\int d\mu_\G(\z)e^{-(\l_*/2)(1-4\r)|\z({\bar x})
+{\cal R}\phi({\bar x})|^2}
e^{4\k\Vert\z+{\cal R}\phi\Vert^2_{X,1,\s}}\le$$
$$[L^{-\a}v_*(L^{-\b/2}{\cal R}\phi({\bar x}))]^{-1}v_*(0)
L^{-\a}e^{-(\l_*/2)(1-4\r/L^{\b/2})|L^{-\b/2}{\cal R}\phi({\bar x})|^2}
e^{8\k\Vert {\cal R}\phi\Vert^2_{X,1,\s}}\Eq(3)$$
and \equ(3) implies trivially \equ(1).

\vglue.5truecm

\\One particular point which we have to take account of in this 
work is the fact that
due to our expression of the large field regulator
the usual relation (see e.g. BDH) 
$G(X,\phi)\ge 1$  is
{\it not} true in our case. 
Therefore also the useful relation
$G(X,\phi)\ge G(Y,\phi)$ if $X\supset Y$ is in general false.
This implies that in many cases
the reblocking step has to be evaluated in some detail.
For the contributions due to small sets (see below, section 5)
the following lemma is often used 
\vglue.5truecm

\\LEMMA 2.3.3

\vglue.3truecm
$$\sum_{X\ {\rm s.\ s.}\atop{\bar X}=LZ}G(L^{-1}X,\phi)
\le O(1)L^D G(Z,\phi)\Eq(22.100)$$
\vglue.3truecm
\\{\it Proof}

$$\sum_{X\ {\rm s.\ s.}\atop{\bar X}=LZ}G(L^{-1}X,\phi)
  =\sum_{X\ {\rm s.\ s.}\atop{\bar X}=LZ}
{1\over |L^{-1}X|}
\int_{L^{-1}X}dxe^{-(\l_*/2)(1-\r)|\phi(x)|^2}
e^{\k\Vert\phi\Vert^2_{L^{-1}X,1,\s}}\Eq(22.101)$$

First we observe that since $X$ is a small set, so is $Z$. We have

$$\sum_{X\ {\rm s.\ s.}\atop{\bar X}=LZ}G(L^{-1}X,\phi)
\le O(1)L^D 
\left[
\sum_{\D \atop{\bar \D}=LZ}
\int_{L^{-1}\D}dxe^{-(\l_*/2)(1-\r)|\phi(x)|^2}
\right.$$

$$\left.
  +
\sum_{X:|X|=2, X\ {\rm conn.}\atop{\bar X}=LZ}
\int_{L^{-1}X}dxe^{-(\l_*/2)(1-\r)|\phi(x)|^2}
\right]e^{\k\Vert\phi\Vert^2_{Z,1,\s}}
\le$$

$$\le O(1)L^D 
\left[
\sum_{\D \atop{\bar \D}=LZ}
\int_{L^{-1}\D}dxe^{-(\l_*/2)(1-\r)|\phi(x)|^2}
\right.$$

$$\left.
  +
\sum_{\D_1 ,\D_2 \ {\rm conn.}\atop{\overline {\D_1 \cup \D_2}}=LZ}
\left[\int_{L^{-1}\D_1}dxe^{-(\l_*/2)(1-\r)|\phi(x)|^2}
+     \int_{L^{-1}\D_2}dxe^{-(\l_*/2)(1-\r)|\phi(x)|^2}
\right]
\right]e^{\k\Vert\phi\Vert^2_{Z,1,\s}}
\le$$

$$\le O(1)L^D 
\left[
\sum_{\D \atop{ \D} \subset LZ}
\int_{L^{-1}\D}dxe^{-(\l_*/2)(1-\r)|\phi(x)|^2}
\right.$$

$$\left.
  +
\sum_{\D_1 \atop{ \D_1} \subset LZ}
\sum_{\D_2 \atop \D_1,\D_2  \ {\rm conn.} }
\int_{L^{-1}\D_1}dxe^{-(\l_*/2)(1-\r)|\phi(x)|^2}
\right]e^{\k\Vert\phi\Vert^2_{Z,1,\s}}
\le$$

$$\le O(1)L^D\int_{Z}dxe^{-(\l_*/2)(1-\r)|\phi(x)|^2}
e^{\k\Vert\phi\Vert^2_{Z,1,\s}} \le O(1)L^D G(Z,\phi) \Eq(22.102)$$

In passing to the last line we have used 

$$\sum_{\D_2 \atop \D_1,\D_2  \ {\rm conn.} }
1 \le 2 $$ 

and the fact that $ |Z| \le 2 $.

Q.E.D
\vglue.3truecm
\\{\it 2.4. Norms.}
\vglue.3truecm
\\Now we have all the ingredients to construct norms on $K$.
We define the norms
$$\Vert K(X)\Vert_{G,1}=
\Vert K(X)\Vert_{G}+\Vert DK(X)\Vert_{G}\Eq(22.14)$$
$$\Vert K\Vert_{G,1,\G}=\Vert\Vert K(X)\Vert_{G,1}\Vert_\G\Eq(22.15)$$

\\However sometimes it will be useful to define $L^\io$ norms
on certain activities in the following way

$$\Vert K(X)\Vert_{\io,1}=
\Vert K(X)\Vert_{\io}+\Vert DK(X)\Vert_{\io}\Eq(22.16)$$

\\where

$$\Vert K(X)\Vert_{\io}=\sup_{\phi\in {\cal C}^r}|K(X,\phi)|\Eq(22.17)$$

\vglue.3truecm
\\{\it 2.5. Basic estimates on generic integrated activities.}
\vglue.3truecm

\\In this subsection we state some bounds valid for
integrated activities
with initial norm small enough. These bounds will be used often
in the next sections. 

\vglue.5truecm
\\LEMMA 2.5.1
\vglue.3truecm
\\Let ${ K}$ be an activity such that $\Vert K\Vert_{G,1,\G}=
O(\e^q)$, with $q\ge 1/10$, and let us define ${\cal S}_{\ge k}{ K}$ by 
restricting the sum on N in \equ(11.22.1) to $ N \ge k $. Then we have

$${\cal S}_{\ge k}{ K}(Z,\phi)=
\sum_{X\atop{\bar X}=LZ}({\cal R}{\bar K}_k)(L^{-1}X,\phi)\Eq(22.41)$$

\\with 
${\bar K}_k$ defined by

$${\bar K}_k(X,\phi)=\sum_{N\ge k}{1\over N!}
\sum_{{X_1,...,X_N\ {\rm disj.}
\atop\cup_jX_j=X}\atop \{{\bar X}_j\}\ {\rm ov\ conn}}
\prod_{j=1}^N{ K}(X_j)\Eq(22.41.1)$$

\\Then one has for any integer $p\ge 0$ and $ \e \ge 0 $ sufficiently small

$$\Vert\left({\cal S}_{\ge k}{ K}\right)
^{\natural}\Vert_{G,1,\G_p}\le
O(1)^k L^{(\b/2+kD)}\Vert{K}\Vert_{G,1,\G}^k\Eq(22.44)$$

\\with $O(1)$ depending on $p$.
\vglue.3truecm
\\{\it Proof}

\\The
action of the fluctuation operator is controlled by
the stability of the large fields regulator:
$$\vert(\mu_{\G}*{\bar K})(X,{\cal 
R}\phi)\vert\le L^{-D}O(1)2^{|X|}G(L^{-1}X,\phi)\Vert{\bar K}(X)\Vert_{G}
\Eq(22.45)$$
then
$$\vert\left({\cal S}_{\ge k}{K}\right)
^{\natural}(Z,\phi)\vert\le\sum_{X\atop{\bar X}=LZ}
L^{-D}O(1)2^{|X|}G(L^{-1}X,\phi)\Vert{\bar K}_k(X)\Vert_{G}\le$$
$$\le L^{-D}O(1)e^{\k\Vert\phi\Vert^2_{Z,1,\s}}|Z|
{1\over |Z|}\int_Zdxe^{-(\l_*/2)(1-\r)|\phi(x)|^2}
\sum_{X\atop{\bar X}=LZ}{2^{|X|}\over |L^{-1}X|}
\chi_{L^{-1}X}(x)\Vert{\bar K}_k(X)\Vert_{G}\Eq(22.46)$$
where $\chi_{L^{-1}X}(x)$ is the characteristic function of the set
$L^{-1}X$.
Using the trivial bounds
$${1\over |L^{-1}X|}
\chi_{L^{-1}X}(x)\le L^D,\qquad |Z|\le 2^{|Z|}$$
we have

$$\Vert\left({\cal S}_{\ge k}{K}\right)
^{\natural}(Z)\Vert_G\le O(1)2^{|Z|}\sum_{X\atop{\bar X}=LZ}
2^{|X|}\Vert{\bar K}_k(X)\Vert_{G}
\Eq(22.47)$$
Performing the same bound on the functional derivative we have
$$\Vert\left({\cal S}_{\ge k}{K}\right)
^{\natural}(Z)\Vert_{G,1}\le O(1)2^{|Z|}
L^{\b/2}\sum_{X\atop{\bar X}=LZ}
2^{|X|}\Vert{\bar K}_k(X)\Vert_{G,1}
\Eq(22.48)$$
and therefore
$$\Vert\left({\cal S}_{\ge k}{K}\right)
^{\natural}(Z)\Vert_{G,1}\G_p(Z)\le O(1)
L^{\b/2}\sum_{X\atop {\bar X}=LZ}
\Vert{\bar K}_k(X)\Vert_{G,1}\G_{p+2}(L^{-1}{\bar X})
\Eq(22.49)$$

\\Defining (see also [BY])

$$\Vert{\bar K}_k\Vert_{G,1,\G_{p+2}}^{(1)}=\sup_\D
\sum_{X\atop L^{-1}{\bar X}\cap\D\ne\emptyset}
\Vert{\bar K}_k(X)\Vert_{G,1}\G_{p+2}(L^{-1}{\bar X})
\Eq(22.49.1)$$

\\we easily obtain from \equ(22.49) 

$$\Vert\left({\cal S}_{\ge k}{K}\right)
^{\natural}\Vert_{G,1,\G_p}\le O(1)
L^{\b/2}\Vert{\bar K}_k\Vert_{G,1,\G_{p+2}}^{(1)}
\Eq(22.50)$$

\\Now we use the BY argument (Lemma 7.1 [BY]) and we obtain

$$\Vert{\bar K}_k\Vert_{G,1,\G_{p+2}}^{(1)}\le\sum_{N\ge k}
O(1)^N(\Vert{K}\Vert_{G,1,\G_{p+3}}^{(1)})^N\Eq(22.51)$$

\\From \equ(22.4) of Lemma 2.1.1

$$\G_{p+3}(L^{-1}{\bar X})\le O(1)\G(X)\Eq(22.51.1)$$

\\and using again BY we obtain finally

$$\Vert{K}\Vert_{G,1,\G_{p+3}}^{(1)}\le O(1)L^D
\Vert{K}\Vert_{G,1,\G}\Eq(22.52)$$

\\hence

$$\Vert\left({\cal S}_{\ge k}{K}\right)
^{\natural}\Vert_{G,1,\G_p}\le O(1)L^{\b/2}\sum_{N\ge k}
O(1)^N(L^D)^N(\Vert{K}\Vert_{G,1,\G})^N\Eq(22.53)$$

\\Using the bound
$\Vert{K}(X)\Vert_{G,1,\G}\le O(1)\e^q$ to control
the sum over $N$
we have the lemma
$$\eqno Q.E.D.$$

\\Lemma 2.5.1 obviously implies the following

\vglue.5truecm
\\COROLLARY 2.5.2
\vglue.3truecm
\\For the linearized scaling operator ${\cal S}_{1}$ 
the following bound holds

$$\Vert\left({\cal S}_{1}{K}\right)
^{\natural}\Vert_{G,1,\G_p}\le
O(1)L^{(\b/2+D)}\Vert{K}\Vert_{G,1,\G}\Eq(22.54)$$

\\for any integer $p\ge 1$, $O(1)$ depends on $p$.
\vglue.8truecm
\\We now define the linearized scaling operator restricted
to contributions from large sets by

$${\cal S}_{1}^{(l.s.)}{K}(Z,\phi)=
\sum_{X\ {\rm conn.\ large\ set}\atop{\bar X}=LZ}
({\cal R}{K})(L^{-1}X,\phi)\Eq(22.55)$$

\\We have the following result.

\vglue.5truecm
\\LEMMA 2.5.3
\vglue.3truecm

$$\Vert\left({\cal S}_{1}^{(l.s.)}{K}\right)
^{\natural}\Vert_{G,1,\G_p}\le
O(1)L^{-(1-\b/2)}\Vert{K}\Vert_{G,1,\G}\Eq(22.56)$$

\\for any integer $p\ge 0$, $O(1)$ depends on $p$.
\vglue.3truecm
\\{\it Proof}

\\Repeat the proof of lemma 2.5.1 using 

$$\G_{p+3}(L^{-1}{\bar X})\le O(1)L^{-D-1}\G(X)\quad{\rm for}\
X\ {\rm large\ set}\Eq(22.51.2)$$

\\which comes from \equ(22.5) of Lemma 2.1.1 instead of \equ(22.51.1).
$$\eqno Q.E.D.$$

\\Let now $F$ be a polymer activity supported on small sets,
such that for every integer $p\ge 0$ and for $q\ge 1/10$

$$\Vert F\Vert_{G,1,\G_p}\le
O(\e^q)\Eq(22.57)$$
and
$$\Vert F\Vert_{\io,1,\G_p}\le
O(\e^q)\Eq(22.58)$$

\\with $O(1)$ depending on $p$.

\\We then have for $\e \ge 0 $ sufficiently small

\vglue.5truecm
\\LEMMA 2.5.4
\vglue.3truecm

$$\Vert e^{-F}-1\Vert_{G,1,\G_p}\le
O(1)\Vert F\Vert_{G,1,\G_p}\Eq(22.59)$$

$$\Vert e^{-F}-1-F\Vert_{G,1,\G_p}\le
O(1)\Vert F\Vert_{G,1,\G_p}\Vert F\Vert_{\io,1,\G_p}\Eq(22.60)$$

\vglue.3truecm
\\{\it Remark:}

\\Lemma 2.5.4 remains true if the $G$-norm is replaced
by the $L^\io$ norm, by the same proof.
\vglue.3truecm
\\{\it Proof}

$$\Vert e^{-F(X)}-\sum_{l=0}^k{1\over l!}(-F(X))^l\Vert_{1}\le
\sum_{N\ge k+1}{1\over N!}\Vert F(X)\Vert_{1}^N\Eq(22.61)$$

\\whence:

$$\Vert e^{-F(X)}-\sum_{l=0}^k{1\over l!}(-F(X))^l\Vert_{G,1}\le
\sum_{N\ge k+1}{1\over N!}\Vert F(X)\Vert_{G,1}
\Vert F(X)\Vert_{\io,1}^{N-1}\le$$
$$\le \sum_{N\ge k+1}{1\over N!}\Vert F(X)\Vert_{G,1}
(\Vert F(X)\Vert_{\io,1}
\G(X))^{N-1}\Eq(22.62)$$

\\Let $\D\subset X$ be any block in $X$. Then

$$\Vert F(X)\Vert_{\io,1}\G(X)\le \!\!
\sum_{Y\cap\D\ne\emptyset}
\Vert F(Y)\Vert_{\io,1}\G(Y)\le
\sup_{\D'}\sum_{Y\cap\D'\ne\emptyset}
\Vert F(Y)\Vert_{\io,1}\G(Y)=\Vert F\Vert_{\io,1,\G}\Eq(22.63)$$

\\From \equ(22.62) and \equ(22.63) we get:

$$\Vert e^{-F(X)}-\sum_{l=0}^k{1\over l!}(-F(X))^l\Vert_{G,1}
\le \sum_{N\ge k+1}{1\over N!}\Vert F(X)\Vert_{G,1}
\Vert F\Vert_{\io,1,\G}
^{N-1}\Eq(22.64)$$

\\or

$$\Vert e^{-F}-\sum_{l=0}^k{1\over l!}(-F)^l\Vert_{G,1,\G}\le$$
$$\le\Vert F\Vert_{G,1,\G}
\Vert F\Vert_{\io,1,\G}^k\sum_{N\ge k+1}{1\over N!}
\Vert F\Vert_{\io,1,\G}^{N-1-k}\Eq(22.65)$$

Because of the smallness of $\Vert F\Vert_{\io,1,\G}$ the
series is bounded by $O(1)$. Hence

$$\Vert e^{-F}-\sum_{l=0}^k{1\over l!}(-F)^l\Vert_{G,1\G}\le
O(1)\Vert F\Vert_{G,1_G}
\Vert F\Vert_{\io,1,\G}^k\Eq(22.66)$$

\\Setting $k=0,1$ in \equ(22.66) we prove the lemma.
$$\eqno Q.E.D.$$

\vglue.5truecm
\\LEMMA 2.5.5
\vglue.3truecm
\\For any integers $k\ge 1$ and $p\ge 0$, and with $O(1)$ 
dependent on $p$

$$\Vert (e^{-F}-1)^+_{\ge k}\Vert_{G,1,\G_p}\le
O(1)^k\Vert F\Vert_{G,1,\G_{p+3}}
\Vert F\Vert_{\io,1,\G_{p+3}}^{k-1}\Eq(22.67)$$

$$\Vert (e^{-F}-1)^+_{\ge k}\Vert_{\io,1,\G_p}\le
O(1)^k
\Vert F\Vert_{\io,1,\G_{p+1}}^{k}\Eq(22.68)$$

\vglue.3truecm
\\{\it Proof}

$$(e^{-F}-1)^+_{\ge k}(X)=\sum_{N\ge k}{1\over N!}
\sum_{X_1,..,X_N:\cup_jX_j=X\atop \{X_j\}\ {\rm overlap\ conn.}}
\prod_{j=1}^N(e^{-F}-1)(X_j)\Eq(22.69)$$

\\then

$$\Vert(e^{-F}-1)^+_{\ge k}(X)\Vert_1\le\sum_{N\ge k}{1\over N!}
\sum_{X_1,..,X_N:\cup_jX_j=X\atop \{X_j\}\ {\rm overlap\ conn.}}
\prod_{j=1}^N\Vert(e^{-F}-1)(X_j)\Vert_1\le$$
$$\sum_{N\ge k}{1\over N!}
\sum_{X_1,..,X_N:\cup_jX_j=X\atop \{X_j\}\ {\rm overlap\ conn.}}
G(X_1)\Vert(e^{-F}-1)(X_1)\Vert_{G,1}
\prod_{j=2}^N\Vert(e^{-F}-1)(X_j)\Vert_{\io,1}\Eq(22.70)$$

\\We estimate for $X_1\subset X$

$$G(X_1)\le \sum_{Y\subset X}G(Y)\Eq(22.71)$$
and then
$$G(Y)={1\over |Y|}\int_Y dx
e^{-{\l_*\over 2}(1-\r)|\phi(x)|^2}e^{\k\Vert\phi\Vert^2_{Y,1,\s}}
\le\int_Y dx
e^{-{\l_*\over 2}(1-\r)|\phi(x)|^2}e^{\k\Vert\phi\Vert^2_{X,1,\s}}\le$$
$$\le|X|{1\over |X|}
\int_X dx
e^{-{\l_*\over 2}(1-\r)|\phi(x)|^2}
e^{\k\Vert\phi\Vert^2_{X,1,\s}}\Eq(22.71.1)$$

\\so that

$$G(Y)\le 2^{|X|}G(X)\Eq(22.72)$$

\\also

$$\sum_{Y\subset X}1\le 2^{|X|}\Eq(22.73)$$

\\Hence from \equ(22.71)-\equ(22.73) we get

$$G(X_1)\le  2^{2|X|}G(X)\le G(X)\prod_{j=1}^N2^{2|X_j|}
\Eq(22.74)$$

\\Putting \equ(22.74) in \equ(22.70) we get

$$\Vert(e^{-F}-1)^+_{\ge k}(X)\Vert_{G,1}\le$$
$$\le\sum_{N\ge k}{1\over N!}
\sum_{X_1,..,X_N:\cup_jX_j=X\atop \{X_j\}\ {\rm overlap\ conn.}}
2^{2|X_1|}\Vert(e^{-F}-1)(X_1)\Vert_{G,1}
\prod_{j=2}^N2^{2|X_j|}\Vert(e^{-F}-1)(X_j)\Vert_{\io,1}\Eq(22.75)$$

\\Since the $\{X_j\}$ are  overlap connected

$$\G_p(X)\le\prod_{j=1}^N\G_p(X_j)$$

\\and hence

$$\Vert(e^{-F}-1)^+_{\ge k}\Vert_{G,1,\G_p}\le\sum_{N\ge k}{1\over N!}
\sup_\D\sum_{X\cap\D\ne\emptyset}
\sum_{X_1,..,X_N:\cup_jX_j=X\atop \{X_j\}\ {\rm overlap\ conn.}}$$
$$\Vert(e^{-F}-1)(X_1)\Vert_{G,1}\G_{p+2}(X_1)
\prod_{j=2}^N\Vert(e^{-F}-1)(X_j)\Vert_{\io,1}
\G_{p+2}(X_j)\Eq(22.76)$$

\\We now estimate the r.h.s. of \equ(22.76) by the spanning
tree argument in the proof of Lemma 5.1 of [BY]. We then get

$$\Vert(e^{-F}-1)^+_{\ge k}\Vert_{G,1,\G_p}\le\sum_{N\ge k}
O(1)^N\Vert(e^{-F}-1)\Vert_{G,1,\G_{p+3}}
\Vert(e^{-F}-1)\Vert_{\io,1,\G_{p+3}}^{N-1}=$$
$$=O(1)^k\Vert(e^{-F}-1)\Vert_{G,1,\G_{p+3}}
\Vert(e^{-F}-1)\Vert_{\io,1,\G_{p+3}}^{k-1}\sum_{N\ge k}
O(1)^{N-k}\Vert(e^{-F}-1)\Vert_{\io,1,\G_{p+3}}^{N-k}\Eq(22.77)$$

\\Now use lemma 2.5.4 and the remark following it
$$\Vert(e^{-F}-1)^+_{\ge k}\Vert_{G,1,\G_p}\le
O(1)^k\Vert F\Vert_{G,1,\G_{p+3}}
\Vert F\Vert_{\io,1,\G_{p+3}}^{k-1}\sum_{N\ge k}
O(1)^{N-k}\Vert F\Vert_{\io,1,\G_{p+3}}^{N-k}\Eq(22.77.1)$$

\\By assumption \equ(22.58) the series converges to $O(1)$.
Hence:

$$\Vert(e^{-F}-1)^+_{\ge k}\Vert_{G,1,\G_p}\le
O(1)^k\Vert F\Vert_{G,1,\G_{p+3}}
\Vert F\Vert_{\io,1,\G_{p+3}}^{k-1}\Eq(22.78)$$

\\This proves \equ(22.67). The proof of \equ(22.68)
is the same, except that we do not need the estimate
\equ(22.74), so that on the r.h.s. of \equ(22.68)
we have the norm with $\G_{p+1}$ instead of $\G_{p+3}$.
$$\eqno Q.E.D.$$
\vglue.3truecm
\\{\it 2.6. Lemmas on increments}
\vglue.3truecm
\\Let $K,\ K'$ be two polymer activities satisfing the hypothesis
of lemma 2.5.1, namely
$$\Vert K\Vert_{G,1,\G}\le O(\e^q)\qquad 
\Vert K'\Vert_{G,1,\G}\le O(\e^q)\Eq(22.79)$$

\\for some $q\ge{1\over 10}$ and $\e \ge 0$ sufficiently small.

\\Define the increments

$$\D K=K-K'\qquad \D({\cal S}_{\ge k}K)=
{\cal S}_{\ge k}K'-{\cal S}_{\ge k}K\Eq(22.80)$$

\\Then we have
\vglue.5truecm
\\LEMMA 2.6.1
\vglue.5truecm
\\For any integer $p\ge 0$

$$\Vert\D({\cal S}_{\ge k}K)^\natural\Vert_{G,1,\G_p}\le O(1)^k
L^{(\b/2+kD)}\e^{(k-1)q}\Vert\D K\Vert_{G,1,\G}\Eq(22.81)$$

\\with $O(1)$ depending on $p$.

\vglue.5truecm

\\Let now $F, F'$ be two polymer activities supported on small sets,
such that for every integer $p\ge 0$

$$\Vert F\Vert_{G,1,\G_p}\le
O(\e^q)\qquad\Vert F'\Vert_{G,1,\G_p}\le
O(\e^q)\Eq(22.82)$$
and
$$\Vert F\Vert_{\io,1,\G_p}\le
O(\e^q)\qquad\Vert F'\Vert_{\io,1,\G_p}\le
O(\e^q)\Eq(22.83)$$

\\for some $q\ge{1\over 10}$, and $\e \ge 0$ sufficiently small.
Define increments as before

\\We then have

\vglue.5truecm
\\LEMMA 2.6.2
\vglue.3truecm

$$\Vert \D(e^{-F}-1)\Vert_{G,1,\G_p}\le
O(1)\Vert\D F\Vert_{G,1,\G_p}\Eq(22.84)$$

$$\Vert\D( e^{-F}-1-F)\Vert_{G,1,\G_p}\le
O(1)\e^q\Vert\D F\Vert_{G,1,\G_p}\Eq(22.85)$$

\vglue.3truecm
\\{\it Remark:}

\\Lemma 2.6.2 remains true if the $G$-norm is replaced
by the $L^\io$ norm, by the same proof.

\vglue.5truecm
\\LEMMA 2.6.3
\vglue.3truecm
\\Given two activities 
satisfying \equ(22.82), \equ(22.83),
for any integers $k\ge 1$ and $p\ge 0$, and with $O(1)$ 
dependent on $p$

$$\Vert \D(e^{-F}-1)^+_{\ge k}\Vert_{G,1,\G_p}\le
O(1)^k\e^{(k-1)q}\Vert\D F\Vert_{G,1,\G_{p+3}}
\Eq(22.86)$$

$$\Vert  \D(e^{-F}-1)^+_{\ge k}\Vert_{\io,1,\G_p}\le
O(1)^k\e^{(k-1)q}
\Vert\D F\Vert_{\io,1,\G_{p+1}}\Eq(22.87)$$

\\We shall only prove lemma 2.6.1, the proofs of lemmas
2.6.2, 2.6.3 being similar
\vglue.5truecm
\\{\it Proof of lemma 2.6.1}

$$\D({\cal S}_{\ge k}K)^\natural=
({\cal S}_{\ge k}(K+\D K))^\natural-({\cal S}_{\ge k}K)^\natural
=$$

$$=\int_0^1dt{\dpr\over\dpr t}({\cal S}_{\ge k}(K+t\D K))^\natural=
\int_0^1dt\left({\dpr\over\dpr t}{\cal S}_{\ge k}(K+t\D K)\right)^\natural
\Eq(22.88)$$

\\From the proof of lemma 2.5.1 we have that

$$K\longrightarrow{\cal S}_{\ge k}(K)$$

\\is an analytic map between the Banach spaces with
norms $\Vert\cdot\Vert_{G,1,\G}$ and $\Vert\cdot\Vert_{G,1,\G_p}$
respectively.

\\Define

$$K(t)=K+t\D K\Eq(22.89)$$

\\Then ${\cal S}_{\ge k}(K(t))$ is analytic in $t$. By the 
Cauchy integral formula

$$\left[{\dpr\over\dpr t}{\cal S}_{\ge k}(K(t))\right]^\natural=
{1\over 2\pi i}\oint_\g dz{\left[{\cal S}_{\ge k}(K(z))\right]^\natural
\over(z-t)^2}\Eq(22.90)$$

\\where we choose the closed contour $\g$ in ${\bf C}$ as follows

$$\g:\ z-t=Re^{i\theta},\quad 0\le\theta\le 2\pi,\quad R=
{\e^q\over\Vert \D K\Vert_{G,1,\G}}\Eq(22.91)$$

\\With this choice of $\g$, and $0\le t\le 1$, for $z\in\g$

$$K(z)=K+\left(t+{\e^q\over\Vert \D K\Vert_{G,1,\G}}e^{i\theta}
\right)\D K\Eq(22.92)$$

\\Clearly $K(z)$ satisfies the hypothesis of lemma 2.5.1
under the hypothesis \equ(22.80)

$$\Vert  K(z)\Vert_{G,1,\G}\le O(\e^q)\Eq(22.93)$$

\\Hence from \equ(22.90) we have the Cauchy estimate

$$\left\Vert\left[{\dpr\over\dpr t}
{\cal S}_{\ge k}(K(t))\right]^\natural
\right\Vert_{G,1,\G_p}\le \Vert\D K\Vert_{G,1,\G}
\e^{-q}\sup_{z\in\g}
\left\Vert\left[
{\cal S}_{\ge k}(K(z))\right]^\natural
\right\Vert_{G,1,\G_p}\le$$

$$\le\Vert\D K\Vert_{G,1,\G}
\e^{-q}\sup_{z\in\g}
\left\Vert K(z)\right\Vert_{G,1,\G}^k
O(1)^kL^{\b/2+kD}\le$$

$$\le O(1)^kL^{\b/2+kD}\e^{(k-1)q}\Vert\D K\Vert_{G,1,\G}\Eq(22.94)$$

\\where in the last two lines we used lemma 2.5.1 and
\equ(22.93) respectively.

\\We use the estimate \equ(22.94) in \equ(22.88) to finish the proof.
$$\eqno Q.E.D.$$

\pagina
\vglue.5truecm\numsec=3\numfor=1\pgn=1
\\\S3. {\it Estimates for a generic RG step. Remainder estimates.}
\vglue.5truecm
\\In this section we present 
the detailed structure of the initial form of the
partition functional and the form of the activities produced by
a RG step.
\vglue.3truecm
\\{\it 3.1. Some manipulation on the starting partition functional }
\vglue.3truecm
\\The starting partition functional $z_0(\L,\phi)$ 
has the initial expression

$$z_0(\L,\phi)=e^{-V_0}{\cal E}xp(\square )\Eq(1.1)$$

\\where the initial activity $V_0$ is

$$V_0(X )=g_0V_*(X)=g_{0}\int_{X} dx\ v_*(\phi (x))=
g_{0}\int_{X} dx\left({\l_{*}\over 2\pi}\right)^{d\over 
2}e^{-{\l_{*}\over 2}\vert\phi (x)\vert^{2}}\Eq(1.2)$$

\\with $\l_{*}={\b\over 2u(0)}$. 

\\However it is convenient to write
\equ(1.1)
in a form suitable for iteration as
$$e^{-V_0}{\cal E}xp(\square\ +  K_0)\Eq(1.4)$$

\\where initially $K_{0}\equiv 0$.

\\We want to show that the structure of \equ(1.4) is reproduced
also after the generic step of RG. We start therefore from an expression
of the form

$$e^{-V(\phi)}{\cal E}xp(\square +K)\Eq(1.10)$$

\\where in \equ(1.10)
the activity $V$ is 
$V=gV_*$, with $g=O(\e)$, and

$$0< \e < L^{-10(D+2)}\Eq(1.5)$$ 

\\with $L$ sufficiently large. The bounds 

$$\Vert V(\D )\Vert_{G,1}= O(\e)\Eq(1.6.1)$$
$$\Vert V(\D )\Vert_{\io,1}= O(\e)\Eq(1.7.1)$$

\\obviously hold.
We assume for the connected activity $K$ the following structure

$$K={\cal I}+r\Eq(1.8)$$

\\where
${\cal I}$ is an activity exactly computed in second order 
perturbation theory, supported only on
small sets and satisfying the following bounds

$$\Vert{\cal I}\Vert_{G,1,\Gamma_6}\le\e^{7/4}\Eq(1.9)$$

$$\Vert({\cal S}_1{\cal I})^\natural\Vert_{G,1,\Gamma}\le
L^{-\b/4}\e^{7/4}\Eq(1.9.1)$$

\\The structure of ${\cal I}$ and the above bounds will be given in  
section 4 devoted to second order perturbation theory.

\\$r$ satisfies
the following inductive bound

$$\Vert r\Vert_{G,1,\G_6}\le \e^{5/2+\eta},\quad 0<\eta\le 1/20
\Eq(1.10.1)$$ 

\\As outlined above (see \equ(11.22))
it is convenient first of all to rewrite the partition functional

$$e^{-gV_*}{\cal E}xp(\square +K)={\cal E}xp(\square +
{\hat K})\Eq(1.13)$$

\\A suitable explicit expression of 
the connected activity ${\hat K}$ is provided by the 
following lemma.

\vglue.5truecm
\\LEMMA 3.1.1
\vglue.3truecm
$${\hat K}=P^+ +K +P^+\vee K\Eq(1.13.1)$$

\\where

$$P(X)=\cases{e^{-gV_*(\D)}-1&for $X=\D$\cr
0&otherwise\cr}\Eq(1.14)$$

\\and, following \equ(11.20.5)

$$P^+(X)=\sum_{N\ge 1}{1\over N!}\sum_{\D_1...\D_N:\ conn.
\atop \cup_j\D_j=X}\prod_{j=1}^NP(\D_j)\Eq(1.15)$$

\vglue.3truecm
\\{\it Proof}

$$e^{-gV_*}=\prod_{\D\in X}((e^{-gV_*(\D)}-1)+1)=
{\cal E}xp(\square +P^+)(X)\Eq(1.18)$$

$$e^{-gV_*}{\cal E}xp(\square +K)={\cal E}xp(\square +P^+)
{\cal E}xp(\square +K)\Eq(1.19)$$

\\and the proof follows from lemma 1.4.1
$$\eqno Q.E.D.$$

\vglue.5truecm
\\{\it 3.2. Extraction of the second order activities}
\vglue.5truecm
\\It is convenient now to separate out the second order
contributions from ${\hat K}$. We write
$${\hat K}={\hat Q}+{\hat r}\Eq(1.20)$$
with
$${\hat Q}(X)=\cases{-gV_*(\D)+
{g^2\over 2}(V_*(\D))^2+{\cal I}(\D)&for  $|X|=1$, $X\equiv \D$\cr
{g^2\over 2}\sum\limits_{\D_1,\D_2\atop
\D_1\cup\D_2=X}V_*(\D_1)V_*(\D_2)+{\cal I}(X)&for  $|X|=2$, $X$ connected\cr
0&otherwise\cr}\Eq(1.21)$$

\\and

$${\hat r}=\Pi+r+P^+\vee K\Eq(1.22)$$

where $\Pi$ is given by

$$\Pi(X)=\cases{
{-g^3\over 2}(V_*(\D))^3\int_0^1ds(1-s)^2 \exp(-sgV_*(\D))
&for  $|X|=1$, $X\equiv \D$\cr
{1\over 2}\sum\limits_{\D_1,\D_2\atop
\D_1\cup\D_2=X}\s(\D_1,\D_2)&for  $|X|=2$, $X$ connected\cr
P^+_{\ge 3}(X)&otherwise\cr}\Eq(1.22.1)$$

with

$$\s(\D_1,\D_2)={-g^3}V_*(\D_1)(V_*(\D_2))^2\int_0^1ds(1-s)
\exp(-sgV_*(\D_2))+(\D_1\rightleftharpoons\D_2)+$$
$$+g^4\prod_{j=1}^2(V_*(\D_j))^2\int_0^1ds(1-s)
\exp(sgV_*(\D_j))\Eq(1.22.2)$$

$$P^+_{\ge 3}(X)=\sum_{N\ge 3}{1\over N!}\sum_{\D_1...\D_N: \ conn.
\atop \cup_j\D_j=X}\prod_{j=1}^NP(\D_j)\Eq(1.22.3)$$

\vglue.5truecm
\\{\it 3.3.  Bound on the remainder} $\hat r$
\vglue.5truecm
\\We prove now the following result.

\vglue.5truecm
\\LEMMA 3.3.1
\vglue.3truecm

$$\Vert {\hat r}\Vert_{G,1,\G}\le O(1)\e^{5/2+\eta}\Eq(1.23)$$ 
\vglue.3truecm
\\{\it Proof}

\\In order to prove lemma 3.3.1 let us list some preliminary
trivial results: recall first that
for any 
polymers $X_1,X_2$ and for any activity $J_1,J_2$

$$\Vert J_1(X_1)J_2(X_2)\Vert_1\le \Vert J_1(X_1)\Vert_1
\Vert J_2(X_2)\Vert_1\Eq(1.23.1)$$
(see section 2)

\\This implies easily:

$$\Vert P(\D)\Vert_1\le O(1)\Vert V(\D)\Vert_1\Eq(1.24)$$
(trivial, from  \equ(1.23.1) and expansion of exponential)
$$\Vert P(\D)\Vert_{\io,1}\le O(1)\Vert V(\D)\Vert_{\io,1}
\le O(1)\e \Eq(1.25)$$
(from \equ(1.24) and \equ(1.7.1))
$$\Vert P(\D)\Vert_{G,1}\le O(1)\Vert V(\D)\Vert_{G,1}
\le O(1)\e \Eq(1.26)$$
(from \equ(1.24) and \equ(1.6.1))
$$\Vert \Pi(X)\Vert_{G,1}\le O(1)\e^3\quad {\rm for}\ |X|\le 2
\Eq(1.27)$$
(from  \equ(1.23.1), \equ(1.7.1) and \equ(1.6.1))

\\To give a bound on $P^+$ and on $P^{+}_{\ge 3}$ we use 
lemmas 2.5.4 and 2.5.5 and we obtain

$$\Vert P^+\Vert_{\io,1,\G}\le O(1)\e^{9/10}
\Eq(1.31.1)$$

\\and

$$\Vert P^{+}_{\ge 3}\Vert_{G,1,\G}
\le {O(1)}\e^{27/10}
\Eq(1.32)$$

\\Therefore we have, using \equ(1.32) and \equ(1.27)
and again the smallness of $\e$ \equ(1.5)

$$\Vert\Pi\Vert_{G,1,\G}\le {O(1)}\e^{27/10}
\le {1\over 8}L^{-1}\e^{5/2+\eta}
\Eq(1.33)$$

\\Finally we consider
$P^+\vee K$. By definition of $\vee$ and using the [BY] 
spanning tree argument we obtain

$$\Vert P^+\vee K\Vert_{G,1,\G}\le
\sum_{N,M\ge 1} (O(1))^{N+M}\Vert P^+\Vert^N_{\io,1,\G_3}
\Vert K\Vert^M_{G,1,\G_3}\Eq(1.38)$$

\\Note that from \equ(1.8), \equ(1.9) and \equ(1.10.1)
it follows that

$$\Vert K\Vert_{G,1,\G_3}\le \e^{7/4}\Eq(1.38.1)$$

\\Then from \equ(1.31.1) and \equ(1.38.1), and summing
the series, we obtain

$$\Vert P^+\vee K\Vert_{G,1,\G}\le O(1)\e^{9/10+7/4}\le
\e^{5/2+\eta}
\Eq(1.39)$$

\\Putting toghether \equ(1.10.1), \equ(1.39), and \equ(1.33) 
we obtain \equ(1.23)
$$\eqno Q.E.D.$$

\\Finally it is useful to note
\vglue.5truecm
\\LEMMA 3.3.2
\vglue.3truecm

$$\Vert {\hat K}\Vert_{G,1,\G}\le O(1)\e^{9/10}\Eq(1.40)$$
$$\Vert{\hat K}+gV_*\Vert_{G,1,\G}\le O(1)\e^{7/4}\Eq(1.40.1)$$

\\The proof is simply obtained by the definition of ${\hat K}$
\equ(1.20), by the estimate \equ(1.6.1) with the smallness
of $\e$, by \equ(1.9) and by the lemma 3.3.1 above.
\vglue.5truecm
\\{\it 3.4. The action of RG }
\vglue.3truecm
\\{\it Reblocking-rescaling}

\\The connected activity 
${\hat K}$ is given by \equ(1.20), \equ(1.21), \equ(1.22).
It is convenient to separate out the second order term ${\cal I}$
from ${\hat Q}$. So we define ${\hat Q}^{(0)}(X)$, supported
on connected sets $|X|\le 2$ by:

$${\hat Q}^{(0)}(X)=\cases{
-gV_*(\D)+
{g^2\over 2}(V_*(\D))^2&for  $|X|=1$, $X\equiv \D$\cr
{g^2\over 2}\sum\limits_{\D_1,\D_2\atop
\D_1\cup\D_2=X}V_*(\D_1)V_*(\D_2)&for  $|X|=2$, $X$ connected\cr
0&otherwise\cr}\Eq(1.41)$$

\\Now it is easy to see that we can express 
the reblocked-rescaled activity ${\cal S}{\hat K}$ as:

$${\cal S}{\hat K}={\cal S}_1{\hat Q}^{(0)}+
g^2{\cal S}_2(V_*)+{\cal S}_1{\cal I}+{\cal S}_1{\hat r}
+{\tilde r}\Eq(1.43)$$

\\where

$${\tilde r}={\cal S}_{\ge 3}{\hat K}+{\cal S}_2({\hat K}+gV_*)
+{\cal S}_2(gV_*,{\hat K}+gV_*)\Eq(1.44)$$

\\{\it Remarks}

\\1) In \equ(1.43), \equ(1.44) $V_*$ is supported on single blocks.

\\2) ${\cal S}_1$ is the linearized reblocking-rescaling,
${\cal S}_2$ is the quadratic part of ${\cal S}$ and ${\cal S}_{\ge 3}$
stands for ${\cal S}-{\cal S}_1-{\cal S}_2$. In the last term of
\equ(1.44), the quadratic reblocking sum has for each term one
factor $gV_*$ and the other ${\hat K}+gV_*$.

\\3) Note that ${\tilde r}$ is formally of $O(\e^3)$. We shall
estimate it after performing the fluctuation integration.

\vglue.3truecm
\\{\it Fluctuation integration}

\\We have

$$\mu_{{\cal R} \G}*{\cal E}xp(\square+{\cal S}{\hat K})=
{\cal E}xp(\square+({\cal S}{\hat K})
^\natural)\Eq(1.45)$$

\\Then

$$({\cal S}{\hat K})
^\natural=-gL^\e V_*+{g^2\over 2}({\cal R}{\tilde Q})^\natural+
({\cal S}_1{\cal I})^\natural+({\cal S}_1{\hat r})^\natural
+{\tilde r}^\natural\Eq(1.46)$$

\\where $V_*$ above is again supported on single blocks
and the connected activity ${\tilde Q}$ 
is supported on $L$ polymers $LZ$ such that  
$|Z|\le 2$ and has the following expression
$${\tilde Q}(L\D)=(V_*(L\D))^2$$
$${\tilde Q}(L\D\cup L\D')=V_*(L\D)V_*(L\D')+
(\D\rightleftharpoons\D')\Eq(1.47)$$

\\The first three terms of \equ(1.46) are 
contributions up to second order
in perturbation theory. They will be treated
in more detail in the next section.
\vglue.3truecm
\\{\it Preliminary extraction}

\\It is convenient to extract the first order term and to rewrite our
partition functional in the following way

$${\cal E}xp(\square+({\cal S}{\hat K})
^\natural)(L^{-1}\L)=e^{-L^\e gV_*(L^{-1}\L)}
{\cal E}xp(\square+{\tilde K})(L^{-1}\L)\Eq(1.48)$$

\\where the activity ${\tilde K}$ is described by the following lemma
\vglue.5truecm
\\LEMMA 3.4.1
\vglue.3truecm
$${\tilde K}=({\cal S}{\hat K})
^\natural+{\tilde P}^++{\tilde P}^+
\vee({\cal S}{\hat K})
^\natural\Eq(1.49)$$

\\where

$${\tilde P}=
\cases{e^{-L^\e gV_*(\D)}-1&for $X=\D$\cr
0&otherwise\cr}\Eq(1.49.1)$$

\vglue.3truecm
\\{\it Proof}

\\By \equ(1.48)

$${\cal E}xp(\square+{\tilde K})=
e^{L^\e gV_*}{\cal E}xp(\square+({\cal S}{\hat K})
^\natural)\Eq(1.50)$$

\\Then \equ(1.49) follows as in lemma 3.1.1
$$\eqno Q.E.D.$$

\\We now isolate out the terms proportional to $g$ and $g^2$
in \equ(1.49). Introduce the notation
${\tilde P}^+_{(\le 2)}$ (to be distinguished from ${\tilde P}^+_{\le 2}$)
to represent the sum of contributions proportional to $g$ and  $g^2$
in ${\tilde P}^+$. Clearly

$${\tilde P}^+_{(\le 2)}(X)=\cases{
L^\e gV_*(\D)+{1\over 2}L^{2\e}g^2
(V_*(\D))^2&for  $|X|=1$, $X\equiv \D$\cr
{1\over 2}L^{2\e}g^2\sum\limits_{\D_1,\D_2\atop
\D_1\cup\D_2=X}V_*(\D_1)V_*(\D_2)&for  $|X|=2$, $X$ connected\cr
0&otherwise\cr}\Eq(1.51)$$

\\Note also

$$({\tilde P}^+\vee({\cal S}{\hat K})
^\natural)_{(\le 2)}(X)=\cases{
-L^{2\e}g^2
(V_*(\D))^2&for  $|X|=1$, $X\equiv \D$\cr
-L^{2\e}g^2\sum\limits_{\D_1,\D_2\atop
\D_1\cup\D_2=X}V_*(\D_1)V_*(\D_2)&for  $|X|=2$, $X$ connected\cr
0&otherwise\cr}\Eq(1.52)$$

\\Adding \equ(1.51) and \equ(1.52) we get

$$({\tilde P}^++{\tilde P}^+\vee({\cal S}{\hat K})
^\natural)_{(\le 2)}=L^\e gV_*-{1\over 2}L^{2\e}g^2{\tilde Q}
\Eq(1.53)$$

\\with $V_*$ supported on single blocks and ${\tilde Q}$ defined in
\equ(1.47). Hence returning to \equ(1.49) we obtain

$${\tilde K}={g^2\over 2}\int_0^1ds{\dpr\over\dpr s}
({\cal R}{\tilde Q})^{s\natural}+
({\cal S}_1{\cal I})^\natural+({\cal S}_1{\hat r})^\natural
+{\tilde r}^\natural+{\bar r}\Eq(1.54)$$

\\where the new remainder ${\bar r}$ is

$${\bar r}=({\tilde P}^+-{\tilde P}^+_{(\le 2)})+
({\tilde P}^+\vee({\cal S}{\hat K})
^\natural
-({{\tilde P}^+\vee({\cal S}{\hat K})
^\natural})_{\le 2}\Eq(1.55)$$

\\{\it Remarks}

\\1) ${\bar r}$ is formally $O(\e^3)$ and 
toghether with
${\tilde r}^\natural$ needs no further extraction. 
Their norms will be estimated
in the following subsection 3.5.

\\2) ${\tilde K}$ needs further extractions, namely from first
and third term in \equ(1.54). The first extracted term, denoted
$F_{{\tilde Q}}$, is the perturbative relevant part of the first
term of \equ(1.54). It will be computed explicitely 
in the following section 4, and its
irrelevant part will be ${\cal I}$. Note that in section 4 it
will be also proved that $({\cal S}_1{\cal I})^\natural$
needs no extraction, since its norm, by exact computations, goes
down by a contracting factor.
The second extracted term, denoted
$F_{{\hat r}}$, is the relevant part of $({\cal S}_1{\hat r})^\natural$.
Althought ${\hat r}$ is $O(\e^{5/2+\eta})$, an extraction 
has to be performed because the linear reblocking for small sets
produces a factor
$L^D$, and therefore a contractive factor has to be obtained.
This extraction, and the control of the obtained remainder,
will be the subject of section 5.

\vglue.5truecm
\\{\it 3.5. Bounds on irrelevant remainders}
\vglue.5truecm
\\We prove now the following results.

\vglue.5truecm
\\LEMMA 3.5.1
\vglue.3truecm
$$\Vert{\bar r}\Vert_{G,1,\G_p}\le L^{-\b/2}\e^{5/2+\eta}\Eq(1.56)$$
\vglue.3truecm
\\{\it Proof}

\\The first addend of ${\bar r}$, namely the term
$({\tilde P}^+-{\tilde P}^+_{(\le 2)})$, has the same form
as $\Pi$ in \equ(1.22.1) with $V_*$ substituted by $- L^\e V_*$.
Therefore, since $L^\e=O(1)$, one can obtain the bound

$$({\tilde P}^+-{\tilde P}^+_{(\le 2)})\le L^{-1}\e^{5/2+\eta}\Eq(1.57)$$

\\along the same lines as for $\Pi$.
To control the term $({\tilde P}^+\vee({\cal S}{\hat K})
^\natural
-({\tilde P}^+\vee({\cal S}{\hat K})^\natural)_{\le 2}$ 
it is enough to estimate
${\tilde P}^+_{\ge 2}\vee({\cal S}{\hat K})
^\natural$ and ${\tilde P}\vee({\cal S}({\hat K}+gV_*))
^\natural$. We have

$$\Vert{\tilde P}^+_{\ge 2}\vee({\cal S}{\hat K})
^\natural\Vert_{G,1,\G_p}\le
\sum_{N,M\ge 1} (O(1))^{N+M}
\Vert {\tilde P}^+_{\ge 2}\Vert^N_{\io,G,1,\G_{p+3}}
\Vert ({\cal S}{\hat K})
^\natural\Vert^M_{G,1,\G_{p+3}}\Eq(1.58)$$

\\By lemma 2.5.1
and using the fact that from the condition of the
smallness of $\e$ \equ(1.5)
and $D=1$ it turns out $L\le \e^{-1/30}$ we have

$$\Vert({\cal S}{\hat K})
^\natural\Vert^M_{G,1,\G_{p+3}}\le O(1) L^{\b/2}L^D
\Vert{\hat K}\Vert^M_{G,1,\G}\le L^2\e^{9/10}
\le\e^{9/10-2/30}\le \e^{8/10}\Eq(1.59)$$

\\It is easy to see, as in the proof of \equ(1.32), that

$$\Vert {\tilde P}^+_{\ge 2}\Vert^N_{G,1,\G_{p+3}}\le
 \e^{18/10}\Eq(1.60)$$

\\This gives

$$\Vert{\tilde P}^+_{\ge 2}\vee({\cal S}{\hat K})
^\natural\Vert_{G,1,\G_p}\le \e^{26/10}\le L^{-1}\e^{5/2+\eta}
\Eq(1.61)$$

\\We obtain in the same way

$$\Vert{\tilde P}\vee({\cal S}({\hat K}+gV_*))
^\natural\Vert_{G,1,\G_p}\le O(1)\e^{9/10}\e^{7/4}L^{D+\b/2}
\le  L^2\e^{53/20}\le L^{-1} \e^{5/2+\eta}\Eq(1.62)$$

From \equ(1.57), \equ(1.61) and \equ(1.62) we obtain the lemma.
$$\eqno Q.E.D.$$
\vglue.5truecm
\\LEMMA 3.5.2
\vglue.3truecm
$$\Vert{\tilde r}^\natural\Vert_{G,1,\G_p}\le 
L^{-\b/2}\e^{5/2+\eta}\Eq(1.63)$$
\vglue.3truecm
\\{\it Proof}

\\${\tilde r}$ is given  in \equ(1.44). There are three terms. By lemma
2.5.1 the contribution of the first is bounded by

$$\Vert({\cal S}_{\ge 3}{\hat K})^\natural\Vert_{G,1,\G_p}\le
O(1)L^{\b/2+3D}\Vert{\hat K}\Vert_{G,1,\G}^3\le$$
$$\le{L^{4-\b/2}\over 3}\e^{27/10}\le {1\over 3}\e^{27/10-4/30}L^{-\b/2}\le
{1\over 3}L^{-\b/2}\e^{5/2+\eta}\Eq(1.64)$$

\\From the second, by lemma 3.3.2, we get

$$\Vert({\cal S}_2({\hat K}+gV_*))^\natural\Vert_{G,1,\G_p}\le
O(1)L^{\b/2+2D}\Vert{\hat K}+gV_*\Vert_{G,1,\G}^2\le$$
$$\le{L^{3-\b/2}\over 3}e^{14/4}\le {1\over 3}L^{-\b/2}\e^{14/4-1/10}\le
{1\over 3}L^{-\b/2}\e^{5/2+\eta}\Eq(1.65)$$

\\whilst from the third we obtain

$$\Vert
({\cal S}_2(gV_*,{\hat K}+gV_*))^\natural\Vert_{G,1,\G_p}\le
O(1)L^{\b/2+2D}\Vert{\hat K}+gV_*\Vert_{G,1,\G}
\Vert{\hat K}\Vert_{G,1,\G}\le$$
$$\le{L^{3-\b/2}\over 3}e^{7/4}e^{9/10}\le {1\over 3}
L^{-\b/2}\e^{53/20-1/10}\le
{1\over 3}L^{-\b/2}\e^{5/2+\eta}\Eq(1.66)$$

\\summing \equ(1.64), \equ(1.65) and \equ(1.66) we obtain the proof.
$$\eqno Q.E.D.$$

\pagina

\vglue.5truecm\numsec=4\numfor=1\pgn=1
\\\S4. {\it RG step to second order. Relevant and irrelevant terms. Estimates.}
\vglue.5truecm
\\{\it 4.1. The starting second order activity}
\vglue.3truecm
\\In this section we consider the contribution to the
partition functional
up to the second order

$${\cal E}xp(\square +
{\hat K}_{\le 2})\equiv{\cal E}xp(\square + {\hat Q})\Eq(2.1)$$

\\As we showed in \equ(1.21) the second order 
activity ${\hat Q}$ is given by

$${\hat Q}(X)=\cases{\displaystyle -gV_*(\D)+
{g^2\over 2}(V_*(\D))^2+{\cal I}(\D)&for  $|X|=1$, $X\equiv \D$\cr
\displaystyle{g^2\over 2}\sum\limits_{\D_1,\D_2\atop
\D_1\cup\D_2=X}V_*(\D_1)V_*(\D_2)+{\cal I}(X)&for  $|X|=2$, $X$ connected\cr
0\vphantom{1\over 2}&otherwise\cr}\Eq(2.2)$$

\\It is convenient in the following computations to use the obvious
representation

$$V_*(\D)=\left({\l_*\over 2\pi}\right)^{d/2}\int_\D
d^Dx\int{d^dk\over(2\pi\l_*)^{d/2}}e^{-{|k|^2\over 2\l_*}}
e^{ik\cdot\phi(x)}\Eq(2.2.1)$$

\\The irrelevant second order activity ${\cal I}(X)=
{\cal I}_k(X)$ depends actually
on the number $k$ of iterations of RG so far performed. We assume here
inductively that ${\cal I}_0(X)=0$ and for $k\ge 1$

$${\cal I}_k(X)=\sum_{l=1}^kg_{k-l}^2{\bar{\cal I}}_l(X)\Eq(2.2.2)$$ 

In the above $g=g_k$ and we assume inductively that $g_j=O(\e)$ for all
$0\le j\le k$. 

\\We will see
in a moment that the action of RG will give us
an irrelevant second order activity of the form ${\cal I}_{k+1}(X)$.
The activities ${\bar{\cal I}}_l(X)$ are supported on polymer $X$
such that $|X|\le 2$, and are defined by the following expression:

$${\bar{\cal I}}_l(\D)=L^{2l\e}{1\over 2}
\left({\l_*\over 2\pi}\right)^{d}
\int_0^1ds{\dpr\over\dpr s}\int_\D
d^Dx_1\int{d^dk_1\over(2\pi\l_*)^{d/2}}
\int_\D
d^Dx_2\int{d^dk_2\over(2\pi\l_*)^{d/2}}$$
$$\int_0^1dt{\dpr\over\dpr t}\exp\left[{-{1\over 2\l_*}(k,I_l(s,t)k)
+i(k_1\cdot \phi(x_1)+tk_2\cdot(\phi(x_2)-\phi(x_1)))}\right]
\Eq(2.3)$$

$${\bar{\cal I}}_l(\D\cup\D')=L^{2l\e}{1\over 2}
\left({\l_*\over 2\pi}\right)^{d}
\int_0^1ds{\dpr\over\dpr s}\int_\D
d^Dx_1\int{d^dk_1\over(2\pi\l_*)^{d/2}}
\int_{\D'}
d^Dx_2\int{d^dk_2\over(2\pi\l_*)^{d/2}}$$
$$\int_0^1dt{\dpr\over\dpr t}\exp\left[{-{1\over 2\l_*}(k,I_l(s,t)k)
+i(k_1\cdot \phi(x_1)+tk_2\cdot(\phi(x_2)-\phi(x_1)))}\right]
+(\D\rightleftharpoons\D')\Eq(2.4)$$

\\where $k\equiv (k_1,k_2)$
and 

$$I_l(s,t)=\pmatrix{\displaystyle 1\vphantom{1\over 2}&-tC_l(s)\cr
\displaystyle -tC_l(s)&2\left[C_l(s)+
(1-t^2)D_1\G(x_2-x_1)\right]\cr}
\Eq(2.4.1)$$

\\with

$$C_l(s)=
\left[{(1-s{\bar \G}(L^{(l-1)}(x_2-x_1)))\over L^{\b(l-1)}}+
\sum_{p=0}^{l-2}{{\bar \G}(0)-
{\bar \G}(L^p(x_2-x_1))\over L^{p\b}}\right]$$

\\where the sums over $p$ are void if $l=1$, and with

$$D_1{\bar \G}(x_2-x_1)=\sum_{j=1}^\io L^{j\b}\left[
{\bar \G}(0)-{\bar \G}((x_2-x_1)/L^j)\right]\Eq(2.18.22)$$

\\Note that $D_1{\bar \G}(x)\ge 0$ and that the series converges by virtue
of the estimate

$$0 \le L^{j\b}({\bar \G}(0)-{\bar \G}(x/L^j)) 
\le O(1)L^{-(j-1)(2-\b)}\left|x\right|^{2}$$

\\We will compute explicitly in the rest of this section
the evolution of these terms under
RG transfomation.
\vglue.5truecm
\\{\it 4.2. Reblocking}
\vglue.3truecm
\\First we consider the reblocking of ${\hat Q}$ up to the
second order: 
it is easy to see that

$$({\cal B}{\hat Q})_{(\le 2)}(LZ)=\sum_{X\atop {\bar X}=LZ}{\hat Q}(X)
+ {1\over 2}\sum_{X_1,X_2\ {\rm disj}\atop{{\bar X_1},{\bar X_2}
{\rm overlap\ conn.}\atop {\bar X_1}\cup{\bar X_2}=LZ}}{\hat Q}(X_1)
{\hat Q}(X_2)=$$
$$=\sum_{X\atop {\bar X}=LZ}{\hat Q}(X)
+ {1\over 2}\sum_{\D_1,\D_2\ {\rm disj}\atop{{\bar \D_1},{\bar \D_2}
{\rm overlap\ conn.}\atop {\bar \D_1}\cup{\bar \D_2}=LZ}}V(\D_1)
V(\D_2)\Eq(2.5)$$

\\This gives, in the case $|LZ|=1$, i.e. $Z\equiv \D$,

$$({\cal B}{\hat Q})_{(\le 2)}(L\D)=-V(L\D)+{V(L\D)^2\over 2}+
\sum_{X\atop {\bar X}=L\D}{\cal I}_k(X)\Eq(2.6)$$

\\while, for $|Z|=2$, i.e. $Z\equiv (\D\cup\D')$,

$$({\cal B}{\hat Q})_{(\le 2)}(L(\D\cup\D'))=+{V(L\D)V(L\D')\over 2}+
(\D\rightleftharpoons\D')+
{\cal I}_k(\D_1\cup\D_2)\Eq(2.7)$$

\\where in \equ(2.7), since $D=1$, $\D_1\subset L\D$ and $\D_2\subset L\D'$ are
uniquely defined by the fact that they have to be overlap connected,
and therefore the relation $\D_1\cap\D_2=L\D\cap L\D'$ has to be 
fullfilled.
No symmetrization is necessary in $\D_1, \D_2$ 
because ${\cal I}_k(\D_1\cup\D_2)$ is already symmetrized.
Note that, exploiting the compactness of the propagator
${\bar\G}$, the reblocking for the irrelevant activities 
${\cal I}_k$ can be written in the following form

$$\sum_{X\atop {\bar X}=L\D}{\cal I}_k(X)={\cal B}{\cal I}_k(L\D)=
\sum_{l=1}^kg_{k-l}^2{\cal B}{\bar{\cal I}}_l(L\D)\Eq(2.7.1)$$

$${\cal I}_k(\D_1\cup\D_2)={\cal B}{\cal I}_k(L(\D\cup\D'))=
\sum_{l=1}^kg_{k-l}^2{\cal B}{\bar{\cal I}}_l(L(\D\cup\D'))\Eq(2.7.2)$$

\\with

$${\cal B}{\bar{\cal I}}_l(L\D)=L^{2l\e}{1\over 2}
\left({\l_*\over 2\pi}\right)^{d}
\int_0^1ds{\dpr\over\dpr s}\int_{L\D}
d^Dx_1\int{d^dk_1\over(2\pi\l_*)^{d/2}}
\int_{L\D}
d^Dx_2\int{d^dk_2\over(2\pi\l_*)^{d/2}}$$
$$\int_0^1dt{\dpr\over\dpr t}\exp\left[{-{1\over 2\l_*}(k,I_l(s,t)k)
+i(k_1\cdot \phi(x_1)+tk_2\cdot(\phi(x_2)-\phi(x_1)))}\right]
\Eq(2.7.3)$$

$${\cal B}{\bar{\cal I}}_l(L(\D\cup\D')))=L^{2l\e}{1\over 2}
\left({\l_*\over 2\pi}\right)^{d}
\int_0^1ds{\dpr\over\dpr s}\int_{L\D}\!\!\!
d^Dx_1\int{d^dk_1\over(2\pi\l_*)^{d/2}}
\int_{L\D'}\!\!\!
d^Dx_2\int{d^dk_2\over(2\pi\l_*)^{d/2}}$$
$$\int_0^1dt{\dpr\over\dpr t}\exp\left[{-{1\over 2\l_*}(k,I_l(s,t)k)
+i(k_1\cdot \phi(x_1)+tk_2\cdot(\phi(x_2)-\phi(x_1)))}\right]
+(\D\rightleftharpoons\D')\Eq(2.7.4)$$

\vglue.5truecm
\\{\it 4.3. Rescaling, integration and preliminary extraction}
\vglue.3truecm
\\Now we rescale and we integrate the fluctuating field
and we obtain immediately \equ(1.46) up to the second order in $g$:

$$({\cal S}{\hat Q})^{\natural}_{(\le 2)}(\D)=
-gL^\e V_*+{g^2\over 2}({\cal R}{\tilde Q})^\natural+
({\cal S}_1{\cal I}_k)^\natural\Eq(2.8)$$

\\where the $V_*$ is supported only on single blocks and
${\tilde Q}$ is defined by

$${\tilde Q}(L\D)=(V_*(L\D))^2$$
$${\tilde Q}(L\D\cup L\D')=V_*(L\D)V_*(L\D')+
(\D\rightleftharpoons\D')\Eq(2.9)$$

\\Following again section 3 we obtain
for the contribution up to the second order ${\tilde K}_{(2)}$
of the activity ${\tilde K}$ defined in \equ(1.54) the following 
expression

$${\tilde K}_{(2)}={g^2\over 2}\int_0^1ds{\dpr\over\dpr s}
({\cal R}{\tilde Q})^{s\natural}+
({\cal S}_1{\cal I})^\natural\equiv{\tilde K}_{\tilde Q}
+({\cal S}_1{\cal I}_k)^\natural\Eq(2.10)$$

\\where, using \equ(11.24.2)  and the representation \equ(2.2.1)
we have

$${\tilde K}_{\tilde Q}(\D)=L^{2\e}{g^2\over 2}
\left({\l_*\over 2\pi}\right)^{d}
\int_0^1ds{\dpr\over\dpr s}\int_\D
d^Dx_1\int{d^dk_1\over(2\pi\l_*)^{d/2}}
\int_\D
d^Dx_2\int{d^dk_2\over(2\pi\l_*)^{d/2}}$$

$$\int d\mu_{{s\cal R}\G}(\z)\exp\left[-{|k_1|^2+|k_2|^2\over 2\l_*}
+i\left(k_1\cdot(\zeta(x_1)+\phi(x_1))+
k_2\cdot(\zeta(x_2)+\phi(x_2))\right)\right]\Eq(2.11)$$

$${\tilde K}_{\tilde Q}(\D\cup\D')=L^{2\e}{g^2\over 2}
\left({\l_*\over 2\pi}\right)^{d}
\int_0^1ds{\dpr\over\dpr s}
\int_\D
d^Dx_1\int{d^dk_1\over(2\pi\l_*)^{d/2}}
\int_{\D'}
d^Dx_2\int{d^dk_2\over(2\pi\l_*)^{d/2}}$$

$$
\int d\mu_{{s\cal R}\G}(\z)\exp\left[-{|k_1|^2+|k_2|^2\over 2\l_*}
+i\left(k_1\cdot(\zeta(x_1)+\phi(x_1))+
k_2\cdot(\zeta(x_2)+\phi(x_2))\right)\right]+$$
$$+(\D\rightleftharpoons\D')\Eq(2.12)$$

\\\equ(2.11), \equ(2.12) are obtained writing ${\tilde Q}$
in terms of the representation \equ(2.2.1) and performing 
the change of variables $k\rightarrow L^{-\b/2}k$,
$x\rightarrow Lx$.

\\The gaussian 
integral with respect to the measure $\mu_{{\cal R}\G(s)}$
appearing in \equ(2.11), \equ(2.12)
is easily done:

$${\tilde K}_{\tilde Q}(\D)=L^{2\e}{g^2\over 2}
\left({\l_*\over 2\pi}\right)^{d}
\int_0^1ds{\dpr\over\dpr s}\int_\D
d^Dx_1\int{d^dk_1\over(2\pi\l_*)^{d/2}}
\int_\D
d^Dx_2\int{d^dk_2\over(2\pi\l_*)^{d/2}}$$

$$\exp\left[-{1\over 2\l_*}(k,\s_1k)
+i(k_1\cdot \phi(x_1)+k_2\cdot\phi(x_2))\right]
\Eq(2.13)$$

$${\tilde K}_{\tilde Q}(\D\cup\D')=L^{2\e}{g^2\over 2}
\left({\l_*\over 2\pi}\right)^{d}
\int_0^1ds{\dpr\over\dpr s}
\int_\D
d^Dx_1\int{d^dk_1\over(2\pi\l_*)^{d/2}}
\int_{\D'}
d^Dx_2\int{d^dk_2\over(2\pi\l_*)^{d/2}}$$
$$\exp\left[-{1\over 2\l_*}(k,\s_1k)
+i(k_1\cdot \phi(x_1)+k_2\cdot\phi(x_2))\right]+
(\D\rightleftharpoons\D')\Eq(2.14)$$

\\where 
the matrix $\s_1$ is given by

$$\s_1=\pmatrix{
1&s{\bar \G}(x_2-x_1)\cr
s{\bar \G}(x_2-x_1)&1\cr}\Eq(2.15.1)$$

\\In order to extract the relevant part from \equ(2.13),
\equ(2.14) it is useful to perform the following change of variables
$k_1\rightarrow k_1-k_2$, $k_2\rightarrow k_2$ obtaining 

$${\tilde K}_{\tilde Q}(\D)=L^{2\e}{g^2\over 2}
\left({\l_*\over 2\pi}\right)^{d}
\int_0^1ds{\dpr\over\dpr s}\int_\D
d^Dx_1\int{d^dk_1\over(2\pi\l_*)^{d/2}}
\int_\D
d^Dx_2\int{d^dk_2\over(2\pi\l_*)^{d/2}}$$
$$\exp\left[-{1\over 2\l_*}(k,T\s_1k)
+i(k_1\cdot \phi(x_1)+k_2\cdot(\phi(x_2)-\phi(x_1)))\right]
\Eq(2.16.1)$$

$${\tilde K}_{\tilde Q}(\D\cup\D')=L^{2\e}{g^2\over 2}
\left({\l_*\over 2\pi}\right)^{d}
\int_0^1ds{\dpr\over\dpr s}
\int_\D
d^Dx_1\int{d^dk_1\over(2\pi\l_*)^{d/2}}
\int_{\D'}
d^Dx_2\int{d^dk_2\over(2\pi\l_*)^{d/2}}$$
$$\exp\left[-{1\over 2\l_*}(k,T\s_1k)
+i(k_1\cdot \phi(x_1)+k_2\cdot(\phi(x_2)-\phi(x_1)))\right]
+(\D\rightleftharpoons\D')\Eq(2.17.1)$$

\\where the matrix $T\s_1$ is given by

$$T\s_1=\pmatrix{
1&-(1-s{\bar \G}(x_2-x_1))\cr
-(1-s{\bar \G}(x_2-x_1))&2(1-s{\bar \G}(x_2-x_1))\cr}\Eq(2.18.1)$$

\\In order to give the explicit expression of 
$({\cal S}_1{\cal I}_k)^\natural$ and to prove the iterative
form of ${\cal I}_k$ \equ(2.2.2) it is useful the following lemma
\vglue.5truecm
\\LEMMA 4.3.1
\vglue.3truecm
$$({\cal S}_1{\bar{\cal I}}_l)^\natural={\bar{\cal I}}_{l+1}\Eq(2.16.2)$$

\vglue.3truecm
\\{\it Proof}

\\Let us write the proof for the single block contribution only.
The proof for the couple of adjacent blocks is identical.
Integrating and rescaling \equ(2.7.3) we obtain

$$({\cal S}_1{\bar{\cal I}}_l)^\natural(\D,\phi)=
(({\cal B}_1{\bar{\cal I}}_l)^\sharp)(L\D,\cal R \phi)=$$
$$=L^{2l\e}{1\over 2}
({\l_*\over 2\pi})^{d}
\int_0^1ds{\dpr\over\dpr s}\int_{L\D}
d^Dx_1\int{d^dk_1\over(2\pi\l_*)^{d/2}}
\int_{L\D}
d^Dx_2\int{d^dk_2\over(2\pi\l_*)^{d/2}}.$$
$$.\int_0^1dt{\dpr\over\dpr t}e^{-{1\over 2\l_*}(k,I_l(s,t)k)}
e^{i(k_1\cdot L^{\b/2}\phi(x_1/L)+tk_2\cdot L^{\b/2}(\phi(x_2/L)-\phi(x_1/L)))}
e^{-{1\over 2}(k,\G(t)k)}$$

\\with the matrix $\G(t)$ given by

$$\G(t)=\pmatrix{
\G(0)&-t(\G(0)-\G(x_2-x_1))\cr
-t(\G(0)-\G(x_2-x_1))&2t^2(\G(0)-\G(x_2-x_1))\cr}$$

\\Adding ${\l_*}^{-1}I_l(s,t)$ and  $\G(t)$, performing 
the change of variables and some elementary manipulations, and
$k\rightarrow L^{-\b/2}k$,
$x\rightarrow Lx$ we obtain the proof.
$$\eqno Q.E.D.$$

\vglue.5truecm
\\{\it 4.4. Extraction}
\vglue.3truecm

\\Now we define the relevant part
$F_{\tilde Q}(Z,\phi)={\cal L}{\tilde K}_{\tilde Q}(Z,\phi)$ 
in the following way
$${\cal L}{\tilde K}_{\tilde Q}(\D)=
L^{2\e}{g^2\over 2}\left({\l_{*}\over 2\pi}\right)^d
\int_\D
d^Dx_1\int{d^dk_1\over(2\pi\l_*)^{d/2}}
\int_\D
d^Dx_2\int{d^dk_2\over(2\pi\l_*)^{d/2}}$$
$$\int_0^1ds{\dpr\over\dpr s}\exp\left[-{1\over 2\l_*}(k,R_1k)
+i(k_1\cdot \phi(x_1))\right]
\Eq(2.15)$$

$${\cal L}{\tilde K}_{\tilde Q}(\D\cup\D',\phi)=
L^{2\e}{g^2\over 2}\left({\l_{*}\over 2\pi}\right)^{d}
\int_\D
d^Dx_1\int{d^dk_1\over(2\pi\l_*)^{d/2}}
\int_{\D'}
d^Dx_2\int{d^dk_2\over(2\pi\l_*)^{d/2}}$$
$$\int_0^1ds{\dpr\over\dpr s}
\exp\left[-{1\over 2\l_*}(k,R_1k)
+i(k_1\cdot \phi(x_1))\right]
+(\D\rightleftharpoons\D')\Eq(2.16)$$

\\where the matrix $R_1$ is given by

$$R_1=\pmatrix{
1&0\cr
0&2[1-s{\bar \G}(x_2-x_1)+D_1{\bar\G}(x_2-x_1)]\cr}\Eq(2.18.2)$$

\\It is convenient to perform the gaussian integral in the
$k$'s variables obtaining for \equ(2.15) and \equ(2.16) the following 
expression

$$F_{\tilde Q}(\D,\phi)
=L^{2\e}{g^2\over 2}\left({\l_{*}\over 2\pi}\right)^d
\int_\D d^Dx_1\int_{\D}
d^Dx_2\int_0^1ds{\dpr\over\dpr s}$$
$$[2(1-s{\bar \G}(x_2-x_1)+D_1{\bar\G}(x_2-x_1))]^{-d/2}
e^{-{\l_{*}\over 2}\vert\phi (x_1)\vert^{2}}\Eq(2.18.3)$$

$$F_{\tilde Q}(\D\cup\D',\phi)
=L^{2\e}{g^2\over 2}\left({\l_{*}\over 2\pi}\right)^d
\int_\D d^Dx_1\int_{\D'}
d^Dx_2\int_0^1ds{\dpr\over\dpr s}$$
$$[2(1-s{\bar \G}(x_2-x_1)+D_1{\bar\G}(x_2-x_1))]^{-d/2}
e^{-{\l_{*}\over 2}\vert\phi (x_1)\vert^{2}}
+(\D\rightleftharpoons\D')\Eq(2.18.4)$$

\\Following [BDH2] , we want to write $F_{\tilde Q}(Z,\phi)$
in terms of the sets where the dependence from the field $\phi$ is
localized. In other words, we want to
write the decomposition

$$F_{\tilde Q}(Z,\phi)=\sum_{\D\subset Z}F_{\tilde Q}(Z,\D,\phi)\Eq(2.17)$$

\\where in $F_{\tilde Q}(Z,\D,\phi)$ appear only fields defined in $\D$.
The explicit expression for the relevant contribution
$F_{\tilde Q}(Z,\D,\phi)$ is therefore

$$F_{\tilde Q}(Z,\D,\phi)=
\int_{\D} dx_1
v_*(\phi (x_1))f_{\tilde Q}(Z,\D)\Eq(2.18)$$

\\where

$$f_{\tilde Q}(\D,\D)=L^{2\e}{g^2\over 2}
\left({\l_{*}\over 4\pi}\right)^{d/2}
\int_{\D} dx_2\int_0^1ds{\partial\over\partial s}
[1-s{\bar \G}(x_2-x_1)+D_1\G(x_2-x_1)]^{-d/2}\Eq(2.19)$$

$$f_{\tilde Q}(\D\cup\D',\D)=L^{2\e}{g^2\over 2}
\left({\l_{*}\over 4\pi}\right)^{d/2}\int_{\D'} dx_2
\int_0^1ds{\partial\over\partial s}
[1-s{\bar \G}(x_2-x_1)+D_1\G(x_2-x_1)]^{-d/2}\Eq(2.20)$$

\\By definition

$$V'_{F_{\tilde Q}}(\D,\phi)=-\sum_{Z\supset\D\atop |Z|\le 2,\ {\rm conn}}
F_{\tilde Q}(Z,\D,\phi)=-
\int_{\D} dx_1
v_*(\phi (x_1))
\sum_{Z\supset\D\atop |Z|\le 2}
f_{\tilde Q}(Z,\D)=$$
$$=-
\int_{\D} dx_1
v_*(\phi (x_1))f_{\tilde Q}(\L,\D)\Eq(2.21)$$

\\where

$$f_{\tilde Q}(\L,\D)=L^{2\e}{g^2\over 2}
\left({\l_{*}\over 4\pi}\right)^{d/2}
\int_{\L} dx_2\int_0^1ds{\partial\over\partial s}
[1-s{\bar \G}(x_2-x_1)+D_1{\bar\G}(x_2-x_1)]^{-d/2}
\Eq(2.22)$$

\\and we have used the fact that ${\bar \G(y)}$
vanishes for $|y|\ge 1$.
Now we can use the translation invariance for $f_{\tilde Q}(\L,\D,\phi)$
and we obtain

$$V'_{F_{\tilde Q}}(\D,\phi)=-V_*(\D,\phi)L^{2\e}{g^2\over 2}
\left({\l_{*}\over 2\pi}\right)^{d/2}\!\!\int dy\int_0^1\!
ds{\partial\over\partial s}
[2(1-s{\bar \G}(y)+D_1\G(y))]^{-d/2}\Eq(2.22.1)$$

\\and therefore

$$V'(\D,\phi)=L^{\e}gV_*(\D,\phi)+V'_{F_{\tilde Q}}(\D,\phi)=
(L^{\e}g-b_1g^2)V_*(\D,\phi)\Eq(2.23)$$

\\with

$$b_1=L^{2\e}{1\over 2}\left({\l_{*}\over 4\pi}\right)^{d/2}
\int dy\int_0^1ds{\partial\over\partial s}
[1-s{\bar \G}(y)+D_1{\bar\G}(y)]^{-d/2}
\Eq(2.24)$$

\\In order to give a good estimate of $b_1$ we have to
study the behaviour for short distances of the covariance
${\bar\G}$.

\vglue.5truecm
\\{\it 4.5. Asymptotic behaviour of the propagator ${\bar\G}$}
\vglue.3truecm

\\The asymptotic behaviour for small $|y|$ of ${\bar\G}(y)$
is described by the following lemma
\vglue.5truecm
\\LEMMA 4.5.1
\vglue.3truecm
\\For $1/L\le y\le 1/2$ we have

$$O(1)y^{\b}\le\vert{\bar\G}(0)-{\bar\G}(y)\vert\le O(1)y^{\b}\Eq(11.10)$$
\vglue.3truecm
\\{\it Proof}
$${\bar\G}(0)-{\bar\G}(y)
=-{L^{-\b} \l_*}\int_1^{L}\!\!{dl\over l}\ l^{\b}\int_0^1dt{yL\over
{l}}u'\left({tyL\over {l}}\right)
=-{y^{\b}\l_*}\int_{1/Ly}^{1/y}\!\!{dl\over l}l^{(\b-1)}
\int_0^1dtu'\left({t\over {l}}\right)$$

\\we have therefore to show that in the region   $1/L\le
y\le 1/2$ 

$$\left|\int_{1/Ly}^{1/y}dl\ {dl\over l}l^{(\b-1)}
\int_0^1dtu'\left({t\over {l}}\right)\right|=O(1)$$

\\We observe first that

$$\int_{1/Ly}^{1/y}dl\ {dl\over l}l^{(\b-1)}
\int_0^1dtu'\left({t\over {l}}\right)=$$
$$=\int_{1/Ly}^{1}dl\ {dl\over l}l^{(\b-1)}
\int_0^1dtu'\left({t\over {l}}\right)+
\int_{1}^{1/y}dl\ {dl\over
l}l^{(\b-1)}
\int_0^1dtu'\left({t\over {l}}\right)$$

\\To obtain the upper bound we bound

$$\left|\int_0^1dtu'\left({t\over {l}}\right)\right|\le O(1){l}\quad {\rm for}
\ l\le 1$$
$$\left|\int_0^1dtu'\left({t\over {l}}\right)\right|\le O(1)\quad {\rm for}
\ l\ge
1$$

\\therefore

$$\left\vert\int_{1/Ly}^{1/y}\ {dl\over l}l^{(\b-1)}
\int_0^1dtu'\left({t\over {l}}\right)\right\vert\le O(1)
\left[\int_{1/Ly}^{1}\ {dl\over l}l^{\b}+
\int_{1}^{1/y}\ {dl\over l}l^{(\b-1)}\right]\le O(1)
$$

\\The lower bound is obtained simply observing that
since $u'(x)<0$ for all $x>0$ the integrand has always
the same sign, and therefore

$$\left\vert\int_{1/Ly}^{1/y}\ {dl\over l}l^{(\b-1)}
\int_0^1dtu'\left({t\over {l}}\right)\right\vert\ge 
\int_{1}^{2}\ {dl\over l}l^{(\b-1)}
\left\vert\int_{1/2}^1dtu'\left({t\over {l}}\right)\right\vert
=O(1)\eqno Q.E.D.$$
\vglue.3truecm
\\It will be useful in the following to define the quantity

$${\tilde a}(L,\e)=\int_{0}^1dy(1-{\bar\G}(y)+D_1{\bar\G}(y))^{-d/2}
\Eq(11.11)$$

\\For such quantity we have
\vglue.5truecm
\\LEMMA 4.5.2
\vglue.3truecm
$${\tilde a}(L,\e)=O(\ln L)\Eq(11.12)$$
\vglue.3truecm
\\{\it Proof}

\\First observe that from (4.8) and the remark following it we have

$$0\le D_1{\bar\G}(y)\le O(1)\left|y \right|^{2}$$

\\Write
$${\tilde a}(L,\e)=\int_{0}^{1/L}dy(1-{\bar\G}(y)+D_1{\bar\G}(y))^{-d/2}+
\int_{1/L}^1dy(1-{\bar\G}(y)+D_1{\bar\G}(y))^{-d/2}\equiv$$
$$\equiv {\tilde a}_<(L,\e)+{\tilde a}_>(L,\e)$$

\\Now we can estimate ${\tilde a}_<(L,\e)$ simply observing that
$1-{\bar\G}(y)\ge L^{-\b}$ and $D_1{\bar\G}(y)\ge 0$

$${\tilde a}_<(L,\e)\le L^{\b d/2}(1/L)=L^{-\e}\le 1$$

\\${\tilde a}_>(L,\e)$ is estimated using lemma 4.5.1 the dominant 
contribution coming from the region $1/L\le y\le 1/2$.

\\Write 
$$ {\tilde a}_>(L,\e)=
\int_{1/L}^1dy(L^{-\b}+{\bar\G}(0)-{\bar\G}(y)+D_1{\bar\G}(y))^{-d/2}$$

\\Then we have
$${\tilde a}_>(L,\e)\le 
\int_{1/L}^1dy(L^{-\b}+{\bar\G}(0)-{\bar\G}(y))^{-d/2}\le$$
$$\le \int_{1/L}^1dy(L^{-\b}+O(1)y^{\b})^{-d/2}$$

\\where in the last step we have used Lemma 4.5.1

\\On the other hand,
$${\tilde a}_>(L,\e)\ge 
\int_{1/L}^1dy(L^{-\b}+{\bar\G}(0)-{\bar\G}(y)+O(1)
\left|y \right|^{2})^{-d/2}\ge$$
$$\ge \int_{1/L}^1dy(L^{-\b}+O(1)y^{\b}+O(1)
\left|y \right|^{2}))^{-d/2}\ge$$
$$\ge \int_{1/L}^1dy(L^{-\b}+O(1)y^{\b})^{-d/2}$$

\\where in the last step we have used $\left|y \right|^{2}\le y^{\b}$ 
in the region of 
integration since $\b$ is positive but very small.

\\Hence for large L,
$${\tilde a}_>(L,\e)= O(1)\int_{1/L}^1{dy\over y^{\b d/2}}=O(1)
{1\over \e}(1-L^{-\e})=O(1){L^{\e}-1\over \e}L^{-\e}=O(\ln L)\eqno Q.E.D.$$

\vglue.5truecm
\\{\it 4.6. Second order coefficient of the relevant part}
\vglue.3truecm

\\$b_1$ is controlled by the following lemma.
\vglue.5truecm
\\LEMMA 4.6.1
\vglue.3truecm
$$b_1=O(\ln L),\ b_1> 0$$
\vglue.3truecm
\\{\it Proof}

\\That $b_1 > 0$ follows from its definition. Note that the function
whose s-derivative is to be taken is is independent of s for $y\ge 1$
since in this region ${\bar\G}(y)$ vanishes. This together with the 
evenness of ${\bar\G}(y)$ implies for $\e$ sufficiently small 

$$b_1=O(1)
\int_{0}^1 dy\int_0^1ds{\partial\over\partial s}
(1-s{\bar\G}(y)+D_1{\bar\G}(y))^{-d/2}=$$
$$=O(1)\int_{0}^1 dy(1-{\bar\G}(y)+D_1{\bar\G}(y))^{-d/2}-O(1)
\int_{0}^1 dy(1+D_1{\bar\G}(y))^{-d/2}$$

\\Since $0\le D_1{\bar\G}(y)\le O(1)y$ , the second integral is bounded by
O(1). The first integral is estimated by Lemma 4.5.2 and we are done. 

$$\eqno Q.E.D.$$

\\From lemma 4.6.1 and the definition \equ(2.18.3), \equ(2.18.4)
it is immediate to show that 
\vglue.5truecm
\\LEMMA 4.6.2
\vglue.3truecm
$$\Vert F_{\tilde Q}\Vert_{G,1,\G_p}\le O(1)\e^{7/4}\qquad
\Vert F_{\tilde Q}\Vert_{\io,1,\G_p}\le O(1)\e^{7/4}$$

\\for any integer $p\ge 1$, with $O(1)$ depending on $p$.

\vglue.5truecm
\\{\it 4.7. Second order irrelevant part }
\vglue.3truecm

\\From the explicit expression of ${\tilde K}_{\tilde Q}$
given in \equ(2.16.1) and \equ(2.17.1) and of $F_{\tilde Q}$
given in \equ(2.18.3) and \equ(2.18.4) we obtain

$$({\tilde K}_{\tilde Q}-F_{\tilde Q})(\D)=L^{2\e}{g^2\over 2}
\left({\l_*\over 2\pi}\right)^{d}
\int_0^1ds{\dpr\over\dpr s}\int_\D
d^Dx_1\int{d^dk_1\over(2\pi\l_*)^{d/2}}
\int_\D
d^Dx_2\int{d^dk_2\over(2\pi\l_*)^{d/2}}$$
$$\int_0^1dt{\dpr\over\dpr t}\exp\left[-{1\over 2\l_*}(k,I_1(s,t)k)
+i(k_1\cdot \phi(x_1)+tk_2\cdot(\phi(x_2)-\phi(x_1)))\right]
\Eq(2.25)$$

$$({\tilde K}_{\tilde Q}-F_{\tilde Q})(\D\cup\D')=L^{2\e}{g^2\over 2}
\left({\l_*\over 2\pi}\right)^{d}
\int_0^1ds{\dpr\over\dpr s}\int_\D
d^Dx_1\int{d^dk_1\over(2\pi\l_*)^{d/2}}
\int_{\D'}
d^Dx_2\int{d^dk_2\over(2\pi\l_*)^{d/2}}$$
$$\int_0^1dt{\dpr\over\dpr t}\exp\left[-{1\over 2\l_*}(k,I_1(s,t)k)
+i(k_1\cdot \phi(x_1)+tk_2\cdot(\phi(x_2)-\phi(x_1)))\right]
+(\D\rightleftharpoons\D')\Eq(2.26)$$

\\Therefore $({\tilde K}_{\tilde Q}-F_{\tilde Q})=g^2{\bar{\cal I}}_1$,
and by lemma 4.3.1 the induction on \equ(2.2.2) is proved.

\\The bounds \equ(1.9) and \equ(1.9.1) are now easy consequences
of the following lemma 4.7.1, lemma 4.3.1 and the smallness of the 
coupling constants ( stated after (4.4)).

\vglue.5truecm
\\LEMMA 4.7.1
\vglue.3truecm
\\For any positive integer $p\ge 1$ and with $O(1)$
$p$-dependent

$$\Vert {\bar{\cal I}}_l\Vert_{G,1,\G_p}\le 
O(1)L^{3\b-l\b/2}L^{2(D+2)}\Eq(2.27)$$

$$\Vert {\bar{\cal I}}_l\Vert_{\io,1,\G_p}\le 
O(1)L^{3\b-l\b/2}L^{2(D+2)}\Eq(2.28)$$

\vglue.3truecm
\\{\it Proof}

\\We perform first of all in (4.5), (4.6) the gaussian integral
in the $k$ variables and we obtain

$${\bar{\cal I}}_l(\D)=L^{2l\e}{1\over 2}
\left({\l_*\over 2\pi}\right)^{d}
\int_0^1ds{\dpr\over\dpr s}
\int_0^1dt{\dpr\over\dpr t}$$
$$\int_\D
d^Dx_1\int_\D d^Dx_2\exp\left[-{1\over 2}\l_*(\phi,I_l(s,t)^{-1}\phi)\right]
(\det I_l(s,t))^{-d/2}
\Eq(2.29)$$

$${\bar{\cal I}}_l(\D\cup\D')=L^{2l\e}{1\over 2}
\left({\l_*\over 2\pi}\right)^{d}
\int_0^1ds{\dpr\over\dpr s}
\int_0^1dt{\dpr\over\dpr t}$$
$$\int_\D
d^Dx_1\int_{\D'} d^Dx_2
\exp\left[-{1\over 2}\l_*(\phi,I_l(s,t)^{-1}\phi)\right]
(\det I_l(s,t))^{-d/2}
+(\D\rightleftharpoons\D')\Eq(2.30)$$

\\where $\phi=(\phi(x_1),\phi(x_2)-\phi(x_1))$.

\\Then we observe that the derivative with respect to $s$
produces a factor ${\bar \G}(L^{(l-1)}(x_2-x_1))$,
and due to the compact support of the covariance this implies
that

$$\left|{\bar{\cal I}}_l(X)\right|\le L^{2l\e}{1\over 2}
\left({\l_*\over 2\pi}\right)^{d}
\int_\D
d^Dx_1\int_{\d_l} 
d^Dx_2 2\sup_{0\le s\le 1}$$
$$\left|\int_0^1dt{\dpr\over\dpr t} \left[ 
\exp\left[-{1\over 2}\l_*(\phi,I_l(s,t)^{-1}\phi)\right]
(\det I_l(s,t))^{-d/2}\right]\right|
\Eq(2.30.1)$$

\\where $\d_l\equiv\{x_2:{|x_2-x_1|\le L^{-(l-1)}}\}$

\\Then we consider the derivative

$${\dpr\over\dpr t}
\exp\left[-{1\over 2}\l_*(\phi,I_l(s,t)^{-1}\phi)\right]
(\det I_l(s,t))^{-d/2}\Eq(2.31)$$

\\We will denote
$I_{t}={\dpr\over\dpr t}\det I_l(s,t)$
and we compute explicitely the inverse matrix $I_l(s,t)^{-1}$:

$$I_l(s,t)^{-1}={1\over \det I_l(s,t)}J_l(s,t)\Eq(2.33)$$

\\with

$$J_l(s,t)=\pmatrix{2\left[C_l(s)+
(1-t^2)D_1\bar\G(x_2-x_1)\right]&
tC_l(s)\cr
tC_l(s)&1\cr}\Eq(2.34)$$

\\obtaining for \equ(2.31)

$${\dpr\over\dpr t}
\exp\left[-{\l_*\over 2}(\phi,I_l(s,t)^{-1}\phi)\right]
(\det I_l(s,t))^{-d/2}=$$

$$={e^{-{\l_*\over 2}(\phi,I_l(s,t)^{-1}\phi)}\over
(\det I_l(s,t))^{d/2+1}}
\left[I_t\left({\l_*\over 2}{(\phi,J_l(s,t)\phi)\over\det I_l(s,t)}-(d/2)
\right)-\left(
{\l_*\over 2}{\dpr\over\dpr t}(\phi,J_l(s,t)\phi)
\right)
\right]\Eq(2.35)$$

\\In order to bound \equ(2.35) 
we compute first of all explicitely

$$\det I_l(s,t)=2\left[C_l(s)+
(1-t^2)D_1\bar\G(x_2-x_1)\right]-t^2(C_l(s))^2
\Eq(2.35.1)$$

\\and we list  
some preliminary useful bounds:

$$\left|\exp\left[-{1\over 2}\l_*(\phi,I_l(s,t)^{-1}\phi)\right]
\right|\le
e^{-{1\over 2}\l_*|\phi(x_1)|^2}\Eq(2.36)$$

\\{\it Proof of \equ(2.36)}

\\From \equ(2.33), \equ(2.34) it is easy to see that

$$I_l(s,t)^{-1}=E_1+{\tilde I}_l(s,t)^{-1}$$

\\where

$$E_1=\pmatrix{1&0\cr 0&0\cr}
\qquad\qquad{\tilde I}_l(s,t)^{-1}={1\over\det I_l(s,t)}\pmatrix
{t^2C_l(s)^2&tC_l(s)\cr tC_l(s)&1\cr} $$

\\and the proof follows since ${\tilde I}_l(s,t)^{-1}$ is positive 
definite. In fact the displayed matrix is manifestly positive
definite and from (4.60) below $\det I_l(s,t) \ge 0$.

$$C_l(s)\ge\cases{\displaystyle\vphantom{1\over 2}
O(1)L^{-\b l}&for $|x_2-x_1|<L^{-l}$\cr
\displaystyle\vphantom{1\over 2}
O(1)|x_2-x_1|^\b&for $L^{-l}\le|x_2-x_1|\le L^{-l+1}$\cr}\Eq(2.37)$$

\\{\it Proof of \equ(2.37)}

\\The first summand in the definition (4.8) of $C_l(s)$ can be bounded by

$${(1-s{\bar \G}(L^{(l-1)}(x_2-x_1)))\over L^{\b(l-1)}}\ge\cases{
O(1)L^{-\b l}&for $|x_2-x_1|<L^{-l}$\cr
O(1)|x_2-x_1|^\b&for $L^{-l}\le|x_2-x_1|\le L^{-l+1}$\cr}$$

\\repeating the proof of lemma 4.5.1. The second summand is positive.
Thus \equ(2.37) is proved.

$$0\le C_l(s)\le
O(1)L^{-\b (l-1)}\quad {\rm for}\ |x_2-x_1|\le L^{-l+1}\Eq(2.37.1)$$

\\and $O(1)=1$ for $l=1$

\\{\it Proof of \equ(2.37.1)}

\\We first observe 
that

$$0\le
{(1-s{\bar \G}(L^{(l-1)}(x_2-x_1)))\over L^{\b(l-1)}}\le 
L^{-\b (l-1)}$$

\\which is trivial

\\Then we observe,
by a simple Taylor expansion

$$0\le {\bar \G}(0)-
{\bar \G}(x)\le O(1)|x|^2 L^{(2-\b)}$$

obtaining

$$0\le {{\bar \G}(0)-
{\bar \G}(L^p(x_2-x_1))\over L^{p\b}}\le
O(1)|x_2-x_1|^2 L^{(2-\b)(p+1)}$$

\\and for $|x_2-x_1|\le L^{-l+1}$ we obtain

$$0\le {{\bar \G}(0)-
{\bar \G}(L^p(x_2-x_1))\over L^{p\b}}\le
L^{-\b(l-1)} L^{-(2-\b)(l-p-2)} $$

\\This proves the convergence of the second sum in 
$C_l(s)$, and the fact that the sum is dominated by the 
therm with $p=l-2$, giving the proof of 
\equ(2.37.1).

$$\det I_l(s,t)\ge\cases{\displaystyle\vphantom{1\over 2}
O(1)L^{-\b l}&for $|x_2-x_1|<L^{-l}$\cr
\displaystyle\vphantom{1\over 2}
O(1)|x_2-x_1|^\b&for $L^{-l}\le|x_2-x_1|\le L^{-l+1}$\cr}\Eq(2.38)$$

\\{\it Proof of \equ(2.38)}

\\We note that

$$D_1\bar\G(x_2-x_1)\ge 0$$

\\so that

$$\det I_l(s,t)\ge C_l(s)(2-C_l(s))$$

\\Now the proof follows from (4.58) and (4.59).

$$|I_t|\le O(1)L^{-2\b (l-1)}
\quad {\rm for}\ |x_2-x_1|\le L^{-l+1}\Eq(2.40)$$

\\{\it Proof of \equ(2.40)}

\\Differentiate (4.56), then use (4.59) and

$$0\le D_1\G(x_2-x_1)\le O(1)|x_2-x_1|^2$$

\\which follows from (4.58) and the remark following it to
obtain the proof.

$$|(\phi,J_l(s,t)\phi)|\!\le  \!
O(1)L^{-\b (l-1)}e^{(\k/2)\Vert\phi\Vert_{1,\s,X}^2}
e^{{1\over 2}\l_*(\r/2)|\phi(x_1)|^2}\quad {\rm for}\ |x_2-x_1|\!\le \! 
L^{-l+1}
\Eq(2.42)$$

\\{\it Proof of \equ(2.42)}

\\From the explicit expression of $J_l(s,t)$ \equ(2.34) it
is easy to find the following bound

$$|(\phi,J_l(s,t)\phi)|\le  
O(1)|\phi(x_1)|^{2} (C_l(s)+D_1\bar\G)+$$
$$+O(1)\left[|\phi(x_1)||\phi(x_2)-\phi(x_1)|
C_l(s)+|\phi(x_2)-\phi(x_1)|^2\right]$$

\\Then from the mean value theorem and Sobolev embedding we bound

$$|\phi(x_2)-\phi(x_1)|\le |x_2-x_1|\Vert\phi\Vert_{1,\s,X}$$

\\Finally, since $k$ and $\r$ are small but $O(1)$, we have

$$\Vert\phi\Vert_{1,\s,X}\le O(1)e^{(\k/4)\Vert\phi\Vert_{1,\s,X}^2}
\qquad |\phi(x_1)|\le O(1)e^{{1\over 2}\l_*(\r/4)|\phi(x_1)|^2}$$

\\and we obtain 

$$|(\phi,J_l(s,t)\phi)|\le  
O(1)((C_l(s)+D_1\bar\G)+|x_2-x_1|)e^{(\k/2)\Vert\phi\Vert_{1,\s,X}^2}
e^{{1\over 2}\l_*(\r/2)|\phi(x_1)|^2}$$

\\that gives the proof of \equ(2.42) by \equ(2.37.1) and the estimate
on $D_1\bar\G$ given in the proof of (4.61) .

$$\left|{\dpr\over\dpr t}(\phi,J_l(s,t)\phi)\right|\le$$
$$
\le 
O(1)L^{-(\b+1)(l-1)}e^{(\k/2)\Vert\phi\Vert_{1,\s,X}^2}
e^{{\l_*\over 2}(\r/2)|\phi(x_1)|^2}\ {\rm for}\ 0
\!\le\!|x_2-x_1|\!\le \! 
L^{-l+1}
\Eq(2.43)$$

\\{\it Proof of \equ(2.43)}

\\We start from the bound

$$\left|{\dpr\over\dpr t}(\phi,J_l(s,t)\phi)\right|\le 
O(1)(C_l(s)|\phi(x_1)||\phi(x_2)-\phi(x_1)|+
D_1\bar\G(x_2-x_1)|\phi(x_1)|^2)$$

\\then we proceed as in \equ(2.42) and we obtain the proof.

\\Using the bounds \equ(2.36)-\equ(2.43) it is now an easy task
to bound the explicit expression \equ(2.35). We have,
defining $g_{\r,\k}(X,\phi)$ by

$$G_{\r,k}(X,\phi)={1\over |X|}\int_X dx_1 g_{\r,k}(x_1,X,\phi)$$

\\the following bound 

$$\left|{\dpr\over\dpr t}
\exp\left[-{1\over 2}\l_*(\phi,I_l(s,t)^{-1}\phi)\right]
(\det I_l(s,t))^{-d/2}\right|\le$$
$$\le \cases{\displaystyle\vphantom{1\over 2}
O(1)L^{\b l(d/2+2)}L^{-3\b(l-1)}
g_{\r,\k}(x_1,X,\phi)
&for $|x_2-x_1|<L^{- l}$\cr
\displaystyle\vphantom{1\over 2}
O(1)|x_2-x_1|^{-\b(d/2+2)}L^{-3\b(l-1)}
g_{\r,\k}(x_1,X,\phi)&for $L^{-l}\!\le\!|x_2\!-\!x_1|\!\le\! L^{-l+1}$\cr}
\Eq(2.48)$$

\\Now we come back to \equ(2.30.1) and obtain

$$|{\bar{\cal I}}_l(X)|\le O(1)
\left[\int_0^{L^{-l}} dy L^{\b l(d/2+2)}L^{-3\b(l-1)}
G_{\r,k}(X,\phi)+\right.$$
$$\left.\int_{L^{-l}}^{L^{-l+1}} dy 
|y|^{-\b(d/2+2)}L^{-3\b(l-1)}G_{\r,k}(X,\phi)\right]\le$$
$$\le O(1)
\left[ L^{l(\b d/2-1)}L^{-\b(l-3)}+\int_{L^{-l}}^{L^{-l+1}} dy 
|y|^{-\b(d/2)}L^{-\b(l-3)}\right]G_{\r,k}(X,\phi)\Eq(2.50)$$

\\Observe that $\b(d/2)=1-\e$. We obtain for the
integral in \equ(2.50) ,for $\e$ sufficiently small,

$$\int_{L^{-l}}^{L^{-l+1}} dy 
|y|^{-\b(d/2)}={1\over \e}L^{-l\e}(L^\e-1)=O(\ln L)\Eq(2.51)$$

\\Hence we have

$$\Vert {\bar{\cal I}}_l(X)\Vert_G\le O(1)L^{3\b-l\b/2}\Eq(2.52)$$

\\Evaluating the functional derivative $D{\bar{\cal I}}_l(X)$ we 
obtain the same bound, so that we have

$$\Vert {\bar{\cal I}}_l(X)\Vert_{G,1}\le O(1)L^{3\b-l\b/2}\Eq(2.53)$$

\\and we obtain finally the lemma by the smallness of the set $X$.
$$\eqno Q.E.D.$$

\pagina
\vglue.5truecm\numsec=5\numfor=1\pgn=1
\\\S5. {\it RG action on the remainder. 
Relevant and irrelevant terms. Estimates.}
\vglue.5truecm
\\This section is devoted to the activity $({\cal S}_1{\hat r})^\natural$, 
which we encountered earlier in \equ(1.54) of section 3. We need to extract
relevant terms from the contributions from small sets. We define the relevant
and irrelevant terms and give suitable bounds. We also control the remainder
contribution to the flow of the effective coupling constant. The contributions
from large sets of course need no subtractions since we easily obtain
a contractive bound for them.

\vglue.5truecm
\\{\it 5.1. Linear reblocking, small set contributions}
\vglue.5truecm
\\Let $X$ be a small set. 
We can write

$${\hat r}(X,\phi)={1\over |X|}\int_Xdx_1{\hat r}(X,\phi)\Eq(5.1)$$

\\and define

$${\hat r}_{*}(X,\phi({\bar x}),\phi)=e^{(\l_*/2)|\phi({\bar x})|^2}
{\hat r}(X,\phi)\Eq(5.2)$$

\\where ${\bar x}$ is the midpoint of the polymer $X$.
We have therefore

$${\hat r}(X,\phi)={1\over |X|}\int_Xdx_1e^{-(\l_*/2)|\phi({\bar x})|^2}
{\hat r}_{*}(X,\phi({\bar x}),\phi)\Eq(5.3)$$

\\Now we consider the contribution to the linear reblocking of the 
activity $\hat r$ restricted to small sets, denoted by
${\hat r}^{s.s}$. We have

$${\cal B}_{1}{\hat r}^{(s.s.)}(LZ,\phi)=
\sum_{X\ {\rm small\ set}\atop{\bar X}=LZ}
{1\over |X|}\int_Xdx_1e^{-(\l_*/2)|\phi({\bar x})|^2}
{\hat r}_{*}(X,\phi({\bar x}),\phi)\Eq(5.4)$$

\\and after rescaling and convolution integration, using the master
formula in Lemma 1.3.1,

$$\left(\mu_{{\bar\G}}*({\cal S}_{1}{\hat r}^{(s.s.)})\right)(Z,\phi)=
\sum_{X\ {\rm small\ set}\atop{\bar X}=LZ}L^{-\a}
{1\over |X|}\int_Xdx_1e^{-(\l_*/2)L^{-\b}|{\cal R}\phi({\bar x})|^2}$$

$$\left(\mu_{\Sigma^{{\bar x}}}*{\hat r}_{*}\right)
(X,L^{-\b}{\cal R}\phi({\bar x}),L^{-\b}T^{{\bar x}}{\cal
R}\phi)\Eq(5.5)$$ 

\\we will write \equ(5.5)  in compact notation

$$\left({\cal S}_{1}{\hat r}^{(s.s.)}\right)^{\natural}(Z,\phi)=
\sum_{X\ {\rm s.\ s.}\atop{\bar X}=LZ}({\cal R}{\hat r})
^{\natural}(L^{-1}X,\phi)=$$

$$=\sum_{X\ {\rm s.\ s.}\atop{\bar X}=LZ}
{L^{-\a}\over |X|}\int_Xdx_1e^{-(\l_*/2)L^{-\b}|{\cal R}\phi({\bar
x})|^2} {\hat r}_{*}^{\#\Sigma_{\bar x}}
(X,L^{-\b}{\cal R}\phi({\bar x}),L^{-\b}T^{{\bar x}}{\cal
R}\phi)\Eq(5.6)$$ 

\\Then we define the relevant part in the
following way

$${\cal L}\left({\cal S}^{(1)}{\hat r}^{(s.s.)}\right)^{\natural}(Z,\phi)=
\sum_{X\ {\rm s.\ s.}\atop{\bar X}=LZ}{\cal L}({\cal R}{\hat r})
^{\natural}(L^{-1}X,\phi)=$$

$$=\sum_{X\ {\rm s.\ s.}\atop{\bar X}=LZ}
{L^{-\a}\over |X|}\int_Xdx_1e^{-(\l_*/2)L^{-\b}|{\cal R}\phi(x_1)|^2}
{\hat r}_{*}^{\#\Sigma_{\bar x}}(X,0,0)
\Eq(5.7)$$

\\so that the irrelevant term is

$$(1-{\cal L})\left({\cal S}^{(1)}{\hat r}^{(s.s.)}\right)^{\natural}(Z,\phi)=
\sum_{X\ {\rm s.\ s.}\atop{\bar X}=LZ}(1-{\cal L})({\cal R}{\hat r})
^{\natural}(L^{-1}X,\phi)\Eq(5.7.1)$$

\\with

$$(1-{\cal L})({\cal R}{\hat r})
^{\natural}(L^{-1}X,\phi)=$$

$$={L^{-\a}\over |X|}\int_Xdx_1e^{-(\l_*/2)L^{-\b}|{\cal R}\phi(x_1)|^2}
\left[-{\hat r}_{*}^{\#\Sigma_{\bar x}}(X,0,0)+\right.$$

$$\left.+e^{-(\l_*/2)L^{-\b}\left(|{\cal R}\phi({\bar x})|^2-
|{\cal R}\phi(x_1)|^2\right)}
{\hat r}_{*}^{\#\Sigma_{\bar x}}
(X,L^{-\b}{\cal R}\phi({\bar x}),L^{-\b}T^{{\bar x}}{\cal R}\phi)
\right]=$$

$$={L^{-\a}\over |X|}\int_Xdx_1e^{-(\l_*/2)L^{-\b}|{\cal R}\phi(x_1)|^2}
\int_0^1dt{\dpr\over\dpr t}$$

$$\left[e^{-{t}^{2}(\l_*/2)L^{-\b}\left(|{\cal R}\phi({\bar x})|^2- |{\cal
R}\phi(x_1)|^2\right)}
{\hat r}_{*}^{\#\Sigma_{\bar x}}
(X,tL^{-\b}{\cal R}\phi({\bar x}),tL^{-\b}T^{{\bar x}}{\cal
R}\phi)\right]\Eq(5.7.2)$$

\\Then we perform the derivative with respect to $t$
obtaining

$$(1-{\cal L})({\cal R}{\hat r}) ^{\natural}(L^{-1}X,\phi)=
{L^{-\a}\over |X|}\int_Xdx_1e^{-{\l_*\over 2}L^{-\b}|{\cal
R}\phi(x_1)|^2}\int_0^1dt$$

$$\left[{\dpr\over\dpr t}
\left(e^{-{\l_*\over 2}t^2L^{-\b}\left(|{\cal R}\phi({\bar
x})|^2- |{\cal R}\phi(x_1)|^2\right)}\right)
{\hat r}_{*}^{\#\Sigma_{\bar x}} (X,tL^{-\b}{\cal R}\phi({\bar
x}),tL^{-\b}T^{{\bar x}}{\cal R}\phi)+\right.$$

$$\left.+\left(e^{-{\l_*\over 2}t^2L^{-\b}\left(|{\cal R}\phi({\bar
x})|^2- |{\cal R}\phi(x_1)|^2\right)}\right){\dpr\over\dpr t}
{\hat r}_{*}^{\#\Sigma_{\bar x}}
(X,tL^{-\b}{\cal R}\phi({\bar x}),tL^{-\b}T^{{\bar x}}{\cal
R}\phi)\right]=$$

$$={L^{-\a}\over |X|}\int_Xdx_1e^{-(\l_*/2)L^{-\b}|{\cal
R}\phi(x_1)|^2}\int_0^1dt$$

$$\left[
{\dpr\over\dpr t}
\left(e^{-{\l_*\over 2}t^2L^{-\b}\left(|{\cal R}\phi({\bar
x})|^2- |{\cal R}\phi(x_1)|^2\right)}\right)
{\hat r}_{*}^{\#\Sigma_{\bar x}} (X,tL^{-\b}{\cal R}\phi({\bar
x}),tL^{-\b}T^{{\bar x}}{\cal R}\phi)+\right.$$

$$+\left(e^{-{\l_*\over 2}t^2L^{-\b}\left(|{\cal R}\phi({\bar
x})|^2- |{\cal R}\phi(x_1)|^2\right)}\right)
\!\left[(D{\hat r}_{*}^{\#\Sigma_{\bar x}})
(X,tL^{-\b}{\cal R}\phi({\bar x}),tL^{-\b}T^{{\bar x}}{\cal R}\phi;
(L^{-\b}{\cal R}\phi({\bar x}),0))+\right.$$

$$\left.\left.+(D{\hat r}_{*}^{\#\Sigma_{\bar x}})
(X,tL^{-\b}{\cal R}\phi({\bar x}),tL^{-\b}T^{{\bar x}}{\cal R}\phi;
(0,L^{-\b}T^{{\bar x}}{\cal R}\phi))
\right]\right]
\Eq(5.8)$$

\\where the derivatives with respect to the field $\phi({\bar x})$
are ordinary partial derivatives, and we will denote them hereafter
with $\dpr$, while variational derivatives with respect to the
field $\phi$ have norms computed in $C^1(X)$ topology. We denote
with $D_{(2)}K$ the first variational derivative in the direction of
$\phi$ in the activity $K(X,\phi({\bar x}),\phi)$.
We can write the following bound for the activity
$(1-{\cal L})({\cal R}{\hat r})
^{\natural}(L^{-1}X,\phi)$:

$$\left\vert(1-{\cal L})({\cal R}{\hat r})
^{\natural}(L^{-1}X,\phi)\right\vert\le
{L^{-\a}\over |X|}\int_Xdx_1e^{-(\l_*/2)L^{-\b}|{\cal R}\phi(x_1)|^2}$$

$$\int_0^1dt
\left[L^{-1/2}e^{(\l_*/2)L^{1/2-\b}\left||{\cal R}\phi({\bar x})|^2-
|{\cal R}\phi(x_1)|^2\right|}
\left|({\hat r}_{*}^{\#\Sigma_{\bar x}}) (X,tL^{-\b}{\cal
R}\phi({\bar x}),tL^{-\b}T^{{\bar x}}{\cal R}\phi)\right|\right.+$$

$$+\left(e^{-{\l_*\over 2}t^2L^{-\b}\left(|{\cal R}\phi({\bar
x})|^2- |{\cal R}\phi(x_1)|^2\right)}\right)
\left[|L^{-\b}{\cal R}\phi({\bar x})|\left|(\dpr{\hat
r}_{*R}^{\#\Sigma_{\bar x}}) (X,tL^{-\b}{\cal R}\phi({\bar
x}),tL^{-\b}T^{{\bar x}}{\cal R}\phi)\right| +\right.$$

$$\left.\left.+\left\Vert L^{-\b}T^{{\bar x}}{\cal
R}\phi\right\Vert_{C^1(X)}
\left\Vert(D_{(2)}{\hat r}_{*}^{\#\Sigma_{\bar x}})
(X,tL^{-\b}{\cal R}\phi({\bar x}),tL^{-\b}T^{{\bar x}}{\cal
R}\phi)\right\Vert
\right]\right]
\Eq(5.9)$$

v\\where in the second line we used the trivial inequality

$$\left|xe^x\right|\le{1\over\a}\ e^{\a|x|}\quad\forall\a>2$$

\\We can give a suitable estimate of \equ(5.9) by means of
the following lemmas
\vglue.5truecm
\\LEMMA 5.1.1
\vglue.3truecm
$$\left||{\cal R}\phi({\bar x})|^2- |{\cal R}\phi(x_1)|^2\right|\le
O(1)L^{-(1-\b)}\left(\Vert\phi\Vert_{L^{-1}X,1,\s}^2+
L^{-\b}|{\cal R}\phi(x_1)|^2\right)$$
\vglue.3truecm

\\{\it Proof }
$$\left||{\cal R}\phi({\bar x})|^2- |{\cal R}\phi(x_1)|^2\right|\le
\left||{\cal R}\phi({\bar x})|- |{\cal R}\phi(x_1)|\right|
\left||{\cal R}\phi({\bar x})|+ |{\cal R}\phi(x_1)|\right|$$

\\Using the mean value theorem, a Sobolev inequality and the 
assumption that X is a small set 

$$\left||{\cal R}\phi({\bar x})|^2- |{\cal R}\phi(x_1)|^2\right|\le\!
\left(|{\bar x}-x_1|L^{-(1-\b/2)}\!\!\sup_{x\in L^{-1}X}\!\nabla\phi(x)\right)
\!\left(2|{\cal R}\phi(x_1)|+2\Vert\phi\Vert_{L^{-1}X,1,\s}\right)\!\le$$

$$\le |X|L^{-(1-\b)}\Vert\phi\Vert_{L^{-1}X,1,\s}
\left(2L^{-\b/2}|{\cal
R}\phi(x_1)|+2\Vert\phi\Vert_{L^{-1}X,1,\s}\right)$$

\\and the lemma easily follows.
\vglue.5truecm
\\LEMMA 5.1.2
\vglue.3truecm

$$\left\Vert L^{-\b}T^{{\bar x}}{\cal R}\phi\right\Vert_{C^1(X)}\le$$

$$\le
O(1)L^{-\b/2}{1\over\sqrt{1-t^2}}\left({1\over\sqrt{k}}+{1\over\sqrt{\r}}
\right)
e^{(\l_*/2)(\r/4)(1-t^2)L^{-\b}|{\cal R}\phi(x_1)|^2}
e^{(\k/4)(1-t^2)\Vert\phi\Vert^2_{L^{-1}X,1,\s}}\Eq(5.10)$$
\vglue.3truecm

\\{\it Proof }

\\Recall that

$$T^{{\bar x}}{\cal R}\phi(x)=L^{\b}{\cal R}\phi(x)-
\l_*\G(x-{\bar x}){\cal R}\phi({\bar x})$$

\\By trivial algebraic manipulation, and  the property
$L^{\b}-1=\l_*\G(0)$ we obtain

$$L^{-\b}T^{{\bar x}}{\cal R}\phi(x)=
[{\cal R}\phi(x)-{\cal R}\phi({\bar x})]+L^{-\b}(1+\l_*(\G(0)-\G(x-{\bar x})))
{\cal R}\phi({\bar x})$$

\\then we observe that because of Lemma 5.1.1

$$|\G(0)-\G(x-{\bar x})|\le |x-{\bar x}|\sup_x|\nabla\G(x)|\le O(1) |X|$$

$$|{\cal R}\phi(x)-{\cal R}\phi({\bar x})|\le
L^{-(1-\b/2)}|x-{\bar x}|\sup_x|\nabla\phi(x/L)|\le |X|L^{-(1-\b/2)}
\sup_{x\in L^{-1}X}|\nabla\phi(x)|$$

\\and we have therefore

$$\sup_{x\in X}|L^{-\b}T^{{\bar x}}{\cal R}\phi(x)|\le|X|[L^{-(1-\b/2)}
\sup_{x\in L^{-1}X}|\nabla\phi(x)|+ O(1)L^{-\b}|{\cal R}\phi({\bar x})|]$$

\\Analogously, for the first derivative

$$L^{-\b}\nabla_xT^{{\bar x}}{\cal R}\phi(x)=L^{-(1-\b/2)}\nabla\phi(x/L)-
\l_*\nabla\G(x-{\bar x}){\cal R}\phi({\bar x})$$

\\and therefore

$$\sup_{x\in X}|L^{-\b}\nabla_xT^{{\bar x}}{\cal R}\phi(x)|\le L^{-(1-\b/2)}
\sup_{x\in L^{-1}X}|\nabla\phi(x)|+ O(1)L^{-\b}|{\cal R}\phi({\bar x})|$$

\\Combining the two relations above we obtain

$$\left\Vert L^{-\b}T^{{\bar x}}{\cal R}\phi\right\Vert_{C^1(X)}
\le|X|[L^{-(1-\b/2)}
\sup_{x\in L^{-1}X}|\nabla\phi(x)|+ O(1)L^{-\b}|{\cal R}\phi({\bar x})|]$$

\\Moreover,

$$|{\cal R}\phi({\bar x})|\le |{\cal R}\phi( x_1)| +
|X|[L^{-(1-\b/2)}
\sup_{x\in L^{-1}X}|\nabla\phi(x)$$

\\Now we exploit the estimates

$$\sup_{x\in L^{-1}X}|\nabla\phi(x)|\le
{1\over\sqrt{1-t^2}}{1\over\sqrt{k}}
e^{(\k/4)(1-t^2)\Vert\phi\Vert^2_{L^{-1}X,1,\s}}$$

\\and

$$L^{-\b/2}|{\cal R}\phi( x_1)|\le O(1){1\over\sqrt{1-t^2}}
{1\over\sqrt{\r}}
e^{(\l_*/2)(\r/4)(1-t^2)L^{-\b}|{\cal R}\phi( x_1)|^2}$$

\\ and X is a small set

\\The lemma has been proved.
$$\eqno Q.E.D$$

\vglue.5truecm
\\LEMMA 5.1.3
\vglue.3truecm
$$|L^{-\b}{\cal R}\phi({\bar x})|\le{O(1)}{1\over\sqrt{\r}}
{1\over\sqrt{k}}
{L^{-\b/2}\over\sqrt{1-t^2}}
e^{(\l_*/2)(\r/4)(1-t^2)L^{-\b}|{\cal R}\phi(x_1)|^2}
e^{(\k/4)(1-t^2)\Vert\phi\Vert^2_{L^{-1}X,1,\s}}
\Eq(5.11)$$

\\The proof follows the above lines.

\\$\r,k$ are chosen sufficiently small but are of O(1) in L.

\\Then we obtain from \equ(5.9) using the above lemmas, 

$$\left\vert(1-{\cal L})({\cal R}{\hat r})
^{\natural}(L^{-1}X,\phi)\right\vert\le O(1)
L^{-\b/2}{L^{-\a}\over |X|}\int_Xdx_1e^{-{\l_*\over 2}L^{-\b}
(1-{\r\over 4})|{\cal R}\phi(x_1)|^2}e^{{k\over
4}\Vert\phi\Vert^2_{L^{-1}X,1,\s}}$$

$$\int_0^1 {dt}{1\over\sqrt{1-t^2}} 
e^{(\l_*/2)L^{-\b}(1-t^2)
(\r/4)|{\cal 
R}\phi(x_1)|^2}e^{(\k/4)(1-t^2)\Vert\phi\Vert^2_{L^{-1}X,1,\s}}$$

$$\left[\left|({\hat r}_{*}^{\#\Sigma_{\bar x}}) (X,tL^{-\b}{\cal
R}\phi({\bar x}),tL^{-\b}T^{{\bar x}}{\cal R}\phi)\right|+\right.$$

$$+
\left|(\dpr{\hat
r}_{*R}^{\#\Sigma_{\bar x}}) (X,tL^{-\b}{\cal R}\phi({\bar
x}),tL^{-\b}T^{{\bar x}}{\cal R}\phi)\right| +$$

$$\left.+
\left\Vert(D_{(2)}{\hat r}_{*}^{\#\Sigma_{\bar x}})
(X,tL^{-\b}{\cal R}\phi({\bar x}),tL^{-\b}T^{{\bar x}}{\cal
R}\phi)\right\Vert
\right]
\Eq(5.11.1)$$

\\In order to bound the activities in
\equ(5.11.1) in terms of the
norms introduced in section 2 we introduce the following 
intermediate regulator

$$G_{*\r}(X,\phi({\bar x}),\phi)=e^{(\l_*/2)(1+\r)|\phi({\bar x})|^2}G(X,\phi)
\Eq(5.12)$$

\\where $G(X,\phi)$ is defined in \equ(22.11).
For $G_{*\r}(X,\phi({\bar x}),\phi)$ the following lemma holds
\vglue.5truecm
\\LEMMA 5.1.4
\vglue.3truecm

\\Let $\b$ be small enough 
but $O(1)$ independent of L. Let $X$ be a
small set. Then

$$G_{*\r}(X,\phi({\bar x}),\phi)\le e^{(\l_*/2)4\r|\phi({\bar x})|^2}
e^{2\k\Vert\phi\Vert^2_{X,1,\s}}\Eq(5.14)$$
\vglue.3truecm

\\{\it Proof }

\\We plug in the definition of $G(X,\phi)$ and observe that

$$e^{-(\l_*/2)(1-\r)|\phi(x)|^2}
e^{(\l_*/2)(1+\r)|\phi({\bar x})|^2}=
e^{-(\l_*/2)(|\phi(x)|^2-|\phi({\bar x})|^2)}
e^{(\l_*/2)\r(|\phi(x)|^2+|\phi({\bar x})|^2)}\le$$

$$\le e^{(\l_*/2)|\phi(x)-\phi({\bar x})||\phi(x)+\phi({\bar x})|}
e^{(\l_*/2)\r(|\phi(x)|^2+|\phi({\bar x})|^2)}$$

\\Recall that x and $\bar x$ belong to X a small set. Then by using the 
Sobolev inequality we have 

$$|\phi(x)-\phi({\bar x})|\le 2\Vert\phi\Vert_{X,1,\s}$$

$$|\phi(x)+\phi({\bar x})|\le 2|\phi({\bar x})|+2\Vert\phi\Vert_{X,1,\s}$$

$$|\phi(x)|^2\le 2|\phi({\bar x})|^2+4\Vert\phi\Vert^2_{X,1,\s}$$

\\Then, using elementary inequalities we get

$$G_{*\r}(X,\phi({\bar x}),\phi)\le e^{(\l_*/2)4\r|\phi({\bar x})|^2}
e^{(\k+\l_*(1/\r+2+2\r))\Vert\phi\Vert^2_{X,1,\s}}$$

\\and lemma 5.1.4 follows choosing $\b$  small enough to give
$\l_*\le k/(1/\r+2+2\r)$
$$\eqno Q.E.D.$$

\\We will also need :
\vglue.5truecm
\\LEMMA 5.1.5
\vglue.3truecm

$$\Vert{\hat r}_{*}(X)\Vert_{G_{*\r}}\le
\Vert{\hat r}(X)\Vert_{G}\Eq(5.15)$$

$$\Vert(D_{(2)}{\hat r}_{*})(X)\Vert_{G_{*\r}}\le
\Vert(D{\hat r})(X)\Vert_{G}\Eq(5.15.1)$$

$$\Vert(\dpr{\hat r}_{*})(X)\Vert_{G_{*\r}}\le
{O(1)\over\sqrt{\r}}\Vert{\hat r}(X)\Vert_{G}\Eq(5.16)$$
\vglue.3truecm

\\{\it Proof }

It follows immediately
from the definition of $G_{*\r}$ and of ${\hat r}_{*}$
$$\eqno Q.E.D.$$
We observe now that the derivatives and the fluctuation integral
commute, and denoting

$$J=\left\{\matrix{{\hat r}_{*}\cr
\dpr{\hat r}_{*}\cr 
D_{(2)}{\hat r}_{*}}\right.$$ 

\\we have

$$\vert(J^{\#\Sigma^{\bar x}})
(X,tL^{-\b}{\cal R}\phi({\bar x}),tL^{-\b}T^{{\bar x}}{\cal R}\phi)\vert\le$$

$$\le\int d\mu_{\Sigma^{{\bar x}}}(\z)G_{*\r}
(X,\z({\bar x})+tL^{-\b}{\cal R}\phi({\bar x}),
\z+tL^{-\b}T^{{\bar x}}{\cal R}\phi)\Vert J(X)\Vert_{G_{*\r}}\le$$

$$\le\int d\mu_{\Sigma^{{\bar x}}}(\z) e^{(\l_*/2)4\r|\z({\bar x})
+tL^{-\b}{\cal R}\phi({\bar x})|^2}
e^{4\k\Vert\z+tL^{-\b}T^{{\bar x}}{\cal R}\phi\Vert^2_{X,1,\s}}
\Vert J(X)\Vert_{G_{*\r}}\Eq(5.17)$$

\\In passing to the last line we have used Lemma 5.1.4

\\Now using the stability of the large fields regulator 
in the form \equ(1) we have

$$\vert(J^{\#\Sigma^{\bar x}})
(X,tL^{-\b}{\cal R}\phi({\bar x}),tL^{-\b}T^{{\bar x}}{\cal R}\phi)\vert\le$$

$$\le e^{(\l_*/2)(\r/8)t^2L^{-\b}|{\cal R}\phi({\bar x})|^2}
e^{(\k/8)t^2\Vert\phi\Vert^2_{L^{-1}X,1,\s}}
\Vert J(X)\Vert_{G_{*\r}}\Eq(5.18)$$

\\Using again lemma 5.1.1 we obtain

$$\vert(J^{\#\Sigma^{\bar x}})
(X,tL^{-\b}{\cal R}\phi({\bar x}),tL^{-\b}T^{{\bar x}}{\cal R}\phi)\vert\le$$

$$\le e^{(\l_*/2)(\r/4)t^2L^{-\b}|{\cal R}\phi({x_1})|^2}
e^{(\k/4)t^2\Vert\phi\Vert^2_{L^{-1}X,1,\s}}
\Vert J(X)\Vert_{G_{*\r}}\Eq(5.18.1)$$

\\Finally returning to \equ(5.11.1) and using 
\equ(5.18.1) together with the fact that
$\int_0^1dt(1-t^2)^{-1/2}=O(1)$ we obtain

$$\left|(1-{\cal L})({\cal R}{\hat r})
^{\natural}(L^{-1}X,\phi)\right|\le$$

$$\le O(1)L^{-\b/2}{L^{-\a}\over |X|}\!\int_X\!dx_1
e^{-(\l_*/2)(1-\r/2)L^{-\b}|{\cal R}\phi(x_1)|^2}
e^{(\k/2)\Vert\phi\Vert^2_{L^{-1}X,1,\s}}$$

$$\left[
\Vert{\hat r}_*(X)\Vert_{G_{*\r}}
+\Vert(\dpr{\hat r}_*)(X)\Vert_{G_{*\r}}+
\Vert(D_{(2)}{\hat r}_*)(X)\Vert_{G_{*\r}}
\right]
\Eq(5.19)$$

\\and then from \equ(5.15)-\equ(5.16), performing the 
rescaling, using $L^{-\a}=O(1)L^{-D}$ for $\e$ small enough, and Lemma 5.1.5
we get

$$\left|(1-{\cal L})({\cal R}{\hat r})
^{\natural}(L^{-1}X,\phi)\right|
\le O(1)L^{-\b/2}L^{-D}G(L^{-1}X,\phi)\Vert{\hat r}(X)\Vert_{G,1}
\Eq(5.20)$$

\bigskip
\\Now we want to obtain an analogous estimate for the derivative
$(D(1-{\cal L})({\cal R}{\hat r})
^{\natural})(L^{-1}X,\phi)$ By definition we have

$$(D(1-{\cal L})({\cal R}{\hat r})
^{\natural})(L^{-1}X,\phi;f)=\left.{d\over ds}\right|_{s=0}
(1-{\cal L})({\cal R}{\hat r})
^{\natural}(L^{-1}X,\phi+sf)=$$

$$={L^{-\a}\over |X|}\int_Xdx_1e^{-(\l_*/2)L^{-\b}|{\cal R}\phi(x_1)|^2}
\left\{-{\l_*\over 2}f(x_1/L)(L^{-\b/2}{\cal R}\phi)(x_1)\right.$$

$$\int_0^1 dt {\dpr\over\dpr t}
\left[
\left(e^{-{\l_*\over 2}t^2L^{-\b}\left(|{\cal R}\phi({\bar
x})|^2- |{\cal R}\phi(x_1)|^2\right)}\right)
{\hat r}_{*}^{\#\Sigma_{\bar x}} (X,tL^{-\b}{\cal R}\phi({\bar
x}),tL^{-\b}T^{{\bar x}}{\cal R}\phi)\right]+$$

$$+\l_*(f({{\bar x}\over L})\phi({{\bar x}\over L})-
f({x_1\over L})\phi({x_1\over L}))
e^{-{\l_*\over 2}L^{-\b}\left(|{\cal R}\phi({\bar
x})|^2- |{\cal R}\phi(x_1)|^2\right)}
{\hat r}_{*}^{\#\Sigma_{\bar x}} (X,\!L^{\!-\b}{\cal R}\phi({\bar
x}),\!L^{\!-\b}T^{{\bar x}}{\cal R}\phi)+$$

$$+e^{-{\l_*\over 2}L^{-\b}\left(|{\cal R}\phi({\bar
x})|^2- |{\cal R}\phi(x_1)|^2\right)}
\left[(D_{(2)}{\hat r}_{*}^{\#\Sigma})
(X,L^{-\b}{\cal R}\phi({\bar x}),L^{-\b}T^{{\bar x}}{\cal R}\phi;
(0,L^{-\b}T^{{\bar x}}{\cal R}\phi))+\right.$$

$$\left.\left.+
L^{-\b}({\cal R} f({\bar x}))(\dpr{\hat r}_{*}^{\#\Sigma^{\bar x}})
(X,L^{-\b}{\cal R}\phi({\bar x}),L^{-\b}T^{{\bar x}}{\cal R}\phi;
(L^{-\b}{\cal R}\phi({\bar x}),0))
\right]\vphantom{1\over 2}\right\}
\Eq(5.21)$$

\\We note that

$$|f({{\bar x}\over L})\phi({{\bar x}\over L})-
f({x_1\over L})\phi({x_1\over L})|\le
|(f({{\bar x}\over L})-f({x_1\over L}))\phi({x_1\over L})|+
|f({{\bar x}\over L})(\phi({{\bar x}\over L})-\phi({x_1\over L}))|\le$$

$$\le {O(1)\over L}\Vert f\Vert_{C^1(L^{-1}X)}(|\phi({x_1\over L})|+
\Vert\phi\Vert_{L^{-1}X,1,\s})
\Eq(5.22)$$

\\and

$$\Vert L^{-\b}T^{x_1}{\cal R} f\Vert_{C^1(X)}\le O(1)L^{-\b/2}
\Vert f\Vert_{C^1(L^{-1}X)}$$

\\which is proved as \equ(5.10). 

\\Now
we can proceed to the estimate of \equ(5.21) along the same lines
as the proof of \equ(5.20). Recalling the definition of the norm of
functional derivatives given in section 2, we obtain 

$$\left\Vert(D(1-{\cal L})({\cal R}{\hat r})
^{\natural})(L^{-1}X,\phi)\right\Vert\le O(1)L^{-\b/2}L^{-D}
G(L^{-1}X,\phi)\left[
\Vert{\hat r}(X)\Vert_{G}+
\Vert(D{\hat r})(X)\Vert_{G}
\right]
\Eq(5.23)$$

\\The results \equ(5.20) and \equ(5.23) can be
written in terms of the norms introduced in section 2
in the following way

$$\left\Vert(1-{\cal L})({\cal R}{\hat r})
^{\natural}(L^{-1}X,\phi)\right\Vert_1
\le O(1)L^{-\b/2}L^{-D}G(L^{-1}X,\phi)
\Vert{\hat r}(X)\Vert_{G,1}
\Eq(5.24)$$

\\Now we go back to \equ(5.7.1). Using \equ(5.24) we obtain

$$\Vert(1-{\cal L})\left({\cal S}_{1}{\hat r}^{(s.s.)}\right)
^{\natural}(Z,\phi)\Vert_1\le O(1)L^{-\b/2}L^{-D}
\sum_{X\ {\rm s.\ s.}\atop{\bar X}=LZ}G(L^{-1}X,\phi)
\Vert{\hat r}(X)\Vert_{G,1}\Eq(5.24.1)$$

\\This implies, exploiting Lemma 2.1.1

$$\Vert(1-{\cal L})\left({\cal S}_{1}{\hat r}^{(s.s.)}\right)
^{\natural}(Z,\phi)\Vert_1\G_p(Z)\le $$

$$\le O(1)L^{-\b/2}L^{-D}
\sum_{X\ {\rm s.\ s.}\atop{\bar X}=LZ}G(L^{-1}X,\phi)
\Vert{\hat r}(X)\Vert_{G,1}\G(X)\le$$

$$\le O(1)L^{-\b/2}L^{-D}\Vert{\hat r}\Vert_{G,1,\G}
\sum_{X\ {\rm s.\ s.}\atop{\bar X}=LZ}G(L^{-1}X,\phi)\Eq(5.25)$$

\\Now we use lemma 2.3.3 to get 

$$\Vert(1-{\cal L})\left({\cal S}_{1}{\hat r}^{(s.s.)}\right)
^{\natural}(Z,\phi)\Vert_1\G_p(Z)\le
O(1)L^{-\b/2}\Vert{\hat r}\Vert_{G,1,\G}G(Z,\phi)\Eq(5.29)$$

\\Observing now that

$$\sum_{Z\ {\rm s.\ s.}\atop Z\supset\D}1=O(1)$$

\\we have proved the following proposition
\vglue.5truecm
\\PROPOSITION 5.1.6
\vglue.3truecm
\\The contribution $(1-{\cal L})\left({\cal S}_{1}{\hat r}^{(s.s.)}\right)
^{\natural}$ 
due to small set activities and linear reblocking
satisfies the following bound for any $p\ge 0$,

$$\Vert(1-{\cal L})\left({\cal S}_{1}{\hat r}^{(s.s.)}\right)
^{\natural}\Vert_{G,1,\G_p}\le
O(1)L^{-\b/2}\Vert{\hat r}\Vert_{G,1,\G}\Eq(5.30)$$
\vglue.5truecm
\\{\it 5.2. Linear reblocking, large set contributions}
\vglue.5truecm

\\Now we consider the contribution 
$\left({\cal S}_{1}{\hat r}^{(l.s.)}\right)
^{\natural}$ 
due to large sets in linear reblocking.
After reblocking and rescaling we obtain for such terms 

$${\cal S}_{1}{\hat r}^{(l.s.)}(Z,\phi)=
\sum_{X\ {\rm large\ set}\atop{\bar X}=LZ}({\cal R}K)(L^{-1}X,\phi)\Eq(5.31)$$
\\For such contributions we have the following lemma :

\vglue.5truecm
\\LEMMA 5.2.1
\vglue.3truecm

$$\Vert\left({\cal S}_{1}{\hat r}^{(l.s.)}\right)
^{\natural}\Vert_{G,1,\G}\le
O(1)L^{-(1-\b/2)}\Vert{\hat r}\Vert_{G,1,\G}\Eq(5.34)$$
\vglue.3truecm
\\{\it Proof}

\\This is nothing but Lemma 2.5.3 of section 2.

\vglue.5truecm
\\{\it 5.3. Bound for the relevant part.}
\vglue.5truecm
\\We prove some preliminary
lemmas on the bounds of the relevant part.
The relevant terms $F_{{\hat r}}(Z,\phi)$ are defined by (see \equ(5.7))

$$F_{{\hat r}}(Z,\phi)=\sum_{X\ {\rm s.\ s.}\atop{\bar X}=LZ}{\cal L}({\cal R}
{\hat r})^{\natural}(L^{-1}X,\phi)\Eq(5.56)$$

Since $|{\bar X}|\le |X|$, $Z$ is a small set.

\\Then we have
\vglue.5truecm
\\LEMMA 5.3.1
\\For any integer $p\ge 0$
\vglue.3truecm
$$\Vert F_{{\hat r}}
\Vert_{G,1,\G_p}\le O(1)\Vert{\hat r} \Vert_{G,1,\G}\Eq(5.57)$$
\\where O(1) depends on $p$.
\vglue.5truecm
\\{\it Proof}

\\X is a small set.
\\From \equ(5.7) we have

$${\cal L}({\cal R}
{\hat r})^{\natural}(L^{-1}X,\phi)={L^{-\a}\over |L^{-1}X|}
\int_{L^{-1}X}dx_1e^{-(\l_*/2)|\phi(x_1)|^2}
{\hat r}_{*}^{\#\Sigma^{\bar x}}(X,0,0)\Eq(5.58)$$

\\and it is easy to see, using \equ(5.18) as well as (from Lemma 5.1.5)

$$\Vert({\hat r}_{*})(X)\Vert_{G_{*\r}}\le
{O(1)}\Vert{\hat r}(X)\Vert_{G}\Eq(5.59)$$

\\that the following inequality holds

$$\Vert{\cal L}({\cal R}
{\hat r})^{\natural}(L^{-1}X)\Vert_{G}\le 
{O(1)}L^{-D}\Vert{\hat r}(X)\Vert_{G}\Eq(5.60)$$

\\Analogously, for the functional derivative

$$\Vert(D{\cal L}({\cal R}
{\hat r})^{\natural})(L^{-1}X)\Vert_{G}\le 
{O(1)}L^{-D}\Vert{\hat r}(X)\Vert_{G}\Eq(5.61)$$

\\and we have therefore

$$\Vert{\cal L}({\cal R}
{\hat r})^{\natural}\Vert_{G,1,\G_p}\le 
{O(1)}L^{-D}\Vert{\hat r}\Vert_{G,1,\G}\Eq(5.62)$$

\\Applying lemma 2.3.3 finishes the proof
$$\eqno Q.E.D$$
\vglue.5truecm
\\LEMMA 5.3.2
\vglue.3truecm
$$\Vert F_{{\hat r}}
\Vert_{\io,1,\G_p}\le O(1)\Vert{\hat r} \Vert_{G,1,\G}\Eq(5.63)$$
\vglue.5truecm
\\{\it Proof}

\\From \equ(5.58) and \equ(5.59) we have

$$\vert{\cal L}({\cal R}
{\hat r})^{\natural}(L^{-1}X,\phi)\vert\le 
{O(1)}L^{-D}\Vert{\hat r}(X)\Vert_{G}\Eq(5.64)$$

\\and similarly, for the functional derivative

$$\Vert(D{\cal L}({\cal R}
{\hat r})^{\natural})(L^{-1}X,\phi)\Vert\le 
{O(1)}L^{-D}\Vert{\hat r}(X)\Vert_{G}\Eq(5.65)$$

\\Applying again lemma 2.3.3 we have the proof.
$$\eqno Q.E.D$$

\\We want to write $F_{\hat r}(Z,\phi)$
in terms of the sets where the dependence from the field $\phi$ is
localized. In other words, we want to
write the decomposition

$$F_{\hat r}(Z,\phi)=\sum_{\D\subset Z}F_{\hat r}(Z,\D,\phi)\Eq(5.66)$$

\\where in $F_{\hat r}(Z,\D,\phi)$ appear only fields defined in $\D$.

$F_{\hat r}(Z,\D,\phi)$ is given by

$$F_{\hat r}(Z,\D,\phi)=
\sum_{\D_1:{\bar\D}_1=L\D}\sum_{{X:\ s.s.\atop{\bar X}=LZ}\atop X\supset\D_1}
L^{-\a}
\int_{\D_1} dx_1
e^{-{\l_{*}\over 2}L^{-\b}\vert{\cal R}\phi (x_1)\vert^{2}}
f_{\hat r}(X)\Eq(5.68)$$

\\where

$$f_{\hat r}(X)=
{1\over |X|}
{\hat r}_{*}^{\#\Sigma_{\bar x}}(X,0,0)
\Eq(5.69)$$
\\From the above expressions it is easy to check that \equ(5.66) is satisfied.

\\Now, following [BDH 1],
 we define the contribution to the local 
effective potential :

$$V'_{F_{\hat r}}(\D,\phi)=-\sum_{Z\supset\D\atop |Z|\le 2,\ {\rm conn}}
F_{\hat r}(Z,\D,\phi)\Eq(5.71)$$

\\and from \equ(5.68)

$$\sum_{Z\supset\D}F_{\hat r}(Z,\D,\phi)=
\sum_{\D_1:{\bar\D}_1=L\D}\sum_{Z\supset\D}
\sum_{{X:\ s.s.\atop{\bar X}=LZ}\atop X\supset\D_1}
L^{-\a}
\int_{\D_1} dx_1
e^{-{\l_{*}\over 2}L^{-\b}\vert{\cal R}\phi (x_1)\vert^{2}}
f_{\hat r}(X)\Eq(5.70)$$

\\It is immediate to see that this can be rewritten as

$$\sum_{Z\supset\D}F_{\hat r}(Z,\D,\phi)=
\sum_{\D_1:{\bar\D}_1=L\D}L^{-\a}
\int_{\D_1} dx_1
e^{-{\l_{*}\over 2}L^{-\b}\vert{\cal R}\phi (x_1)\vert^{2}}
\sum_{X:\ s.s.\atop X\supset\D_1}
f_{\hat r}(X)\Eq(5.70.1)$$

\\Now we want to prove the following lemma

\vglue.5truecm
\\LEMMA 5.3.3
\vglue.3truecm

$$V'_{F_{\hat r}}(\D,\phi)=\xi_{\hat r}V_*(\D,\phi)\Eq(5.72)$$

\\with

$$|\xi_{\hat r}|\le O(1)\Vert{\hat r}\Vert_{G,1,\G}
\le O(1)\e^{5/2+\eta}\Eq(5.73)$$

\vglue.5truecm
\\{\it Proof}

\\By translation invariance

$$\sum_{X:\ s.s.\atop X\supset\D_1}
f_{\hat r}(X)=\sum_{X:\ s.s.\atop X\supset\D_1}{1\over |X|}
{\hat r}_{*}^{\#\Sigma_{\bar x}}(X,0,0)\Eq(5.74)$$

\\with ${\bar x}$ midpoint of $X$ is independent of $\D_1$,
i.e. $\D_1$ can be taken any unit block and the sum does not
change. Therefore we define

$$\sum_{X:\ s.s.\atop X\supset\D_1}{1\over |X|}
{\hat r}_{*}^{\#\Sigma_{\bar x}}(X,0,0)=
\xi_{\hat r}L^{-\e}$$

\\and \equ(5.70.1) can be rewritten

$$\sum_{Z\supset\D}F_{\hat r}(Z,\D,\phi)=
L^{-\a}
\int_{L\D} dx_1
e^{-{\l_{*}\over 2}L^{-\b}\vert{\cal R}\phi (x_1)\vert^{2}}
\xi_{\hat r}L^{-\e}\Eq(5.75)$$

\\Performing the rescaling in  \equ(5.75) we obtain \equ(5.72). 

\\To prove the bound \equ(5.73)
we observe

$$|\xi_{\hat r}|\le O(1)\sum_{X:\ s.s.\atop X\supset\D_1}{1\over |X|}
|{\hat r}_{*}^{\#\Sigma_{\bar x}}(X,0,0)|\le O(1)
\sum_{X:\ s.s.\atop X\supset\D_1}O(1)\Vert{\hat r}(X)\Vert_G 
\Eq(5.76)$$

\\where in the last step we used \equ(5.18.1) followed by
Lemma 5.1.5, \equ(5.15).

\\From \equ(5.76) we have

$$|\xi_{\hat r}|
\le O(1)(\sum_{Z\ {\rm s.\ s.}\atop Z\supset\D_{1}}1)
\Vert{\hat r}\Vert_{G,1,\G}\le$$
$$\le O(1)\Vert{\hat r}\Vert_{G,1,\G}$$

\\and by lemma 3.3.1 we obtain the proof.
$$\eqno Q.E.D.$$

\pagina
\vglue1.5truecm\numsec=6\numfor=1\pgn=1
\\\S6 {\it Extraction estimates.}
\vglue.5truecm
From lemma 3.1.1, then reblocking-rescaling followed by
fluctuation integration, and then the preliminary extraction
provided by lemma 3.4.1, we obtained a single step of RG in the form

$$e^{-gV_*(\L)}{\cal E}xp(\square + K)(\L)\longrightarrow
e^{-gL^\e V_*(L^{-1}\L)}{\cal E}xp(\square + {\tilde K})(L^{-1}\L)\Eq(6.1)$$

\\with ${\tilde K}$ given by \equ(1.49). In sections 4-5 respectively we
obtained the relevant parts $F_{\tilde Q}$ (see
\equ(2.15)) and $F_{\hat r}$ (see \equ(5.56)). They are supported on small
sets and have the following local representations:

$$F_{\tilde Q}(Z)=\sum_{\D\subset Z}F_{\tilde Q}(Z,\D)\Eq(6.2)$$
$$F_{\hat r}(Z)=\sum_{\D\subset Z}F_{\hat r}(Z,\D)\Eq(6.3)$$

\\The corresponding local potentials are

$$V'_{\tilde Q}(L^{-1}\L)=
\sum_{\D\subset L^{-1}\L}V'_{\tilde Q}(\D)\Eq(6.4)$$
$$V'_{\tilde Q}(\D)=\sum_{Z\supset \D\atop Z\subset L^{-1}\L}
F_{\tilde Q}(Z,\D)=-L^\e b_1g^2V_*(\D)\Eq(6.5)$$

\\with $b_1=O(\ln L)>0$ (see lemma 4.6.1)

$$V'_{\hat r}(L^{-1}\L)=
\sum_{\D\subset L^{-1}\L}V'_{\hat r}(\D)\Eq(6.6)$$
$$V'_{\hat r}(\D)=\sum_{Z\supset \D\atop Z\subset L^{-1}\L}
F_{\hat r}(Z,\D)=\xi_{\hat r}V_*(\D)\Eq(6.7)$$

\\with $|\xi_{\hat r}|\le O(1)\e^{5/2+\eta}$ (see lemma 5.3.3).
These lemmas are valid under the assumptions \equ(1.5) and \equ(1.10.1)
on $g$ and on the remainder $ r$.

\\Define the total contribution to the local effective
potential, due to $F_{\tilde Q}$ and $F_{\hat r}$ as 

$$V'_F=V'_{\tilde Q}+V'_{\hat r}\Eq(6.8)$$

\\where the total relevant part is

$$F=F_{\tilde Q}+F_{\hat r}\Eq(6.9)$$

\\Then the local effective potential $gL^\e V_*(L^{-1}\L)$
can be replaced by

$$V'=gL^\e V_*+V'_F\Eq(6.10)$$

\\provided we replace the activity ${\tilde K}$ with a
new activity ${\cal E}({\tilde K})$. This procedure known as
{\it extraction} (adopting the terminology of BDH)
is given by the following proposition:

\vglue.5truecm
\\PROPOSITION 6.1 (Extraction)
\vglue.5truecm
$$e^{-gL^\e V_*(L^{-1}\L)}{\cal E}xp
(\square + {\tilde K})(L^{-1}\L)=
e^{-V'(L^{-1}\L)}{\cal E}xp
(\square + {\cal E}({\tilde K}))(L^{-1}\L)\Eq(6.11)$$

\\where

$$ {\cal E}({\tilde K})=({\tilde K}-F)+
(e^{-F}-1-F)+(e^{-F}-1)^+_{\ge 2}+
(e^{-F}-1)^+\vee{\tilde K}\Eq(6.12)$$

\\Before we prove this proposition let us observe:

$$V'=g'V_*\Eq(6.13)$$

\\where

$$g'=L^\e g(1-b_1g) + \xi_{\hat r}\Eq(6.14)$$

as follows from \equ(6.8) and \equ(6.10).

\vglue.5truecm
\\{\it Proof of proposition 6.1}

\\$ {\cal E}({\tilde K})$ is determined by

$${\cal E}xp
(\square + {\cal E}({\tilde K}))(L^{-1}\L)=
e^{V'_F(L^{-1}\L)}
{\cal E}xp
(\square + {\tilde K})(L^{-1}\L)
\Eq(6.15)$$

\\From \equ(6.2), \equ(6.3) and \equ(6.9) we have

$$F(Z)=\sum_{\D\subset Z}F(Z,\D)\Eq(6.16)$$

\\Observe that

$$\sum_{Z\subset L^{-1}\L}F(Z)=\sum_{Z\subset L^{-1}\L}
\sum_{\D\subset Z}F(Z,\D)=\sum_{\D\subset L^{-1}\L}
\sum_{Z\supset \D\atop Z\subset L^{-1}\L}F(Z,\D)=$$
$$=\sum_{\D\subset L^{-1}\L}-V'_F(\D)=-V'_F(L^{-1}\L)\Eq(6.17)$$

\\Hence

$$e^{V'_F(L^{-1}\L)}=\prod_{Z\subset L^{-1}\L}e^{-F(Z)}=
\prod_{Z\subset L^{-1}\L}((e^{-F(Z)}-1)+1)=
{\cal E}xp(\square + (e^{-F}-1)^+)(L^{-1}\L)\Eq(6.18)$$

Using this, from \equ(6.15)

$${\cal E}xp
(\square + {\cal E}({\tilde K}))=
{\cal E}xp(\square + (e^{-F}-1)^+)
{\cal E}xp
(\square + {\tilde K})
\Eq(6.19)$$

and from lemma 1.4.1

$$ {\cal E}({\tilde K})=
{\tilde K}+(e^{-F}-1)^++
(e^{-F}-1)^+\vee{\tilde K}=$$
$$=({\tilde K}-F)+
(e^{-F}-1-F)+(e^{-F}-1)^+_{\ge 2}+
(e^{-F}-1)^+\vee{\tilde K}\Eq(6.20)$$
$$\eqno Q.E.D.$$

Note that from the definition of ${\tilde K}$ given in
\equ(1.54) and from \equ(5.56)

$$ {\cal E}({\tilde K})={\cal I}_{k+1}+r'\Eq(6.21)$$

\\where

$$r'=(1-{\cal L})\left({\cal S}_{1}{\hat r}^{(s.s.)}\right)
^{\natural}+\left({\cal S}_{1}{\hat r}^{(l.s.)}\right)
^{\natural}+{\tilde r}^{\natural}+{\bar r}+$$

$$+(e^{-F}-1-F)+(e^{-F}-1)^+_{\ge 2}+
(e^{-F}-1)^+\vee{\tilde K}\Eq(6.22)$$

\\We wish to estimate $r'$. This is provided by the following 
proposition.

\vglue.5truecm
\\PROPOSITION 6.2
\vglue.5truecm
$$\Vert r'\Vert_{G,1,\G_6}\le {1\over L^{\b/4}}\e^{5/2+\eta}\Eq(6.23)$$
\vglue.5truecm
\\{\it Proof }

We recall that we have already proved

$$\Vert(1-{\cal L})\left({\cal S}^{(1)}{\hat r}^{(s.s.)}\right)
^{\natural}\Vert_{G,1,\G_6}\le
O(1)L^{-\b/2}\Vert{\hat r}\Vert_{G,1,\G}\Eq(6.24)$$

\\(see proposition 5.1.6)

$$\Vert\left({\cal S}_{1}{\hat r}^{(l.s.)}\right)
^{\natural}\Vert_{G,1,\G_6}\le
O(1)L^{-(1-\b/2)}\Vert{\hat r}\Vert_{G,1,\G}
\le O(1)L^{-\b/2}\Vert{\hat r}\Vert_{G,1,\G}\Eq(6.25)$$

\\(see lemma 5.2.1)

$$\Vert{\tilde r}^\natural\Vert_{G,1,\G_6}\le 
L^{-\b/2}\e^{5/2+\eta}\Eq(6.26)$$

\\(see lemma 3.5.2)

$$\Vert{\bar r}\Vert_{G,1,\G_6}\le L^{-\b/2}\e^{5/2+\eta}\Eq(6.27)$$

\\(see lemma 3.5.1)

\\We prove the following lemmas before returning
to the proof of proposition 6.2
\vglue.5truecm
\\LEMMA 6.3
\vglue.5truecm
$$\Vert e^{-F}-1-F\Vert_{G,1,\G_6}\le L^{-\b/2}\e^{5/2+\eta}\Eq(6.28)$$

\\{\it Proof}

\\From lemmas 4.6.2, 5.3.1, 5.3.2 we have

$$\Vert F\Vert_{G,1,\G_p}\le O(1)\e^{7/4}\qquad
\Vert F\Vert_{\io,1,\G_p}\le O(1)\e^{7/4}\Eq(6.29)$$

\\for any integer $p\ge 0$, with $O(1)$ depending on $p$.

\\From lemma 2.5.4

$$\Vert e^{-F}-1-F\Vert_{G,1,\G_6}\le
O(1)\Vert F\Vert_{G,1,\G_6}\Vert F\Vert_{\io,1,\G_6}\Eq(6.30)$$

\\putting in \equ(6.30) the bounds \equ(6.29) we have

$$\Vert e^{-F}-1-F\Vert_{G,1,\G_6}\le O(1)\e^{14/4}\Eq(6.31)$$

\\and the lemma follows from the smallness of $\e$.
$$\eqno Q.E.D.$$

\vglue.5truecm
\\LEMMA 6.4
\vglue.5truecm
$$\Vert (e^{-F}-1)^+_{\ge 2}\Vert_{G,1,\G_6}\le 
L^{-\b/2}\e^{5/2+\eta}\Eq(6.32)$$

\\{\it Proof}

\\From lemma 2.5.5, following the same lines of lemma 6.3
$$\eqno Q.E.D.$$

\vglue.5truecm
\\LEMMA 6.5
\vglue.5truecm
$$\Vert (e^{-F}-1)^+\vee{\tilde K}\Vert_{G,1,\G_6}\le 
L^{-\b/2}\e^{5/2+\eta}\Eq(6.33)$$

\\{\it Proof}

$$\Vert (e^{-F}-1)^+\vee{\tilde K}\Vert_{G,1,\G_6}\le 
\sum_{N,M\ge 1}O(1)^{N+M}\Vert (e^{-F}-1)^+\Vert_{\io,1,\G_9}^N
\Vert{\tilde K}\Vert_{G,1,\G_9}^M\Eq(6.34)$$

\\From the expression of ${\tilde K}$ \equ(1.54), corollary 2.5.2,
bound \equ(1.9) proved by means of lemma 4.7.1, and lemmas 3.5.1,
3.5.2 we have

$$\Vert{\tilde K}\Vert_{G,1,\G_9}\le O(1)\e^{7/4}\Eq(6.35)$$

\\From lemma 2.5.5, and \equ(6.29)

$$\Vert (e^{-F}-1)^+\Vert_{\io,1,\G_9}\le O(1)\e^{7/4}\Eq(6.36)$$

\\Then

$$\Vert (e^{-F}-1)^+\vee{\tilde K}\Vert_{G,1,\G_6}\le 
O(1)\e^{14/4}\Eq(6.37)$$

\\and the lemma follows from the smallness of $\e$.
$$\eqno Q.E.D.$$

\\Now we come back to the proof of proposition 6.2.
 
\\From \equ(6.24) - \equ(6.27)  and lemmas 6.3 - 6.5 we have

$$\Vert r'\Vert_{G,1,\G_6}\le O(1)
{1\over L^{\b/2}}\e^{5/2+\eta}\Eq(6.38)$$

\\and proposition 6.2 follows for $L$ large enough.
$$\eqno Q.E.D.$$

\\From \equ(6.14) the evolved coupling constant $g'$
at the end of one RG step is

$$g'=L^\e g(1-b_1g) + \xi_{\hat r}$$

\\In the absence of the remainder contribution $\xi_{\hat r}$,
the approximate flow has the fixed point ${\bar g}$

$${\bar g}=L^\e{\bar g}(1-b_1{\bar g})\Eq(6.39)$$

\\whence

$${\bar g}={L^\e-1\over L^\e b_1}\Eq(6.40)$$

\\From lemma 4.6.1 we have

$$0<b_1=O(\ln L)\Eq(6.41)$$

\\From the smallness of $\e$ \equ(1.5) we have

$${\bar g}=O(\e)\Eq(6.42)$$

\\Assume

$$|g-{\bar g}|\le \e^{3/2}\Eq(6.43)$$

\\Then, since ${\bar g}=O(\e)$, we have $g=O(\e)$ and the
hypotesis at the beginning of section 3 is satisfied.
We have, under the assumption \equ(6.43) above

\vglue.5truecm
\\PROPOSITION 6.6
\vglue.5truecm
\\The evolved coupling constant $g'$ satisfies

$$|g'-{\bar g}|\le \e^{3/2}\Eq(6.44)$$

\\so that the closed ball centered at ${\bar g}$ of radius
$\e^{3/2}$ is stable under RG iteration.

\\Also,
$$|g'- g|\le \e^{1/2}\e^{3/2}\Eq(6.45)$$

\vglue.5truecm
\\{\it Proof}

\\From \equ(6.14) and 
an elementary calculation using property \equ(6.39)
we have

$$g'-{\bar g}=(g'-{\bar g})(1-L^\e b_1g) + \xi_{\hat r}$$

\\$g$ is $O(\e)$, $b_1$ is $O(\ln L)$, hence for $\e$ 
sufficiently small

$$0<1-L^\e b_1g\le 1-O(\e)$$

\\By assumption \equ(6.43) above and by lemma 5.3.3

$$|\xi_{\hat r}|\le \e^{5/2+\eta}$$

\\we have

$$|g'-{\bar g}|\le \e^{3/2}(1-O(\e)+O(\e^{1+\eta}))\le \e^{3/2}$$

\\for $\e$ sufficiently small.

\\The next part of the proposition follows by writing

$$ g'- g=(g-{\bar g})(-L^\e b_1g)+\xi_{\hat r}$$

\\ and estimating as before.
$$\eqno Q.E.D.$$

\pagina
\vglue1.5truecm\numsec=7\numfor=1\pgn=1
\\\S7 {\it Convergence to a non Gaussian fixed point}
\vglue.5truecm
\\The partition functional at the $n$-th step of the RG can be
parametrized by
$$(g_n,{\cal I}_n, r_n)$$
in volume $L^{N+1-n}$. A further RG transformation is a map
$$(g_n,{\cal I}_n, r_n)\longrightarrow (g_{n+1},{\cal I}_{n+1}, r_{n+1})
\Eq(7.1)$$
\\where the volume changes to $L^{N+1-(n+1)}$. 
In order to discuss the convergence
of the sequence of RG transformations we shall
consider $N,n$ very large with $N\gg n$, so that we are
effectively in ``infinite'' volume.
With this hypothesis, the sequence of transformations \equ(7.1)
are iterations of a fixed mapping.

\\Let us now recall some results of the preceeding sections.

\\By hypothesis ${\cal I}_0=r_0=0$ and $|g_0-{\bar g}|\le \e^{3/2}$.
By the structure of irrelevant terms in second order perturbation 
theory given in section 4, and by results established there,
we can write for any $n$

$${\cal I}_n=\sum_{l=1}^n g^2_{n-l}{\bar{\cal I}}_l\Eq(7.2)$$

$${\bar{\cal I}}_l=({\cal S}_1{\bar{\cal I}}_{l-1})^\natural\Eq(7.3)$$

$$\Vert {\bar{\cal I}}_l\Vert_{G,1,\G_p}\le 
O(1)L^{3\b -l\b/2}L^{2(D+2)}\Eq(7.4)$$

\\for all integers $p\ge 1$, with $O(1)$ depending on $p$.

\\Note that for $n\ge 1$:

$$\Vert{\cal I}_n-{\cal I}_{n-1}\Vert_{G,1,\G_p}\le 
L^{-n\b/4}\e^{7/4}\Eq(7.5)$$

\\where we have used the smallness of $\e$.

\\By proposition 6.2, 6.6, the closed ball:
$$B=\left\{g,r\left\vert\ |g-{\bar g}|\le \e^{3/2},\Vert r\Vert_{G,1,\G_6}\le 
\e^{5/2+\eta}\right.\right\}\Eq(7.6)$$

\\is stable under RG iteractions. Moreover from \equ(7.2), \equ(7.4),
and using $g_n=O(\e)$ for all $n$, we get easily

$$\Vert {{\cal I}}_n\Vert_{G,1,\G_p}\le O(\e^2)
L^{2(D+2)}\sum_{l=1}^n L^{-(l-1)\b/4}\le \e^{7/4}\Eq(7.7)$$

\\We thus have
\vglue.5truecm
\\PROPOSITION 7.1
\vglue.5truecm
\\For any $n$
$$|g_n-{\bar g}|\le \e^{3/2}\Eq(7.8)$$
$$\Vert r_n\Vert_{G,1,\G_6}\le 
\e^{5/2+\eta}\Eq(7.9)$$
$$\Vert {{\cal I}}_n\Vert_{G,1,\G_6}\le \e^{7/4}\Eq(7.10)$$

\vglue.5truecm
\\Define, for any sequence $\{a_n\}$, the increments
$\D a_n=a_{n+1}-a_n$, and make the following
{\it inductive hypothesis}:
\vglue.3truecm
\\For all $j=1,2,...,n$,

$$|\D g_{j-1}|\le k_*^j\e^{3/2}\Eq(7.11)$$
$$\Vert \D r_{j-1}\Vert_{G,1,\G_6}\le  k_*^j
\e^{5/2+\eta}\Eq(7.12)$$

\\where

$$k_*=1-\e\ln L+2\e^{1+\eta/2}\Eq(7.13)$$

\\Clearly $0<k_*<1$.

\\Note that the inductive hypothesis is true for $j=1$. In fact, 
for $j=1$
\equ(7.11) follows from proposition 6.6, and
\equ(7.12) follows from proposition 6.2 if we use
$L^{-\b/4}\le k_*$, for sufficiently large $L$. Remember
$r_0=0,\ \eta>0$ and sufficiently small, say $\eta=1/20$
as in section 3.
Our task will be to prove that \equ(7.11) and \equ(7.12)
are true for $j=n+1$. To this end we first note some 
preliminary estimates. These are increment version of lemma
3.3.1, \equ(6.35), and lemmas 3.5.1, 3.5.2. 
They are summarized in the following lemma:

\vglue.5truecm
\\LEMMA 7.2
\vglue.5truecm
\\For any integer $p\ge 0$ and for $O(1)$ depending on $p$
$$\Vert \D {\hat r}_{n-1}\Vert_{G,1,\G}\le O(1) k_*^n
\e^{5/2+\eta}\Eq(7.14)$$
$$\Vert \D {\tilde K}_{n-1}\Vert_{G,1,\G_p}\le  O(1)
k_*^n\e^{7/4}\Eq(7.15)$$
$$\Vert \D {\bar r}_{n-1}\Vert_{G,1,\G_p}\le  O(1){k_*^n\over L^{\b/2}}
\e^{5/2+\eta}\Eq(7.16)$$
$$\Vert (\D {\tilde r}_{n-1})^\natural\Vert_{G,1,\G_p}\le  O(1)
{k_*^n\over L^{\b/2}}
\e^{5/2+\eta}\Eq(7.17)$$
The quantities ${\hat r},\ {\tilde K},\ {\bar r},\ {\tilde r}$
are those introduced in section 3.

\\We omit the proof of lemma 7.2. The proof is straightforward.
We have to use proposition 7.1 and the inductive hypothesis.
We use lemmas 2.6.1, 2.6.2, 2.6.3. For \equ(7.15) we use also
\equ(7.5) and $L^{-\b/4}\le k_*$. Then follow the lines of the proofs
of lemma 3.3.1, \equ(6.35), and lemmas 3.5.1, 3.5.2.

\\Now turn to the extracted activities. By definition 
(see \equ(6.22))

$$r_{n+1}=(1-{\cal L})\left({\cal S}_{1}{\hat r}_n^{(s.s.)}\right)
^{\natural}+\left({\cal S}_{1}{\hat r}_n^{(l.s.)}\right)
^{\natural}+{\tilde r}_n^{\natural}+{\bar r}_n+$$

$$+(e^{-F_n}-1-F_n)+(e^{-F_n}-1)^+_{\ge 2}+
(e^{-F_n}-1)^+\vee{\tilde K}_n\Eq(7.18)$$

\\Hence

$$\D r_{n}=(1-{\cal L})\left({\cal S}_{1}\D{\hat r}_{n-1}^{(s.s.)}\right)
^{\natural}+\left({\cal S}_{1}\D{\hat r}_{n-1}^{(l.s.)}\right)
^{\natural}+(\D{\tilde r}_{n-1})^{\natural}+\D{\bar r}_{n-1}+$$

$$+\D(e^{-F_{n-1}}-1-F_{n-1})+\D(e^{-F_{n-1}}-1)^+_{\ge 2}+$$

$$+
(\D(e^{-F_{n-1}}-1)^+)\vee{\tilde K}_{n}
+(e^{-F_{n-1}}-1)^+)\vee \D{\tilde K}_{n-1}\Eq(7.19)$$

\vglue.5truecm
\\PROPOSITION 7.3
\vglue.5truecm

$$\Vert\D r_{n}\Vert_{G,1,\G_6}\le  O(1)
{k_*^{n+1}}
\e^{5/2+\eta}\Eq(7.20)$$
\vglue.3truecm
\\{\it Proof}

\\We estimate each term on the r. h. s. of \equ(7.19).

\\By proposition 5.1.6 and linearity

$$\Vert(1-{\cal L})\left({\cal S}_{1}\D{\hat r}_{n-1}^{(s.s.)}\right)
^{\natural}\Vert_{G,1,\G_6}\le  {O(1)\over L^{\b/2}}
\Vert\D{\hat r}_{n-1}\Vert_{G,1,\G}\le  {O(1)\over L^{\b/2}}
k_*^n\e^{5/2+\eta}\Eq(7.21)$$

\\where in the last step we have used lemma 7.2.

$$\Vert\left({\cal S}_{1}\D{\hat r}_{n-1}^{(l.s.)}\right)
^{\natural}\Vert_{G,1,\G_6}\le  {O(1)\over L^{\b/2}}
\Vert\D{\hat r}_{n-1}\Vert_{G,1,\G}\le  {O(1)\over L^{\b/2}}
k_*^n\e^{5/2+\eta}\Eq(7.22)$$

\\by lemmas 2.5.3 and 7.2.

$$\Vert (\D {\tilde r}_{n-1})^\natural\Vert_{G,1,\G_6}\le  O(1)
{k_*^n\over L^{\b/2}}
\e^{5/2+\eta}\Eq(7.23)$$

\\by lemma 7.2.

$$\Vert\D(e^{-F_{n-1}}-1-F_{n-1}) \Vert_{G,1,\G_6}\le  O(1)
{k_*^n\over L^{\b/2}}
\e^{5/2+\eta}\Eq(7.24)$$

\\{\it Proof of \equ(7.24)}

$$F_n = F_{{\tilde Q}_n}+F_{{\hat r}_n}$$

\\Consider first $F_{{\tilde Q}_n}$ with $g=g_n=O(\e)$ by
proposition 7.1.
From lemma 4.6.2, for $j=n-1, n$ and any integer $p\ge 0$

$$\Vert F_{{\tilde Q}_j}\Vert_{G,1,\G_p}\le  O(1)\e^{7/4}\qquad
\Vert F_{{\tilde Q}_j}\Vert_{\io,1,\G_p}\le  O(1)\e^{7/4}$$

\\Then, by lemmas 5.3.1, 5.3.2, for $j=n-1, n$ and any integer $p\ge 0$

$$\Vert F_{{\hat r}_j}\Vert_{G,1,\G_p}\le  O(1)
\Vert{\hat r}_j\Vert_{G,1,\G}\le  O(1)\e^{5/2+\eta}\qquad
\Vert F_{{\hat r}_j}\Vert_{\io,1,\G_p}\le  O(1)
\Vert{\hat r}_j\Vert_{G,1,\G}\le  O(1)\e^{5/2+\eta}$$

\\where in the last step we used lemma 3.3.1

\\Hence

$$\Vert F_j\Vert_{G,1,\G_p}\le  O(1)\e^{7/4}\qquad
\Vert F_j\Vert_{\io,1,\G_p}\le  O(1)\e^{7/4}\Eq(7.24.1)$$

\\Hence, by lemma 2.6.2

$$\Vert\D(e^{-F_{n-1}}-1-F_{n-1}) \Vert_{G,1,\G_6}\le  O(1)
\e^{7/4}\Vert\D F_{n-1}\Vert_{G,1,\G_6}\Eq(7.25)$$

\\By \equ(2.18.3), \equ(2.18.4)

$$\Vert\D F_{{\tilde Q}_{n-1}}\Vert_{G,1,\G_p}\le  O(\ln L)
L^{2(D+2)}|g_n^2-g_{n-1}^2|\le O(1) k_*^n\e^{7/4}\Eq(7.26)$$

\\where we used in the last step the inductive hypothesis.
By lemma 5.3.2, linearity, and lemma 7.2

$$\Vert\D F_{{\hat r}_{n-1}}\Vert_{G,1,\G_p}\le  O(1)
\Vert\D {\hat r}_{n-1}\Vert_{G,1,\G}
\le O(1) k_*^n\e^{5/2+\eta}\Eq(7.27)$$

\\From \equ(7.26), \equ(7.27) we obtain

$$\Vert\D F_{n-1}\Vert_{G,1,\G_6}
\le O(1) k_*^n\e^{7/4}\Eq(7.28)$$

\\Putting \equ(7.28) in \equ(7.25) and using the smallness of $\e$
 we obtain the proof of
\equ(7.24)
\bigskip
\\We have now the estimate

$$\Vert\D (e^{-F_{n-1}}-1)^+_{\ge 2}\Vert_{G,1,\G_6}
\le {O(1)\over L^{\b/2}} k_*^n\e^{5/2+\eta}\Eq(7.29)$$

\\This follows from \equ(7.24.1), \equ(7.28) and lemma 2.6.3
\bigskip
\\Next we have the estimate:
$$\Vert(\D(e^{-F_{n-1}}-1)^+)\vee{\tilde K}_{n}\Vert_{G,1,\G_6}\le
O(1)\Vert(\D(e^{-F_{n-1}}-1)^+)\Vert_{\io,1,\G_9}\Vert
{\tilde K}_{n}\Vert_{G,1,\G_9}\le $$
$$\le O(1)\Vert\D F_{n-1}\Vert_{\io,1,\G_9}\Vert
{\tilde K}_{n}\Vert_{G,1,\G_9}\le O(1)k_*^n\e^{14/4}
\le {O(1)\over L^{\b/2}} k_*^n\e^{5/2+\eta}\Eq(7.30)$$

\\where we have used lemma 2.6.3 with $k=1$, and then \equ(7.28),
\equ(6.35).
\bigskip
\\Finally, we have:
$$\Vert(e^{-F_{n-1}}-1)^+\vee \D{\tilde K}_{n-1}\Vert_{G,1,\G_6}\le
O(1)\Vert(e^{-F_{n-1}}-1)^+\Vert_{\io,1,\G_9}
\Vert\D{\tilde K}_{n-1}\Vert_{G,1,\G_9}\le$$ 
$$\le O(1)k_*^n\e^{14/4}
\le {O(1)\over L^{\b/2}} k_*^n\e^{5/2+\eta}\Eq(7.31)$$

\\where we have used \equ(6.36) and lemma 7.2

\\Adding up the estimates given in \equ(7.21)-\equ(7.24)
and \equ(7.29)-\equ(7.31)
we obtain

$$\Vert\D r_n \Vert_{G,1,\G_6}
\le {O(1)\over L^{\b/2}} k_*^n\e^{5/2+\eta}\Eq(7.32.1)$$

\\and from this we obtain the proof for $L$ large enough.
$$\eqno Q.E.D.$$

\vglue.5truecm
\\PROPOSITION 7.4
\vglue.5truecm

$$\vert\D g_{n}\vert\le 
{k_*^{n+1}}
\e^{5/2+\eta}\Eq(7.32)$$
\vglue.3truecm
\\{\it Proof}

$$g_{n+1}=L^\e g_n (1-b_1g_n)+\xi_{{\hat r}_n}$$

\\whence

$$\D g_{n}=\D g_{n-1}(L^\e-L^\e b_1(g_n+g_{n-1}))+\D\xi_{{\hat r}_{n-1}}
\Eq(7.33)$$

\\Using the definition of the approximate fixed point

$${\bar g}={L^\e-1\over L^\e b_1}$$

\\We have

$$L^\e-L^\e b_1(g_n+g_{n-1})=2-L^\e-L^\e b_1(g_n-{\bar g})
-L^\e b_1(g_{n-1}-{\bar g})$$

\\Using the fact that $L^\e-L^\e b_1(g_n+g_{n-1})>0$ for
$\e$ sufficiently small we have

$$0\le L^\e-L^\e b_1(g_n+g_{n-1})\le 1-\e\ln L +L^\e b_1|g_n-{\bar g}|
+L^\e b_1|g_{n-1}-{\bar g}|\le  1-\e\ln L +2L^\e b_1\e^{3/2}
\Eq(7.34)$$

\\We have also by the inductive hypothesis

$$|\D g_{n-1}|\le k_*^n\e^{3/2}\Eq(7.35)$$

$$|\D\xi_{{\hat r}_{n-1}}|\le 
O(1)\Vert\D {\hat r}_{n-1} \Vert_{G,1,\G}
\le {O(1)} k_*^n\e^{5/2+\eta}\Eq(7.36)$$

\\where we also used linearity and lemma 5.3.3.

\\Putting \equ(7.34), \equ(7.35) and \equ(7.36) in
\equ(7.33) we get

$$|\D g_n|\le \e^{3/2}(1-\e\ln L+2L^\e b_1\e^{3/2}
+O(1)\e^{1+\eta})k_*^n \le$$

$$\le k_*^{n+1}\e^{3/2}$$
which gives the proof of
\equ(7.32).
$$\eqno Q.E.D.$$

\\Propositions 7.3 and 7.4 imply that the inductive hypothesis
\equ(7.11), \equ(7.12) is actually true for all $j\ge 1$.
This, together with \equ(7.5) and proposition 7.1
implies:

\vglue.5truecm
\\THEOREM 7.5
\vglue.5truecm
\\Let $\e >0$ and sufficiently small.  Then the  sequence
$(g_n,{\cal I}_n, r_n)$
converges, as $n
\longrightarrow\io$ to the fixed point 
$(g_\io,{\cal I}_\io, r_\io)$. Moreover

$$|g_\io-{\bar g}|\le \e^{3/2}$$
$$\Vert r_\io\Vert_{G,1,\G_6}\le 
\e^{5/2+\eta}$$
$$\Vert {{\cal I}}_\io\Vert_{G,1,\G_6}\le \e^{7/4}$$

\\Since ${\bar g}=O(\e)$, we have $g_\io\ne 0$ 
so that the fixed point is non Gaussian.

\vglue.5cm
\\{\it Acknowledgements}
\vglue.5cm
\\We are grateful to  David Brydges for many interesting conversations 
and suggestions during the course of this work. We thank Marzio Cassandro
for having participated at an early stage of this project and for his 
continued interest. We thank Francois David
and Kay Wiese for interesting comments
and questions as well as bringing to our attention
some relevant references. P.K.M thanks Gerard Menessier for his mathematical
vigilance and for having cooperated on an earlier version of the proof of
lemma 2.3.2. He thanks the INFN, sezione di Roma, for having supported
his visits to Rome during the course of this work.

\pagina
\vglue1.5truecm\numsec=1\numfor=1\pgn=1
\\{\it Appendix A}
\vglue.5truecm
\\
In this appendix we will prove lemma 2.3.2 on the stability of the
large field regulator.
\vglue.5truecm
\\{\it Proof}

\\Using the master formula given in Lemma 1.3.1, we have

$$(\mu_\G*G_{\r,\k})(X,\phi)={L^{-\a}\over |X|}
\int_Xdx_1e^{-(\l_*/2)L^{-\b}|\phi(x_1)|^2}$$
$$\int d\m_{\Sigma^{x_1}}(\z)
e^{(\r\l_*/2)|\z(x_1)+L^{-\b}\phi(x_1)|^2}
e^{\k\Vert\z+L^{-\b}T^{x_1}\phi\Vert^2_{X,1,\s}}\le$$
$$\le{L^{-\a}\over |X|}
\int_Xdx_1e^{-(\l_*/2)L^{-\b}|\phi(x_1)|^2}$$
$$\left(
\int d\m_{\Sigma^{x_1}}(\z)
e^{\r\l_*|\z(x_1)+L^{-\b}\phi(x_1)|^2}\right)^{1/2}
\left(
\int d\m_{\Sigma^{x_1}}(\z)
e^{2\k\Vert\z+L^{-\b}T^{x_1}\phi\Vert^2_{X,1,\s}}\right)^{1/2}\Eqa(A.1)$$

\\Observing that

$$\s=\Sigma^{x_1}(x_1,x_1)=\g-\l_*L^{-\b}\g^2$$

\\so that

$$\l_*\s=1-{1\over L^\b}$$

\\we easily obtain the bound

$$\int d\m_{\Sigma^{x_1}}(\z)
e^{\r\l_*|\z(x_1)+L^{-\b}\phi(x_1)|^2}\le$$

$$\le e^{2\r\l_*L^{-2\b}|\phi(x_1)|^2}\int d\m_{\Sigma^{x_1}}(\z)
e^{2\r\l_*|\z(x_1)|^2}\le$$

$$\le(1-4\r)^{-d/2}e^{2\r\l_*L^{-2\b}|\phi(x_1)|^2}
\Eqa(A.2)$$

\\hence for $0<\r<{1\over 8}$ we get from \equ(A.1)

$$(\mu_\G*G_{\r,\k})(X,\phi)\le 2^{d/4}
{L^{-\a}\over |X|}
\int_Xdx_1e^{-(\l_*/2)\left(1-{2\r\over L^\b}\right)
L^{-\b}|\phi(x_1)|^2}\cdot$$
$$\cdot\left(
\int d\m_{\Sigma^{x_1}}(\z)
e^{2\k\Vert\z+L^{-\b}T^{x_1}\phi\Vert^2_{X,1,\s}}\right)^{1/2}\Eqa(A.3)$$

\\We shall now estimate the last integral in \equ(A.3).
First we observe:

$$\z+L^{-\b}T^{x_1}\phi=\z+\phi-\l_*L^{-\b}\G(\cdot,x_1)\phi(x_1)$$

\\Then we have

$$\int d\m_{\Sigma^{x_1}}(\z)
e^{2\k\Vert\z+L^{-\b}T^{x_1}\phi\Vert^2_{X,1,\s}}
\le e^{4\k\l_*^2L^{-2\b}\Vert\G(\cdot,x_1)\Vert^2_{X,1,\s}
|\phi(x_1)|^2}\int d\m_{\Sigma^{x_1}}(\z)
e^{4\k\Vert\z+\phi\Vert^2_{X,1,\s}}
\Eqa(A.4)$$

\\We now make the following claim.

\vglue.5truecm
\\{\it Claim A.1}

\\For $\k>0$ sufficiently small, independent of $L$,
and any $x_1\in X$

$$\int d\m_{\Sigma^{x_1}}(\z)
e^{4\k\Vert\z+\phi\Vert^2_{X,1,\s}}\le
2^{|X|}e^{8\k\Vert\ \phi\Vert^2_{X,1,\s}}
\Eqa(A.6)$$

\\Observe also that, from lemma 1.1.1,
$$\Vert\G(\cdot,x_1)\Vert^2_{X,1,\s}\le
\sum_{1\le\a\le\s}\int_Xdx|\dpr^\a\G(x-x_1)|^2\le
\sum_{1\le\a\le\s}\int_{\bf R}dx|\dpr^\a\G(x-x_1)|^2\le O(1)\Eqa(A.7)$$

\\Using \equ(A.6) and \equ(A.7) we get from \equ(A.4)

$$\int d\m_{\Sigma^{x_1}}(\z)
e^{2\k\Vert\z+L^{-\b}T^{x_1}\phi\Vert^2_{X,1,\s}}
\le 2^{|X|}e^{O(1)\k\l_*^2L^{-2\b}
|\phi(x_1)|^2}
e^{8\k\Vert\phi\Vert^2_{X,1,\s}}
\Eqa(A.8)$$

\\and using \equ(A.8) we get from \equ(A.3)

$$(\mu_\G*G_{\r,\k})(X,\phi)\le O(1)
{L^{-\a}\over |X|}
\int_Xdx_1e^{-(\l_*/2)\left(1-{2\r\over L^\b}(1+O(1){\k\over\r}\l_*)\right)
L^{-\b}|\phi(x_1)|^2}2^{|X|}e^{4\k\Vert\phi\Vert^2_{X,1,\s}}\Eqa(A.9)$$

\\We have chosen $\k>0$, $O(1)$ in $L$, sufficiently small and
$0<\r<1/8$. We choose $\r\ge\k$. Then we get \equ(22.13) for 
$L$ sufficiently large.
So lemma 2.3.2 will have been proved provided we prove
the claim A.1.

\vglue.5truecm
\\{\it Proof of Claim A.1}

\\The proof is along the lines of that of Lemma 3 in [BDH2],
the only difference being that we have the ``covariance''
$\Sigma^{x_1}$ instead of the covariance $\G$.

\\Recall

$$\Sigma^{x_1}(x,y)=\G(x-y)-\l_*L^{-\b}\G(x-x_1)\G(y-x_1)\Eqa(A.10)$$

\\Define for $t\in[0,1]$

$${\cal G}_t(X,\phi)=e^{U_t(X,\phi)}$$

\\where

$$U_t(X,\phi)=t\ln 2|X|+4\k(1+t)\Vert\phi\Vert^2_{X,1,\s}\Eqa(A.11)$$

\\We have to prove

$$\int_0^1ds{\dpr\over\dpr s}\left(\m_{(1-s)\Sigma^{x_1}}*{\cal G}_s\right)
(X,\phi)\ge 0$$

\\It is sufficient to prove that the integrand is non-negative.

\\Thus we have to prove, for $s\in[0,1]$

$${\dpr\over\dpr s}U_s-{1\over 2}\D_{\Sigma^{x_1}}U_s
-{1\over 2}\Sigma^{x_1}\left({\dpr U_s\over\dpr \phi},
{\dpr U_s\over\dpr \phi}\right)\ge 0\Eqa(A.12)$$

\\Here

$$\D_{\Sigma^{x_1}}U_s=\int_Xdx\int_Xdy\Sigma^{x_1}(x,y)
{\d\over\d\phi(x)}{\d\over\d\phi(y)}U_s=$$
$$=\D_\G U_s-\l_*L^{-\b}\left(\int_Xdx\G(x-x_1)
{\d\over\d\phi(x)}\right)^2U_s
\Eqa(A.13)$$

\\and

$$\Sigma^{x_1}\left({\dpr U_s\over\dpr \phi},
{\dpr U_s\over\dpr \phi}\right)
=\int_Xdx\int_Xdy\Sigma^{x_1}(x,y)
{\d U_s\over\d\phi(x)}{\d U_s\over\d\phi(y)}=$$
$$=\G \left({\dpr U_s\over\dpr \phi},
{\dpr U_s\over\dpr \phi}\right)-\l_*L^{-\b}\left(\int_Xdx\G(x-x_1)
{\d U_s\over\d\phi(x)}\right)^2
\Eqa(A.14)$$

$${\dpr U_s\over\dpr s}=\ln 2|X|+4\k\Vert\phi\Vert^2_{X,1,\s}\Eqa(A.15)$$

\\We have

$$\left|{1\over 2}\D_\G U_s\right|\le O(1)\k|X|\Eqa(A.16)$$

\\where $O(1)$ is independent of $L$. The latter follows from the fact
that the Sobolev norm starts with one derivative and lemma 1.1.1.

$$\left(\int_Xdx\G(x-x_1)
{\d\over\d\phi(x)}\right)^2U_s\le 8\k(1+s)\Vert\G(\cdot,x_1)\Vert^2_{X,1,\s}
\le O(1)\k\Eqa(A.17)$$

\\with  $O(1)$ independent of $L$, by \equ(A.7). Using \equ(A.16)
and \equ(A.17), we get from \equ(A.13)

$$\left|{1\over 2}\D_{\Sigma^{x_1}}U_s\right|\le O(1)\k|X|\Eqa(A.18)$$

\\It is easy to see that

$$\left|\G \left({\dpr U_s\over\dpr \phi},
{\dpr U_s\over\dpr \phi}\right)\right|\le
32\k^2\sum_{1\le\a_1,\a_2\le\s}
\int_Xdy|\dpr^{\a_1}\phi(y)|\ 
|((\dpr^{\a_1+\a_2}\G)*\dpr^{\a_2}\phi)(y)|\le$$

$$\le
32\k^2\sum_{1\le\a_1,\a_2\le\s}\Vert\dpr^{\a_1}\phi\Vert_{L^2(X)}
\Vert(\dpr^{\a_1+\a_2}\G)*\dpr^{\a_2}\phi\Vert_{L^2(X)}\le$$

$$\le 32\k^2\left(\sup_{2\le j\le 2\s}\Vert\dpr^j\G\Vert_{L^1({\bf R})}
\right)
\left(\sum_{1\le\a\le\s}\Vert\dpr^{\a}\phi\Vert_{L^2(X)}\right)^2\le$$

$$\le O(1)\k^2\left(\sup_{2\le j\le 2\s}\Vert\dpr^j\G\Vert_{L^1({\bf R})}
\right)\Vert\phi\Vert^2_{X,1,\s}\Eqa(A.18.1)$$

\\where to pass to the next from the last line we have used
Young's convolution inequality. Now, using the compact support of the kernel
function $u$

$$\sup_{2\le j\le 2\s}\Vert\dpr^j\G\Vert_{L^1({\bf R})}\le
\sup_{2\le j\le 2\s}\int_1^L{dl\over l}l^{\b-j}
\int_{-l}^ldx\ |(\dpr^ju)(x/l)|\le$$
$$\le 2\sup_{2\le j\le 2\s}\int_1^L{dl\over l}l^{\b-j+1}
\Vert\dpr^ju\Vert_\io\le O(1)\Eqa(A.18.2)$$

\\where $O(1)$ is independent on $L$.

\\From \equ(A.18.1) and \equ(A.18.2) we get

$$\left|\G \left({\dpr U_s\over\dpr \phi},
{\dpr U_s\over\dpr \phi}\right)\right|\le
O(1)\k^2\Vert\phi\Vert^2_{X,1,\s}\Eqa(A.19)$$

\\Next we have

$$\left(\int_Xdx\G(x-x_1)
{\d U_s\over\d\phi(x)}\right)^2\le
32\k^2\left(\sum_{1\le\a\le\s}\int_Xdx
\dpr^\a\G(x-x_1)\dpr^\a\phi(x)\right)^2\le$$
$$\le 32\k^2\left(\sum_{1\le\a\le\s}
\left(\int_Xdx(\dpr^\a\G(x-x_1))^2\right)^{1/2}
\left(\int_Xdx|\dpr^\a\phi(x)|^2\right)^{1/2}\right)^2\le$$
$$\le O(1)\k^2\left(\sup_{1\le\a\le\s}\int_{\bf R}(\dpr^\a\G(x))^2\right)
\Vert\phi\Vert^2_{X,1,\s}\le$$
$$\le O(1)\k^2\Vert\phi\Vert^2_{X,1,\s}\Eqa(A.20)$$

\\where $O(1)$ is independent on $L$ and we have used lemma 1.1.1.
Using \equ(A.19) and \equ(A.20) we get from \equ(A.14)

$$\left| \Sigma^{x_1}\left({\dpr U_s\over\dpr \phi},
{\dpr U_s\over\dpr \phi}\right)\right|
\le O(1)\k^2\Vert\phi\Vert^2_{X,1,\s}\Eqa(A.21)$$

\\Putting together the estimates \equ(A.15), \equ(A.18) and
\equ(A.21) we get

$${\dpr\over\dpr s}U_s-{1\over 2}\D_{\Sigma^{x_1}}U_s
-{1\over 2}\Sigma^{x_1}\left({\dpr U_s\over\dpr \phi},
{\dpr U_s\over\dpr \phi}\right)\ge$$
$$\ge(\ln 2-O(1)\k)|X|+(4\k-O(1)\k^2)\Vert\phi\Vert^2_{X,1,\s}
\ge 0\Eqa(A.22)$$

\\for $\k>0$ small enough and independent of $L$, since
the $O(1)$ are independent of $L$. This completes the proof of
the claim, and hence of lemma 2.3.2.
$$\eqno Q.E.D.$$
\vglue2.truecm

\pagina
\\{\bf References}

\vglue.3truecm
\\[AL] J.A. Aronowitz and T.C. Lubensky, Europhys Lett. {\bf 4}, 395 (1987)

\vglue.3truecm
\\[B] D. Brydges, Functional integrals and their applications, Technical
report, Ecole Polytechnique Federale de Lausanne (1992).

\vglue.3truecm
\\[BDH1] D. Brydges, J. Dimock and T.R. Hurd, Comm. Math. Phys., {\bf 172},
143-186 (1995).

\vglue.3truecm
\\[BDH2] D. Brydges, J. Dimock and T.R. Hurd, Estimates on Renormalization
Group Transformation, Preprint (1996)

\vglue.3truecm
\\[BDH3] D. Brydges, J. Dimock and T.R. Hurd, Comm. Math. Phys., {\bf 198}, 
111-156 (1998)

\vglue.3truecm
\\[BY] D. Brydges and H.T. Yau, Comm. Math. Phys., {\bf 129}, 351-392 (1990).

\vglue.3truecm
\\[CM] M. Cassandro and P.K. Mitter, Nucl. Phys., {\bf B422}, 634-674
(1994).

\vglue.3truecm
\\[D1] B. Duplantier, Phys. Rev. Lett. {\bf 58} (1987) 2733

\vglue.3truecm
\\[D2] B. Duplantier, Phys. Rev. Lett. {\bf 62} (1989) 2337

\vglue.3truecm
\\[DDG1] F. David, B. Duplantier and E. Guitter, Phys. Rev. Lett.
{\bf 70}, 2205 (1993).

\vglue.3truecm
\\[DDG2] F. David, B. Duplantier and E. Guitter, Nucl. Phys.
{\bf B394}, 555 (1993).

\vglue.3truecm
\\[DDG3] F. David, B. Duplantier and E. Guitter, Phys. Rev. Lett.
{\bf 72}, 311 (1994).

\vglue.3truecm
\\[DDG4] F. David, B. Duplantier  and E. Guitter,  
Renormalization theory for  
the self-avoiding polymerized  membranes,  cond-mat 9702136 (1997).

\vglue.3truecm
\\[DH1] J. Dimock and T.R. Hurd, J. Stat. Phys,
{\bf 66}, 1277-1318 (1992).

\vglue.3truecm
\\[DH2] J. Dimock and T.R. Hurd, 
Comm. Math. Phys., {\bf 156}, 547-580 (1993).

\vglue.3truecm
\\[DHK] B. Duplantier, T Hwa and  M. Kardar, Phys. Rev. Lett.
{\bf 64}, 2022 (1990).

\vglue.3truecm
\\[DW1] K.J. Wiese and F. David, Nucl. Phys. {\bf B 487}, 529 (1997)

\vglue.3truecm
\\[DW2] F. David and K.J. Wiese, Phys. Rev. Lett. {\bf 76}, 4564 (1996)

\vglue.3truecm
\\[DW3] K.J. Wiese and F. David, Nucl. Phys. {\bf B 450}, 495 (1995)

\vglue.3truecm
\\[DW4] F. David and K.J. Wiese, Nucl. Phys. {\bf B 535}, 555 (1998)
\vglue.3truecm

\\[GK] K. Gawedski and A. Kupiainen, Nucl. Phys., 
{\bf B262}, 33 (1985).

\vglue.3truecm
\\[H] T.Hwa, Phys.Rev., {\bf A 41}, 2733 (1987)

\vglue.3truecm
\\[KN] M. Kardar and D.R. Nelson, Phys. Rev. Lett. {\bf 58} 1298,
2280 (1987) 

\vglue.3truecm
\\[KW] J.Kogut and K.G. Wilson, Phys. Rep., {\bf 12}, 75 (1974).

\vglue.3truecm
\\[NPW] D.R. Nelson, T. Piron and S. Weinberg, eds., Statistical
mechanics of membranes and surfaces (World Scientific, Singapore, 1989)
\ciao